\def\figsiz{8.5cm}

\def\rn{}
\def\nn#1 #2{#2. #1}				
\def\nnn#1 #2 #3{#2. #3. #1}			
\def\nnnn#1 #2 #3 #4{#2. #3. #4 #1}		
\def\nnnnn#1 #2 #3 #4 #5{#2. #3. #4 #5. #1}	

\def\rf#1;#2;#3;#4;#5 {{\frenchspacing\par\rn#1, #3 {\bf #4}, #5 (#2). \par}}
\def\rg#1;#2;#3;#4;#5;#6 {{\frenchspacing\par\rn#1, #3 {\bf #4}, #5 (#2). \par}}
\def\rfbook#1;#2;#3;#4;#5 {{\frenchspacing\par\rn#1, {\it #3} (#5, #4, #2).\par}}
\def\rfprep#1;#2;#3 {{\par\frenchspacing\rn#1, #3 (#2).\par}}
\def\rfproc#1;#2;#3;#4;#5;#6 {{\frenchspacing\par\rn#1 #2, in {\it #3}, ed. #4 (#5: #6)\par}}
\def\rfprocp#1;#2;#3;#4;#5;#6;#7 {{\frenchspacing\par\rn#1 #2, in {\it #3}, ed. #4 (#5: #6), p#7\par}}

\def\rg#1;#2;#3;#4;#5;#6 {\par\rn#1 #2, {\it #3}, {\bf #4}, #5 (``#6'') \par}
\def\rf#1;#2;#3;#4;#5 {\par\rn#1, {\it #3}, {\bf #4}, #5 (#2)\par}
\def\rfbook#1;#2;#3;#4;#5 {{\frenchspacing\par\rn#1, {\it #3} (#4: #5, #2)\par}}
\def\rfproc#1;#2;#3;#4;#5;#6 {{\frenchspacing\par\rn#1 #2, in {\it #3}, ed. #4 (#5: #6)\par}}
\def\rfprocp#1;#2;#3;#4;#5;#6;#7 {{\frenchspacing\par\rn#1 #2, in {\it #3}, ed. #4 (#5: #6), p#7\par}}
\def\rfprep#1;#2;#3  {{\par\rn#1, #3, #2\par}}
\def\rfprepp#1;#2;#3 {{\par\rn#1 #2, #3\par}}

\def\etal{{\frenchspacing\it et al.}}
\def\ie{i.e.\ }
\def\eg{{\frenchspacing\it e.g.}}

\def\ben{\begin{equation}}
\def\een{\end{equation}}
\def\beq#1{\begin{equation}\label{#1}}
\def\eeq{\end{equation}}
\def\bena{\begin{eqnarray}}
\def\eena{\end{eqnarray}}
\def\beqa#1{\begin{eqnarray}\label{#1}}
\def\eeqa{\end{eqnarray}}

\def\Eq#1{Eq.~(\ref{#1})}

\def\Eqnopar#1{Eq.~\ref{#1}}

\def\fig#1{Figure~\ref{#1}}
\def\Fig#1{Figure~\ref{#1}}

\def\Sec#1{Section~\ref{#1}}

\def\spose#1{\hbox to 0pt{#1\hss}}
\def\simlt{\mathrel{\spose{\lower 3pt\hbox{$\mathchar"218$}}
     \raise 2.0pt\hbox{$\mathchar"13C$}}}
\def\simgt{\mathrel{\spose{\lower 3pt\hbox{$\mathchar"218$}}
     \raise 2.0pt\hbox{$\mathchar"13E$}}}
\def\simpropto{\mathrel{\spose{\lower 3pt\hbox{$\mathchar"218$}}
     \raise 2.0pt\hbox{$\propto$}}}

\def\ed{\end{document}}

\def\beq#1{\begin{equation}\label{#1}}
\def\eeq{\end{equation}}
\def\beqa#1{\begin{eqnarray}\label{#1}}
\def\eeqa{\end{eqnarray}}

\def\Eq#1{Eq.~(\ref{#1})}

\def\x{{\bf x}}

\def\I{{\bf I}}
\def\J{{\bf J}}
\def\R{{\bf R}}

\def\R{{\bf R}}

\def\ignore#1{}

\def\simless{\mathbin{\lower 3pt\hbox
        {$\,\rlap{\raise 5pt\hbox{$\char'074$}}\mathchar"7218\,$}}} 
\def\simgreat{\mathbin{\lower 3pt\hbox
        {$\,\rlap{\raise 5pt\hbox{$\char'076$}}\mathchar"7218\,$}}} 

\documentclass[twocolumn,showpacs,amsmath,nofootinbib,superscriptaddress,smartcite]{revtex4} 
\usepackage{url}
\usepackage{graphicx}
\usepackage{hyperref}
\usepackage{hypernat}

\begin{document}
\input{epsf.sty}

\def\mitctp{1}
\def\mitastro{2}

\def\affilmrk#1{$^{#1}$}
\def\affilmk#1#2{$^{#1}$#2}

\title{Constraining Torsion with Gravity Probe B\footnote{This is the ``director's cut'' version of the article published in Phys.~Rev.~D November Issue, including extra bonus derivations in sections $\S$II, $\S$V and Appendix C.}}

\author{Yi Mao}
\affiliation{Dept.~of Physics, Massachusetts Institute of Technology, Cambridge, MA 02139}
\author{Max Tegmark}
\affiliation{Dept.~of Physics, Massachusetts Institute of Technology, Cambridge, MA 02139}
\affiliation{MIT Kavli Institute for Astrophysics and Space Research, Cambridge, MA 02139}
\author{Alan H.~Guth}
\affiliation{Dept.~of Physics, Massachusetts Institute of Technology, Cambridge, MA 02139}
\author{Serkan Cabi}
\affiliation{Dept.~of Physics, Massachusetts Institute of Technology, Cambridge, MA 02139}

\date{Submitted to Phys.~Rev.~D. 1/8-07; Revised 9/18-07; Accepted 9/27-07}

\begin{abstract}
It is well-entrenched folklore that 
all torsion gravity theories predict observationally 
negligible torsion in the solar system, since torsion (if it exists) 
couples only to the intrinsic spin of elementary particles, not to rotational angular momentum.  
We argue that this assumption has a logical loophole which can and should be tested experimentally, 
and consider non-standard torsion theories in which torsion can be generated by macroscopic rotating objects.
In the spirit of action$=$reaction, if a rotating mass like a planet can generate torsion, then a gyroscope would be expected to feel torsion.  
An experiment with a gyroscope (without nuclear spin) such as Gravity Probe B (GPB) can test theories where this is the case.

Using symmetry arguments, we show that to lowest order, 
any torsion field around a uniformly rotating spherical mass is determined by seven dimensionless
parameters. These parameters effectively generalize the PPN formalism and
provide a concrete framework for further testing GR. 
We construct a parametrized Lagrangian that includes both 
standard torsion-free GR and Hayashi-Shirafuji maximal torsion gravity as 
special cases. We demonstrate that classic solar system tests rule out the latter and constrain two observable parameters.
We show that Gravity Probe B is an ideal experiment for further constraining  
non-standard
torsion theories, and work out the most general torsion-induced precession of its gyroscope in terms of our torsion parameters.

\end{abstract}


\maketitle

\setcounter{footnote}{0}

\section{Introduction}

Einstein's General Theory of Relativity (GR) has emerged as the hands down most popular candidate for a relativistic theory of 
gravitation, owing both to its elegant structure and to its impressive agreement with a host of experimental tests 
since it was first proposed about ninety years ago \cite{Will:2005yc,will2,Will:2005va}. 
Yet it remains worthwhile to subject GR to further tests whenever possible, since these can either build further confidence in the theory or
uncover new physics. Early efforts in this regard focused on weak-field solar system tests, and efforts to test GR have since been extended
to probe stronger gravitational fields involved in binary compact objects, black hole accretion and cosmology \cite{Hulse:1974eb,Weisberg:2002qg,Weisberg:2004hi,Champion:2004hc,Peters:1963ux,Stairs:1997kz,Stairs:1999dv,Hotan:2004ua,Hotan:2004ub,vanStraten:2001zk,Armitage:2004ga,Chakrabarti:2004uz,Merloni:2002gx,Menou:2001ga,Peldan:1993hi,Sotiriou:2006hs,Sotiriou:2006qn,Sotiriou:2005cd,Akbar:2006er,Koivisto:2006ie,Amarzguioui:2005zq,Hwang:2001pu,Meng:2003en,Esposito-Farese:1999pa,Esposito-Farese:2004cc,Damour:1996xx,Puetzfeld:2004yg,Kasper:1994xv,Biswas:1999fa,Mukherjee:2005zt,Beesham:1987gd,Mahato:2006gi}.

\subsection{Generalizing general relativity}

\begin{figure}
\centerline{\epsfxsize=\figsiz\epsffile{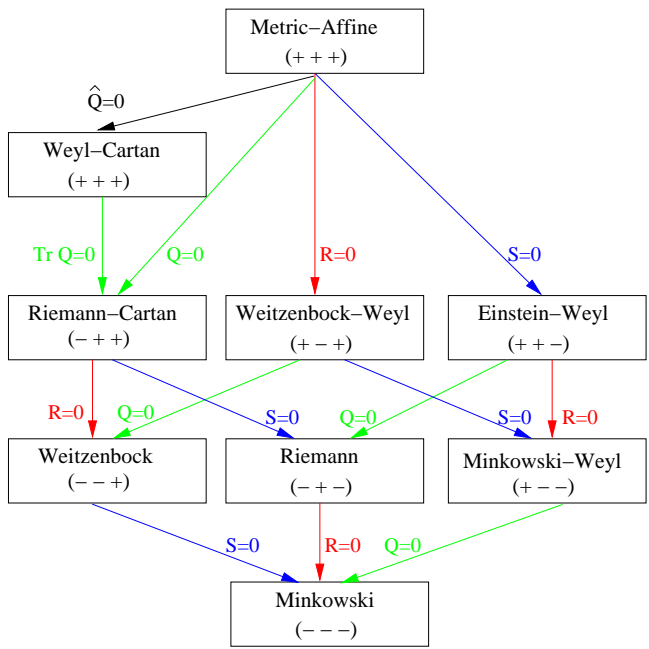}}
\caption[1]{\label{fig:spaces}\footnotesize%
Classification of spaces (Q,R,S) and the reduction flow.  Metric-Affine spacetime is a manifold endowed with Lorentzian
metric and linear affine connection without any restrictions.  All spaces below it except the Weyl-Cartan space are special cases obtained
from it by imposing three
types of constraints: vanishing non-metricity tensor $Q_{\mu\nu\rho}$ ($Q$ for short), 
vanishing Riemann curvature tensor
$R_{\mu\nu\rho\sigma}$ ($R$ for short), or vanishing torsion tensor 
$S_{\mu\nu}^{\phantom{01}\rho}$ ($S$ for short).  A plus
sign in a parenthesis indicates a non-vanishing quantity from the set $(Q,R,S)$, and a minus sign a vanishing quantity.  
For example, Riemann
spacetime $(-+-)$ means that $Q=S=0$ but $R\ne 0$.  Weyl-Cartan space is a Metric-Affine space with vanishing 
``tracefree nonmetricity'' $\hat{Q}_{\mu\nu\rho}$ ($\hat{Q}$ for short), 
defined by $\hat{Q}_{\mu\nu\rho}\equiv Q_{\mu\nu\rho}-\frac{1}{4} ({\rm tr}\>Q)_\mu g_{\nu\rho}$.  
The trace of the nonmetricity is defined by $({\rm tr}\>Q)_{\mu}\equiv g^{\nu\rho}Q_{\mu\nu\rho}$; thus $\hat{Q}$ automatically satisfies that $({\rm tr}\hat{Q})_{\mu}=0$ (tracefree).  
Subsets of the classification scheme are shown in Fig.~2 of \cite{Hehl:1994ue}, Fig.~1 of \cite{Puetzfeld:2004yg} 
and Fig.~5 of \cite{Gronwald:1995em}.  Among the terms, {\it Einstein-Weyl, Weitzenb\"ock} and  {\it Minkowski} spaces are standard, 
{\it Metric-Affine, Weyl-Cartan, Riemann-Cartan} and {\it Riemann} spaces follow \cite{Hehl:1994ue}, and we here introduce the 
terms {\it Weitzenb\"ock-Weyl} and {\it Minkowski-Weyl} space by symmetry.}
\end{figure}

The arguably most beautiful aspect of GR is that it 
geometrizes gravitation, with Minkowski spacetime being deformed by the matter (and energy) inside it.
As illustrated in \Fig{fig:spaces}, for the most general manifold with a metric $g$ and a connection $\Gamma$, departures from Minkowski
space are characterized by three geometrical entities: non-metricity ($Q$), curvature ($R$) and torsion ($S$), defined as follows:
\bena
Q_{\mu\nu\rho}&\equiv&\nabla_{\mu} g_{\nu\rho}\,,\label{eqn:nonmetricity}\\
R^{\rho}_{\phantom{1}\lambda\nu\mu} &\equiv& \Gamma^{\rho}_{\phantom{1}\mu\lambda,\nu}-\Gamma^{\rho}_{\phantom{1}\nu\lambda,\mu}+\Gamma^\rho_{\phantom{1}\nu\alpha}\Gamma^\alpha_{\phantom{1}\mu\lambda}-\Gamma^\rho_{\phantom{1}\mu\alpha}\Gamma^\alpha_{\phantom{1}\nu\lambda}\,,\\
S_{\mu\nu}^{\phantom{\mu\nu}\rho}&\equiv&\frac{1}{2}(\Gamma^{\rho}_{\phantom{\rho}\mu\nu}-\Gamma^{\rho}_{\phantom{\rho}\nu\mu})\,.\label{TorsionDefEq}
\eena
GR is the special case where the non-metricity and torsion are assumed to vanish identically ($Q=S=0$, i.e., Riemann spacetime), 
which determines the connection in terms of the metric and leaves 
the metric as the only dynamical entity. 
However, as \Fig{fig:spaces} illustrates, this is by no means the only possibility, and many alternative geometric gravity
theories have been discussed in the literature \cite{Blagojevic:book,hammond,Gronwald:1995em,Hehl:1997bz,DeAndrade,aldrovandi,hehl,nester,watanabe,capozziello,Gasperini:1986mv,Shapiro:2001rz,Sotiriou:2006qn,Baekler:2006de,Poltorak:2004tz,Mielke:2004gg,Kleyn:2004yj,Vassiliev:2003dk,Minkevich:2003it,Obukhov:2002tm,Tresguerres:1995un,King:2000ha,Gronwald:1997bx,Hehl:1999sb,Poberii:1994rz,Hehl:1994ue,Lord:1978qz,Hehl:1976my,Rajaraman:2003st,Vollick:2003ic,Chiba:2003ir,Gruver:2001tt,Berthias:1993aa,Lam:2002ve,Saa:1993fx,Antoniadis:1992ep,Bytsenko:1993qn,Fabbri:2006xq,Carroll:1994dq} corresponding to 
alternative deforming geometries where other subsets of $(Q,R,S)$ vanish. Embedding GR in a broader parametrized class of theories allowing 
non-vanishing torsion and non-metricity, and experimentally constraining these parameters would provide a natural generalization of
the highly successful parametrized post-Newtonian (PPN) program for GR testing, which assumes vanishing torsion \cite{Will:2005yc,will2,Will:2005va}.

For the purposes of this paper, a particularly interesting generalization of Riemann spacetime is Riemann-Cartan Spacetime (also known as $U_4$),
which retains $Q=0$ but is characterized by non-vanishing torsion. In $U_4$,
torsion can be dynamical and consequently play a role in gravitation alongside the metric. 
Note that gravitation theories including torsion retain what are often regarded as the most beautiful aspects of
General Relativity, \ie general covariance and the idea that ``gravity is geometry''. 
Torsion is just as geometrical an entity as curvature, and torsion theories can be consistent with the Weak Equivalence 
Principle (WEP).

\subsection{Why torsion testing is timely}

Experimental searches for torsion have so far been rather limited \cite{hammond}, in part because most published torsion theories predict
a negligible amount of torsion in the solar system. First of all, many torsion Lagrangians imply that torsion is related to its source 
via an algebraic equation rather than via a differential equation, so that (as opposed to curvature), torsion must vanish in vacuum. 
Second, even within the subset of torsion theories where torsion propagates and can exist in vacuum, it is usually assumed that it
couples only to intrinsic spin, not to rotational angular momentum \cite{hehl,sy1,sy2}, and is therefore negligibly small far from extreme objects
such as neutron stars. This second assumption also implies that even if torsion were present in the solar system, 
it would only affect particles with intrinsic spin (\eg\ a
gyroscope with net magnetic polarization) \cite{sy1,sy2,NSH,Hojman,Hojman2,cognola,kopczynski,pereira1}, while having no influence on the precession of a gyroscope 
without nuclear spin
\cite{sy1,sy2,NSH} such as a gyroscope in Gravity Probe B.  

Whether torsion does or does not satisfy these pessimistic assumptions depends on what the Lagrangian is, which is of course one of the
things that should be tested experimentally rather than assumed. 
Taken at face value, the Hayashi-Shirafuji Lagrangian \cite{HS1} 
provides an explicit counterexample to both assumptions, with even a static massive body generating a torsion field --- indeed, 
such a strong one that the gravitational forces are due entirely to torsion, not to curvature.
As another illustrative example, we will develop in
\Sec{subsec:linearinterpol} a family of tetrad theories in Riemann-Cartan space which linearly interpolate
between GR and the Hayashi-Shirafuji theory.  
Although these particular Lagrangeans come with important caveats to which we return below (see also \cite{Flanagan:2007dc}),
they show that one cannot dismiss out of hand the possibility that angular momentum sources non-local torsion
(see also Table \ref{tab:torsion-theory}).
Note that the proof\cite{sy1,sy2,NSH} of the oft-repeated assertion that a
gyroscope without nuclear spin cannot feel torsion crucially relies on the assumption that orbital angular momentum cannot be
the source of torsion. This proof is therefore not generally applicable in the context of non-standard torsion theories. 

More generally, in the spirit of action$=$reaction, if a (non-rotating or rotating) mass like a planet can generate torsion, then a gyroscope
without nuclear spin could be expected feel torsion, 
so the question of whether a non-standard gravitational Lagrangian causes 
torsion in the solar system is one which can and should be addressed experimentally.  

\begin{table*}
\noindent 
\footnotesize{
\begin{center}
\begin{tabular}{lcccc|p{4.5cm}}
\hline
Theory & Dynamical DOF & Vacuum & Source & Ref. & Notes \\ \hline\hline

$U_4$ theory & $g_{\mu\nu}$, $S_{\mu\nu}^{\phantom{12}\rho}$ & N & Spin & \cite{hehl} & \\ \hline
Pagels theory & $O(5)$ gauge fields $\omega_{\mu}^{\phantom{1}AB}$  & N & Spin & \cite{Pagels:1983pq} & an $O(5)$ gauge theory of gravity \\ \hline 
Metric-affine gravity & general gauge fields & P & Spin & \cite{Hehl:1994ue} & gauge theory of gravity in the metric-affine space\\ \hline
Stelle-West & $SO(3,2)$ gauge fields $\omega_{\mu}^{\phantom{1}AB}$ & P & Spin, Gradient of the Higgs field & \cite{Stelle:1979aj} & a $SO(3,2)$ gauge theory of gravity spontaneously broken to $SO(3,1)$ \\ \hline 
Hayashi-Shirafuji & tetrads $e^{\,k}_{\,\phantom{1}\mu}$ & P & Spin, Rotational & \cite{HS1} & a theory in Weitzenb\"ock space\\ \hline 
Einstein-Hayashi-Shirafuji & tetrads $e^{\,k}_{\,\phantom{1}\mu}$ & P & Spin, Rotational & This paper & a class of theories in Riemann-Cartan space \\ \hline
Teleparallel gravity & tetrads $e^{\,k}_{\,\phantom{1}\mu}$ & P & Spin, Rotational & \cite{DeAndrade,aldrovandi} &  \\ \hline 

\end{tabular}
\caption{A short list of torsion theories of gravity.  The ``DOF'' in the second column is short for ``degrees of freedom''.  In the column \emph{Vacuum}, ``N'' refers to non-propagating torsion in the vacuum while ``P'' means propagating torsion.  In the column \emph{Source}, ``spin'' refers to intrinsic spin while ``rotational'' means rotational angular momentum.}
\label{tab:torsion-theory}
\end{center}
}
\end{table*}

This experimental question is timely because the Stanford-led
gyroscope satellite experiment, Gravity Probe B\footnote{http://einstein.stanford.edu/} (GPB), was launched in April 2004 and has successfully been taking 
data.
Preliminary GPB results, released in April 2007, have confirmed the geodetic precession to better than 1\%, 
and the full results, which are highly relevant to this paper, are due to be released soon. 
GPB contains a set of four extremely spherical gyroscopes and flies in a circular polar orbit with altitude 640 kilometers, and we will show that it has the potential
to severely constrain a broad class of previously allowed torsion theories.
GPB was intended to test the GR prediction \cite{schiff,Will:2002ma,Adler:1999yt,biemond:2004,Ashby:1990,Barker:1970zr} that a gyroscope in this orbit precesses about 6,614.4 milli-arcseconds per year around its orbital angular momentum vector 
(geodetic precession) and about 40.9 milli-arcseconds per year about Earth's angular momentum vector (frame-dragging)\footnote{These numerical
precession rates are taken from the GPB website.}.  Most impressively, GPB should convincingly observe the frame-dragging effect, an arguably still undetected
effect of the off-diagonal metric elements that originate from the rotation of Earth. Of particular interest to us is that GPB can reach a
precision of $0.005\%$ for the geodetic precession, which as we will see enables precision discrimination\footnote{GPB 
also has potential for constraining other GR extensions \cite{Moffat:2004cv} than those we consider in this paper.}  
between GR and a class of torsion theories.  

\subsection{How this paper is organized}

In general, torsion has 24 independent components, each being a function of time and position. Fortunately, symmetry arguments and a perturbative expansion
will allow us to greatly simplify the possible form of any torsion field of Earth, a nearly
spherical slowly rotating massive object. We will show that the most general possibility can be elegantly parametrized by merely seven 
numerical constants to be constrained experimentally. We then derive the effect of torsion on 
the precession rate of a gyroscope in Earth orbit and work out how the anomalous precession that GPB would register depends on these seven parameters.

The rest of this paper is organized as follows.  In Section \ref{sec:u4}, we review the basics of Riemann-Cartan spacetime. In \Sec{sec:param}, we derive
the results of parametrizing the torsion field around Earth. 
In \Sec{sec:eom}, we discuss the equation of motion for the precession of a gyroscope and the world-line of its center of mass.  
We use the results to calculate the instantaneous precession rate in \Sec{sec:instan-precess}, 
and then analyze the Fourier moments for the particular orbit of GPB in \Sec{sec:mom-ana}.  
In \Sec{sec:general-torsion-constraints}, we show that GPB can constrain two linear combinations of 
the seven torsion parameters, given the constraints on the PPN parameters $\gamma$ and $\alpha_1$ from other solar system tests. 
To make our discussion less abstract, we study Hayashi-Shirafuji torsion gravity as an 
explicit illustrative example of an alternative gravitational theory that can be tested within our framework.
In \Sec{sec:counterexample}, we review the basics of Weitzenb\"ock spacetime and
Hayashi-Shirafuji theory, and then give the torsion-equivalent of the linearized Kerr solution.  
In \Sec{subsec:linearinterpol}, we generalize 
the Hayashi-Shirafuji theory to a two-parameter family of gravity theories, 
which we will term Einstein-Hayashi-Shirafuji (EHS) theories, interpolating between torsion-free GR and the Hayashi-Shirafuji 
maximal torsion theory. In \Sec{sec:constrain-torsion}, we apply the precession rate results to the EHS theories and discuss the observational 
constraints that GPB, alongside other solar system tests, will be able to place on the parameter space of the family of EHS theories.   
We conclude in \Sec{sec:conclusion}.  Technical details of torsion parametrization (\ie \Sec{sec:param}) are given in 
Appendices \ref{appendix:spher-symm} \& \ref{appendix:axisymm}. Derivation of solar system tests are given in Appendix \ref{appendix:solar-tests}. 
We also demonstrate in Appendix \ref{appendix:photon-mass} that current ground-based experimental upper bounds on the photon mass do not place more
stringent constraints on the torsion parameters $t_1$ or $t_2$ than GPB will.

After the first version of this paper was submitted, Flanagan and Rosenthal showed that the Einstein-Hayashi-Shirafuji Lagrangian 
has serious defects \cite{Flanagan:2007dc}, while leaving open the possibility that there may be other viable Lagrangians
in the same class (where spinning objects generate and feel propagating torsion).
The EHS Lagrangian should therefore not be viewed as a viable physical model, but as a pedagogical toy model 
giving concrete illustrations of the various effects and constraints that we discuss.
 
Throughout this paper, we use natural gravitational units where $c=G=1$.  Unless we explicitly state otherwise, a Greek letter denotes
an index running from 0 to 3 and a Latin letter an index from 1 to 3. We use the metric signature convention $(- + + +)$.

\section{Riemann-Cartan spacetime}\label{sec:u4}

We review the basics of Riemann-Cartan spacetime only briefly here, and refer the interested reader to 
Hehl {\etal} \cite{hehl} for a more comprehensive discussion of spacetime with torsion. 
Riemann-Cartan spacetime is a connected $C^{\infty}$ four-dimensional
manifold endowed with metric $g_{\mu\nu}$ of Lorentzian signature and an affine 
connection $\Gamma^{\mu}_{\phantom{\mu}\nu\rho}$
such that the non-metricity defined by \Eq{eqn:nonmetricity} with respect to the full connection identically vanishes. 
In other words, the connection in Riemann-Cartan spacetime may have torsion, but it must still be 
compatible with the metric ($g_{\mu\nu;\lambda}=0$).
The covariant derivative of a vector is given by
\bena
\nabla_{\mu}V^{\nu} &=& \partial_{\mu}V^{\nu} + \Gamma^{\nu}_{\phantom{1}\mu\rho}V^{\rho}\,,\label{eqn:aff-conn-def1}\\
\nabla_{\mu}V_{\nu} &=& \partial_{\mu}V_{\nu} - \Gamma^{\rho}_{\phantom{1}\mu\nu}V_{\rho}\,,\label{eqn:aff-conn-def2}
\eena
where the first of the lower indices on $\Gamma^{\lambda}_{\phantom{1}\mu\sigma}$ always corresponds to the index on $\nabla_\mu$.

The full connection has 64 independent components.  The condition of vanishing non-metricity $\nabla_{\mu} g_{\nu\rho} = 0$ 
gives 40 constraints, and the remaining 24 components are the degrees of freedom of the torsion tensor.  

In the more familiar case of Riemann spacetime, the two conditions $S_{\mu\nu}^{\phantom{\mu\nu}\rho}=0$ and $Q_{\mu\nu\rho}=0$ imply that
the connection must be the so-called Levi-Civita connection (Christoffel symbol), uniquely determined by the metric as 
\begin{equation} 
\left\{ \begin{array}{c} \rho \\ \mu\nu \end{array} \right\} =\frac{1}{2} g^{\rho\lambda}(\partial_{\mu}g_{\nu\lambda}+\partial_\nu g_{\mu\lambda}-\partial_{\lambda}g_{\mu\nu})\,.
\end{equation}
In the more general case when torsion is present, the connection must depart from the Levi-Civita connection in order to be metric-compatible 
($\nabla_{\mu} g_{\nu\rho} = 0$), 
and this departure is (up to a historical minus sign) called the \emph{contorsion}, defined as 
\beq{eqn:fullconn1}
K_{\mu\nu}^{\phantom{\mu\nu}\rho} \equiv  \left\{ \begin{array}{c} \rho \\ \mu\nu \end{array} \right\} - \Gamma^{\rho}_{\phantom{\rho}\mu\nu} \,.
\eeq
Using the fact that the torsion is the part of the connection that is antisymmetric in the first two indices (Eq.~\ref{TorsionDefEq}), one readily shows that 
\begin{equation}
K_{\mu\nu}^{\phantom{\mu\nu}\rho} =  -S_{\mu\nu}^{\phantom{\mu\nu}\rho}-S^\rho_{\phantom{\rho}\nu\mu}-S_{\phantom{\rho}\mu\nu}^{\rho}\,.
\end{equation}
In Riemann-Cartan spacetime, the metric is used to raise or lower the indices as usual. 

The curvature tensor is defined as usual, in terms of the full connection rather than the Levi-Civita connection:
\beq{eqn:curvature}
R^{\rho}_{\phantom{1}\lambda\nu\mu}= \partial_{\nu}\Gamma^{\rho}_{\phantom{1}\mu\lambda}-\partial_{\mu}\Gamma^{\rho}_{\phantom{1}\nu\lambda}+\Gamma^{\rho}_{\phantom{1}\nu\alpha}\Gamma^{\alpha}_{\phantom{1}\mu\lambda}-\Gamma^{\rho}_{\phantom{1}\mu\alpha}\Gamma^{\alpha}_{\phantom{1}\nu\lambda}\,.
\eeq
As in Riemann spacetime, one can prove that $R^{\rho}_{\phantom{1}\lambda\nu\mu}$ is a tensor by showing that for any vector $V^{\mu}$,
\ben
\nabla_{[\nu}\nabla_{\mu]}V^{\rho}=\frac{1}{2}R^{\rho}_{\phantom{1}\lambda\nu\mu}V^{\lambda}-S_{\nu\mu}^{\phantom{12}\alpha}\nabla_{\alpha}V^{\rho}\,.
\een
The Ricci tensor and Ricci scalar are defined by contraction the Riemann tensor just as in Riemann spacetime.

\section{Parametrization of the Torsion and Connection}\label{sec:param}

The torsion tensor has twenty-four independent components since it is antisymmetric in its first two
indices.  However, its form can be greatly simplified by the fact that Earth is well approximated as a uniformly
rotating spherical object. 
Throughout this paper, we will therefore Taylor expand all quantities with respect to the dimensionless mass parameter
\beq{lambdaM}
\varepsilon_m\equiv {m\over r}, 
\eeq
 and the dimensionless angular momentum parameter 
\beq{lambdaDefEq}
\varepsilon_a\equiv {a\over r}, 
\eeq
where $a\equiv J/m$ is the specific angular momentum , which has units of length, and $r$ is the distance of the field point from the central gravitating body.  Here $m$ and $J$ are Earth's mass and rotational angular momentum, respectively.  
Since Earth is slowly rotating ($\varepsilon_a\ll 1$), we will only need to keep track of zeroth and first order terms in $\varepsilon_a$.
We will also Taylor expand with respect to $\varepsilon_m$ to first order, since we are interested in objects with orbital radii vastly exceeding
Earth's Schwarzschild radius ($\varepsilon_m\ll 1$).\footnote{These two approximations $\varepsilon_m \ll 1$ and $\varepsilon_a \ll 1$ are highly
accurate for the GPB satellite in an Earth orbit with altitude about 640 kilometers:  $\varepsilon_m \simeq 6.3 \times 10^{-10}$ and
$\varepsilon_a \simeq 5.6 \times 10^{-7}$.}  All calculations will be to first order in $\varepsilon_m$, because to zeroth order in
$\varepsilon_m$, \ie in Minkowski spacetime, there is no torsion.  Consequently, we use the terms ``zeroth order'' and ``first order'' below
with respect to the expansion in $\varepsilon_a$.

We start by studying in section \ref{subsec:spher-symm} the zeroth order part: the static,
spherically and parity symmetric case where Earth's rotation is ignored.  The first correction will be treated in
section \ref{subsection:axisymm}: the stationary and spherically axisymmetric contribution caused by Earth's
rotation.  For each case, we start by giving the symmetry constraints that apply for \emph{any} quantity. We then give the most general 
parametrization of torsion and connection that is consistent with these symmetries, as derived in the appendices.  
The Kerr-like torsion solution of Hayashi-Shirafuji Lagrangian given in \Sec{sec:counterexample} is an explicit
example within this parametrized class. In \Sec{sec:instan-precess}, we will apply these results to the precession
of a gyroscope around Earth. 

\subsection{Zeroth order: the static, spherically and parity symmetric case}\label{subsec:spher-symm}

This is the order at which Earth's slow rotation is neglected ($\varepsilon_a=0$). For this, three convenient
coordinate systems are often employed -- isotropic rectangular coordinates, isotropic spherical coordinates, and standard
spherical coordinates. In the following, we will find it most convenient to work in isotropic rectangular coordinates 
to set up and solve the problem, and then transform the result to standard spherical coordinates.

\subsubsection{Symmetry Principles}\label{subsubsec:general-setup1}

Tetrad spaces with spherical symmetry have been studied by Robertson \cite{robertson} and Hayashi and
Shirafuji \cite{HS1}.  Our approach in this section essentially follows their work.

Given spherical symmetry, one can naturally find a class of isotropic rectangular coordinates $(t,x,y,z)$. Consider
a general quantity $\mathcal{O}(x)$ that may bear upper and lower indices. It may or may not be a tensor. In
either case, its transformation law $\mathcal{O}(x)\to\mathcal{O}\,'(x')$ under the general coordinate transformation
$x\to x'$ should be given. By definition, a quantity $\mathcal{O}$ is static, spherically and parity symmetric if it
has the \emph{formal functional invariance} 
\[
\mathcal{O}\,'(x') = \mathcal{O}(x')
\]
under the following coordinate transformations (note that $\mathcal{O}(x')$ denotes the original function 
$\mathcal{O}(x)$ evaluated at the coordinates $x'$):
\begin{enumerate}

\item{Time translation:} $t\to t'\equiv t+t_0$ where $t_0$ is an arbitrary constant.

\item{Time reversal:} $t\to t'\equiv -t$.

\item{Continuous rotation and space inversion:}
\beq{xform:spherical}
\x\to\x'\equiv\R\x\,,
\eeq
where $\R$ is any $3\times 3$ constant orthogonal ($\R^t\R=\I$) matrix.  
Note that the parity symmetry allows $\R$ to be an improper rotation.

\end{enumerate}

\subsubsection{Parametrization of torsion}\label{subsubsec:param-tor}

It can be shown (see Appendix A) that, under the above conditions, there are only two independent components
of the torsion tensor.  The non-zero torsion components can be parametrized in isotropic rectangular coordinates
as follows: 
\bena
S_{0i}^{\phantom{0i}0} &=& t_1\frac{m}{2r^3}x^i\, ,\label{eqn:t1}\\
S_{jk}^{\phantom{0i}i} &=& t_2\frac{m}{2r^3}(x^j \delta_{ki}-x^k \delta_{ji})\, ,\label{eqn:t2}
\eena
where $t_1$ and $t_2$ are dimensionless constants.  
It is of course only the two combinations $t_1 m$ and $t_2 m$ that correspond to the physical parameters;
we have chosen to introduce a third redundant quantity 
$m$ here, with units of mass, to keep $t_1$ and $t_2$
dimensionless. Below we will see that in the context of specific torsion Lagrangians,
$m$ can be naturally identified with the mass of the object generating the torsion, up to a numerical factor close to unity.

We call $t_1$ the ``anomalous geodetic torsion'' and $t_2$ the ``normal geodetic torsion'', because both
will contribute to the geodetic spin precession of a gyroscope, the former ``anomalously'' and the latter
``regularly'', as will become clear in \Sec{sec:instan-precess} and \ref{sec:mom-ana}.

\subsubsection{Torsion and connection in standard spherical coordinates}

In spherical coordinates, the torsion tensor has the following non-vanishing components:
\beq{eqn:torsion-spher-symm1}
S_{tr}^{\phantom{01}t}(r) = t_1\frac{m}{2r^2}\quad ,\quad S_{r\theta}^{\phantom{12}\theta}(r)=S_{r\phi}^{\phantom{13}\phi}(r) = t_2\frac{m}{2r^2}\, ,
\eeq
where $t_1$ and $t_2$ are the same torsion constants as defined above.

The above parametrization of torsion was derived in isotropic coordinates, but it is also valid in other 
spherical coordinates as far as the linear perturbation around the Minkowski spacetime is concerned.  
The decomposition formula (\Eqnopar{eqn:fullconn1}), derived from $\nabla_{\mu} g_{\nu\rho}=0$, enables one 
to calculate the full connection exactly. However, for that purpose the coordinates with a metric must be specified.  
In general, a spherically symmetric coordinate system has the line element \cite{MTW-Weinberg-Islam:book}
\[
\mathrm{d}s^2 = -h(r)\mathrm{d}t^2+f(r)\mathrm{d}r^2+\alpha(r)r^2\left[\mathrm{d}\theta^2+\sin^2\theta\mathrm{d}\phi^2\right]\,.
\]
There is freedom to rescale the radius, so-called isotropic spherical coordinates corresponding to the
choice $\alpha(r)=f(r)$. Throughout this paper, we make the common choice $\alpha(r)=1$, where $r$ can be interpreted as
$(2\pi)^{-1}$ times the circumference of a circle.
To linear order, 
\begin{eqnarray*}
h(r) &=& 1 + \mathcal{H}\frac{m}{r}\, ,\label{eqn:h}\\
f(r) &=& 1 + \mathcal{F}\frac{m}{r}\, ,\label{eqn:f}
\end{eqnarray*}
where $\mathcal{H}$ and $\mathcal{F}$ are dimensionless constants.

It is straightforward to show that, in the linear regime, the most general
connection that is static, spherically and parity symmetric in Riemann-Cartan spacetime with standard spherical coordinates is as follows:
\beqa{eqn2:conn}
\Gamma^{t}_{\phantom{0}tr} &=& \left(t_1-\frac{\mathcal{H}}{2}\right)\frac{m}{r^2}\,, \nonumber\\
\Gamma^{t}_{\phantom{0}rt} &=& -\frac{\mathcal{H}}{2}\frac{m}{r^2}\,, \nonumber\\
\Gamma^{r}_{\phantom{0}tt} &=& \left(t_1-\frac{\mathcal{H}}{2}\right)\frac{m}{r^2}\,, \nonumber\\
\Gamma^{r}_{\phantom{0}rr} &=& -\frac{\mathcal{F}}{2}\frac{m}{r^2}\,, \nonumber\\
\Gamma^{r}_{\phantom{0}\theta\theta} &=& -r+(\mathcal{F}+t_2)m \,, \\
\Gamma^{r}_{\phantom{0}\phi\phi} &=& -r\sin^2\theta+(\mathcal{F}+t_2)m\sin ^2 \theta \,, \nonumber\\
\Gamma^{\theta}_{\phantom{0}r\theta} &=& \Gamma^{\phi}_{\phantom{0}r\phi}\quad = \quad \frac{1}{r}\,, \nonumber\\
\Gamma^{\theta}_{\phantom{0}\theta r} &=& \Gamma^{\phi}_{\phantom{0}\phi r}\quad = \quad \frac{1}{r} - t_2\frac{m}{r^2} \,, \nonumber\\
\Gamma^{\theta}_{\phantom{0}\phi\phi} &=& -\sin \theta \cos \theta \,, \nonumber\\
\Gamma^{\phi}_{\phantom{0}\theta\phi} &=& \Gamma^{\phi}_{\phantom{0}\phi\theta}\quad = \quad \cot \theta \,. \nonumber
\eeqa
By ``the most general'' we mean that any other connections are related to the one in \Eq{eqn2:conn} by the nonlinear coordinate transformation law
\beq{eqn:gamma-law}
\Gamma\,'^{\mu}_{\phantom{\rho}\nu\lambda}(x')= \frac{\partial x\,'^{\mu}}{\partial x^{\alpha}}\frac{\partial x^{\beta}}{\partial x\,'^{\nu}}\frac{\partial x^{\gamma}}{\partial x\,'^{\lambda}}\Gamma ^{\alpha}_{\phantom{\rho}\beta\gamma}(x)+\frac{\partial x\,'^{\mu}}{\partial x^{\alpha}}\frac{\partial ^2 x^{\alpha}}{\partial x\,'^{\nu}\partial x\,'^{\lambda}}\,.
\eeq

Note that the terms independent of metric and torsion merely reflect the spherical coordinate system and do not represent
a deformation of spacetime --- in other words, the special case 
$t_1=t_2=\mathcal{H}=-\mathcal{F}=0$ corresponds to the connection for Minkowski spacetime.
The case $t_1=t_2=0$ and $\mathcal{H}=-\mathcal{F}=-2$ corresponds to the standard connection for Schwarzschild spacetime in the linear regime ($r\gg m$).

\subsection{First-order: stationary, spherically axisymmetric case}\label{subsection:axisymm}

The terms added at this order are due to Earth's rotation. 
Roughly speaking, ``{\it spherically axisymmetric}'' refers to the property
that a system is spherically symmetric except for symmetries broken by an angular momentum vector.  The rigorous mathematical definition is given in \Sec{subsubsec:prob-setup-axisym}.  Subtleties related to coordinate system choices at this order fortunately do not matter in the $\varepsilon_m \ll 1$ and $\varepsilon_a \ll 1$ limit that we are interested in.

\subsubsection{Symmetry Principles}
\label{subsubsec:prob-setup-axisym}

Suppose we have a field configuration which depends explicitly on the angular momentum $\J$ of the central spinning body.  We can
denote the fields generically as $\mathcal{O}(x|\J)$, which is a function of coordinates $x$ and the value of the angular momentum
vector $\J$.  We assume that the underlying laws of physics are symmetric under rotations, parity, time translation, and time
reversal, so that the field configurations for various values of $\J$ can be related to each other. Specifically, we assume that $\J$
rotates as a vector, reverses under time-reversal, and is invariant under time translation and parity.  It is then possible to define
transformations for the field configurations, $\mathcal{O}(x|\J)\to\mathcal{O}\,'(x'|\J)$, for these same symmetry operations.  Here
$\mathcal{O}\,'(x'|\J)$ denotes the transform of the field configuration that was specified by $\J$ before the transformation;
$\mathcal{O}$ may or may not be a tensor, but its transformation properties are assumed to be specified. The symmetries of the
underlying laws of physics then imply that the configurations $\mathcal{O}(x|\J)$ are stationary and {\it spherically axisymmetric} in
the sense that the transformed configuration is identical to the configuration that one would compute by transforming $\J \to \J'$. 
That is,
\[  \mathcal{O}\,'(x'|\J)=\mathcal{O}(x'|\J')  \]
under the following coordinate transformations:

\begin{enumerate}

\item{time translation:} $t\to t'\equiv t+t_0$ where $t_0$ is an arbitrary constant.

\item{Time reversal:} $t\to t'\equiv -t$.

\item{Continuous rotation and space inversion:} $\x \to \x'\equiv \R (\x)\,$, \ie $\x'$ is related to $\x$ by any proper or improper rotation.

\end{enumerate}
  
Below we will simplify the problem by keeping track only of terms linear in $J/r^2=\varepsilon_m \varepsilon_a$.

\subsubsection{Parametrization of metric}
\label{subsubsec:metric-axisy} 

With these symmetries, it can be shown that the first-order contribution to the metric is
\ben
g_{ti}=g_{it}= \frac{\mathcal{G}}{r^2}\epsilon_{ijk}J^j \hat{x}^{k}
\een
in rectangular coordinates $x^\mu = (t,x^i)$, where $\mathcal{G}$ is a constant, or
\ben
g_{t\phi}= g_{\phi t}= \mathcal{G}\frac{J}{r}\sin^2\theta
\een
in spherical coordinates $x^{\mu}=(t,r,\theta,\phi)$ where the polar angle $\theta$ is the angle 
with respect to the rotational angular momentum $\J$.  The details of the 
derivation are given in Appendix \ref{appendix:axisymm}.

\subsubsection{Parametrization of torsion}
\label{subsubsec:torsion-axisy}

In Appendix \ref{appendix:axisymm}, we 
show that, in rectangular coordinates, the first-order correction to the torsion is
\begin{eqnarray*}
S_{ij}^{\phantom{01}t} &=& \frac{f_1}{2r^3}\epsilon_{ijk}J^k+\frac{f_2}{2r^3}J^k \hat{x}^l (\epsilon_{ikl}\hat{x}^j-\epsilon_{jkl}\hat{x}^i)\,,\\
S_{tij} &=& \frac{f_3}{2r^3}\epsilon_{ijk}J^k+\frac{f_4}{2r^3}J^k \hat{x}^l \epsilon_{ikl}\hat{x}^j+\frac{f_5}{2r^3}J^k \hat{x}^l \epsilon_{jkl}\hat{x}^i\,.
\end{eqnarray*}
In spherical coordinates, these first-order torsion terms are
\begin{eqnarray*}
S_{r\phi}^{\phantom{0i}t} &=& w_1\frac{ma}{2r^2}\sin^2\theta\,,\\
S_{\theta\phi}^{\phantom{0i}t} &=& w_2\frac{ma}{2r}\sin\theta\cos\theta\,,\\
S_{t\phi}^{\phantom{0i}r} &=& w_3\frac{ma}{2r^2}\sin^2\theta\,, \\
S_{t\phi}^{\phantom{0i}\theta} &=& w_4\frac{ma}{2r^3}\sin\theta\cos\theta\,,\\
S_{tr}^{\phantom{0i}\phi} &=& w_5 \frac{ma}{2r^4}\,,\\
S_{t\theta}^{\phantom{0i}\phi} &=& -w_4\frac{ma}{2r^3}\cot\theta\,.
\end{eqnarray*}
Here $f_1,\ldots,f_5$ and $w_1,\ldots,w_5$ are constants.  The latter are linear combinations of the former.  
The details of the derivation are given
in Appendix \ref{appendix:axisymm}.  We call $w_1$,\ldots,$w_5$ the ``frame-dragging torsion'', since they will 
contribute the frame-dragging spin precession of a gyroscope as will become clear in \Sec{sec:instan-precess}.

\subsection{Around Earth}

We now summarize the results to linear order. We have computed the parametrization perturbatively
in the dimensionless parameters $\varepsilon_m\equiv m/r$ and $\varepsilon_a\equiv a/r$.
The zeroth order ($\varepsilon_a=0$) solution, where Earth's slow rotation is ignored, is simply the solution around 
a static spherical body, i.e. the case studied in \Sec{subsec:spher-symm}.  
The first order correction, due to Earth's rotation, is stationary and spherically axisymmetric as derived 
in \Sec{subsection:axisymm}. 
A quantity $\mathcal{O}$ to linear order is the sum of these two orders. 
In spherical coordinates, a general line element thus takes the form  
\bena
\mathrm{d}s^2 &=&  -\left[1+\mathcal{H}\frac{m}{r}\right]\mathrm{d}t^2 + \left[1+\mathcal{F}\frac{m}{r}\right]\mathrm{d}r^2+{}\nonumber\\
& & \!\!\!\!+r^2 d\Omega^2 +2\,\mathcal{G}\frac{ma}{r}\sin ^2 \theta\mathrm{d}t\mathrm{d}\phi \, ,\label{eqn:metricaxisy}
\eena
where $d\Omega^2=\mathrm{d}\theta^2+\sin ^2 \theta \mathrm{d}\phi^2$.  Here $\mathcal{H}$, $\mathcal{F}$ and $\mathcal{G}$ are
dimensionless constants.  In GR, the Kerr metric \cite{kerr,boyer-lind} at large distance gives the constants
$\mathcal{H}=-\mathcal{F}=\mathcal{G}=-2$.  
The result $\mathcal{G}=-2$ can also be derived more generally as shown by de Sitter \cite{deSitter:1916} and Lense \& Thirring \cite{Lense-Thirring:1918}. 
As above, $J=ma$ denotes the magnitude of Earth's rotational angular
momentum. 

Combining our 0th and 1st order expressions from above for the torsion around Earth, we obtain
\begin{eqnarray}
S_{tr}^{\phantom{01}t} &=& t_1\frac{m}{2r^2}\,,\nonumber\\
S_{r\theta}^{\phantom{12}\theta} &=& S_{r\phi}^{\phantom{13}\phi} = t_2\frac{m}{2r^2}\,,\nonumber\\
S_{r\phi}^{\phantom{0i}t} &=& w_1\frac{ma}{2r^2}\sin^2\theta\,,\nonumber\\
S_{\theta\phi}^{\phantom{0i}t} &=& w_2\frac{ma}{2r}\sin\theta\cos\theta\,,\label{eqn:w2-axi}\\
S_{t\phi}^{\phantom{0i}r} &=& w_3\frac{ma}{2r^2}\sin^2\theta\,,\nonumber \\
S_{t\phi}^{\phantom{0i}\theta} &=& w_4\frac{ma}{2r^3}\sin\theta\cos\theta\,,\nonumber\\
S_{tr}^{\phantom{0i}\phi} &=& w_5 \frac{ma}{2r^4}\,,\nonumber\\
S_{t\theta}^{\phantom{0i}\phi} &=& -w_4\frac{ma}{2r^3}\cot\theta\,.\nonumber
\end{eqnarray}
All other components vanish. Again, $t_1,t_2,w_1$,$w_2$, $w_3$, $w_4$, $w_5$ are dimensionless constants.

The calculation of the corresponding connection is straightforward by virtue of \Eq{eqn:fullconn1}. It is not hard to show that,
to linear order in a Riemann-Cartan spacetime in spherical coordinates, the connection around Earth has the following
non-vanishing components:
\bena
\Gamma^{t}_{\phantom{0}tr} &=& \left(t_1-\frac{\mathcal{H}}{2}\right)\frac{m}{r^2}\,, \nonumber\\
\Gamma^{t}_{\phantom{0}rt} &=& -\frac{\mathcal{H}}{2}\frac{m}{r^2}\,, \nonumber\nonumber\\
\Gamma^{t}_{\phantom{0}r\phi} &=& (3\mathcal{G}+w_1-w_3-w_5)\frac{ma}{2r^2}\sin^2\theta\,,\nonumber\\
\Gamma^{t}_{\phantom{0}\phi r} &=& (3\mathcal{G}-w_1-w_3-w_5)\frac{ma}{2r^2}\sin^2\theta\,,\nonumber\\
\Gamma^{t}_{\phantom{0}\theta\phi} &=& w_2\frac{ma}{2r}\sin\theta\cos\theta\,,\nonumber\\
\Gamma^{t}_{\phantom{0}\phi\theta} &=& -w_2\frac{ma}{2r}\sin\theta\cos\theta\,,\nonumber\\
\Gamma^{r}_{\phantom{0}tt} &=& \left(t_1-\frac{\mathcal{H}}{2}\right)\frac{m}{r^2}\,, \nonumber \\ 
\Gamma^{r}_{\phantom{0}rr} &=& -\frac{\mathcal{F}}{2}\frac{m}{r^2}\,, \nonumber\\
\Gamma^{r}_{\phantom{0}\theta\theta} &=& -r+(\mathcal{F}+t_2)m \,, \label{eqn:full-conn323}\\
\Gamma^{r}_{\phantom{0}\phi\phi} &=& -r\sin^2\theta+(\mathcal{F}+t_2)m\sin ^2 \theta \,, \nonumber\\
\Gamma^{r}_{\phantom{0}t\phi} &=& (\mathcal{G}-w_1+w_3-w_5)\frac{ma}{2r^2}\sin^2\theta\,,\nonumber\\
\Gamma^{r}_{\phantom{0}\phi t} &=& (\mathcal{G}-w_1-w_3-w_5)\frac{ma}{2r^2}\sin^2\theta\,,\nonumber\\
\Gamma^{\theta}_{\phantom{0}t\phi} &=& (-2\mathcal{G}-w_2+2w_4)\frac{ma}{2r^3}\sin\theta\cos\theta\,,\nonumber\\
\Gamma^{\theta}_{\phantom{0}\phi t} &=&(-2\mathcal{G}-w_2)\frac{ma}{2r^3}\sin\theta\cos\theta\,,\nonumber\\
\Gamma^{\theta}_{\phantom{0}r\theta} &=& \Gamma^{\phi}_{\phantom{0}r\phi}\; =\; \frac{1}{r} \,,\nonumber\\
\Gamma^{\theta}_{\phantom{0}\theta r} &=& \Gamma^{\phi}_{\phantom{0}\phi r}\; =\; \frac{1}{r}-t_2\frac{m}{r^2} \,,\nonumber
\eeqa
\bena
\Gamma^{\theta}_{\phantom{0}\phi\phi} &=& -\sin \theta \cos \theta \,,\nonumber\\
\Gamma^{\phi}_{\phantom{0}tr} &=&  (-\mathcal{G}+w_1-w_3+w_5)\frac{ma}{2r^4}\,,\nonumber\\
\Gamma^{\phi}_{\phantom{0}rt} &=&  (-\mathcal{G}+w_1-w_3-w_5)\frac{ma}{2r^4}\,,\nonumber\\
\Gamma^{\phi}_{\phantom{0}t\theta} &=& (2\mathcal{G}+w_2-2w_4)\frac{ma}{2r^3}\cot\theta\,,\nonumber\\
\Gamma^{\phi}_{\phantom{0}\theta t} &=& (2\mathcal{G}+w_2)\frac{ma}{2r^3}\cot\theta\,,\nonumber\\
\Gamma^{\phi}_{\phantom{0}\theta\phi} &=& \Gamma^{\phi}_{\phantom{0}\phi\theta} \;=\; \cot \theta\,.\nonumber
\eeqa

\section{Precession of a gyroscope I: fundamentals}
\label{sec:eom}

\subsection{Rotational angular momentum}

There are two ways to covariantly quantify the angular momentum of a spinning object, in the literature denoted
$S^{\mu}$ and $S^{\mu\nu}$, respectively.  (Despite our overuse of the letter $S$, they can be distinguished by the
number of indices.)  In the rest frame of the center of mass of a gyroscope, the 4-vector $S^{\mu}$ 
is defined as 
\ben
S^{\mu}=(0,\vec{S}_0)\,,
\een
and the 4-tensor $S^{\mu\nu}$ 
is defined to be antisymmetric and have the components
\beq{eqn:cyclic-smunu}
S^{0i}=S^{i0}=0, \qquad  \qquad S^{ij}=\epsilon^{ijk}S_0^{\phantom{1}k}\,,
\eeq
where $i=x,y,z$.  $\quad\vec{S}_0=S_0^{\phantom{1}x} \hat{x} + S_0^{\phantom{1}y} \hat{y} + S_0^{\phantom{1}z} \hat{z}\quad $ is the rotational angular momentum of a gyroscope observed by an observer co-moving with the center of mass of the gyroscope.  The relation between $S^{\mu}$ and $S^{\mu\nu}$ can be written in the local (flat) frame as 
\beq{eqn:ss1}
S^{\mu}=\epsilon^{\,\mu\nu\rho\sigma}u_{\nu}S_{\rho\sigma}\,,
\eeq
where $u^{\mu}=\mathrm{d}x^{\mu}/\mathrm{d}\tau$ is the 4-velocity.  

In curved spacetime, the Levi-Civita symbol is generalized to $\bar{\epsilon}^{\,\mu\nu\rho\sigma}=\epsilon^{\mu\nu\rho\sigma}/\sqrt{-g}$ where $g=\det g_{\mu\nu}$.  It is easy to prove that $\bar{\epsilon}^{\,\mu\nu\rho\sigma}$ is a 4-tensor.  Then \Eq{eqn:ss1} becomes a covariant relation
\beq{eqn:ss2}
S^{\mu}=\bar{\epsilon}^{\,\mu\nu\rho\sigma}u_{\nu}S_{\rho\sigma}\,.
\eeq
In addition, the vanishing of temporal components of $S^{\mu}$ and $S^{\mu\nu}$ can be written as covariant conditions 
as follows:
\bena
S^{\mu}u_{\mu} &=& 0 \,,\label{eqn:condition-smu}\\
S^{\mu\nu}u_{\nu} &=& 0\,.\label{eqn:condition-smunu}
\eena
In the literature \cite{schiff}, \Eq{eqn:condition-smunu} is called Pirani's supplementary condition.  
 
Note, however, that unlike the flat space case, the spatial vectors of $S^\mu$ and $S^{\mu\nu}$ 
(denoted by $\vec{S}$ and $\vec{S}'$ respectively) do not coincide in the curved spacetime. 
The former is the spatial component of the 4-vector $S^\mu$, while the latter is historically 
defined as  $\vec{S}^{\,'\,i} \equiv \epsilon^{ijk}S_{jk}$.  
It follows \Eq{eqn:ss2} that $\vec{S}$ and $\vec{S}'$ differ 
by $\vec{S} = \vec{S}' \left[1+\mathcal{O}(m_E/r)+\mathcal{O}(v^2)\right]$ for a gyroscope moving around Earth.

\subsection{Equation of motion for precession of a gyroscope}
\label{subsec:eom4precession}

To derive the equation of motion for $S^{\mu}$ (or $S^{\mu\nu}$) of a small extended object that may have either
rotational angular momentum or net spin, Papapetrou's method \cite{papapetrou} should be generalized to 
Riemann-Cartan spacetime.  This generalization has been studied by Stoeger \& Yasskin \cite{sy1,sy2} 
as well as Nomura, Shirafuji \& Hayashi \cite{NSH}. The starting point of this method is the Bianchi identity 
or Noether current in a
gravitational theory whose derivation strongly relies on an assumption of what sources torsion. 
Under the common assumption that only intrinsic spin sources torsion, both \cite{sy1,sy2} and
\cite{NSH} drew the conclusion that whereas a particle with net intrinsic spin will precess according to the full
connection, the rotational angular momentum of a gyroscope will \emph{not} feel the background
torsion, \ie it will undergo parallel transport by the Levi-Civita connection along the free-falling orbit --- the
same prediction as in GR.  

These results of \cite{sy1,sy2,NSH} have the simple intuitive interpretation that
if angular momentum is not coupled to torsion, then torsion is not coupled to angular momentum.
In other words, for Lagrangians where the angular momentum of a rotating object cannot generate a torsion field, the 
torsion field cannot affect the angular momentum of a rotating object, in the same spirit as Newton's dictum 
``action $=$ reaction''.  

The Hayashi-Shirafuji theory of gravity, which we will discuss in detail in \Sec{sec:counterexample}, raises an objection to the common assumption that
only intrinsic spin sources torsion, in that in this theory even a non-rotating massive body can generate torsion in the vacuum nearby
\cite{HS1}. This feature also generically holds for teleparallel theories. 
It has been customary to assume that spinless test
particles follow metric geodesics (have their momentum parallel transported by the Levi-Civita connection), \ie, that spinless particles decouple
from the torsion even if it is nonzero. For a certain class of Lagrangians, this can follow from using the conventional variational principle. However, Kleinert and Pelster
\cite{Kleinert:1996yi,Kleinert:1998as} argue that the closure failure of parallelograms in the 
presence of torsion adds an additional term to the
geodesics which causes spinless test particles to follow autoparallel worldlines (have their momentum parallel transported by the full
connection). This scenario thus respects the ``action $=$ reaction'' principle, since a spinless test particle can both generate and feel torsion.
As a natural extension, we explore the possibility that in these theories, a rotating body also generates torsion through its rotational angular momentum,
and the torsion in turn affects the motion of spinning objects such as gyroscopes.

An interesting first-principles derivation of how torsion affects a gyroscope in a specific  theory might involve 
generalizing the matched asymptotic expansion method of \cite{D'Eath1975a,D'Eath1975b}, and match
two generalized Kerr-solutions in the weak-field limit to obtain the gyroscope equation of motion.
Since such a calculation would be way beyond the scope of the present paper, 
we will simply limit our analysis to exploring some obvious possibilities for laws of motion, based on 
the analogy with spin precession. 

The exact equation of motion for the precession of net spin is model dependent,
depending on the way the matter fields couple to the metric and torsion in the Lagrangian
(see
\cite{sy1,sy2,NSH,Hojman,Hojman2,cognola,kopczynski,pereira1,adamowicz-trautman}).  However, in the linear regime that we are
interested in here, many of the cases reduce to one of the following two equations if there is no external non-gravitational 
force acting on the test particle:
\bena
\frac{\mathrm{D}S^{\mu}}{\mathrm{D}\tau} &=& 0\,,\label{eqn:eom4smu}\\
\textrm{or}\quad \frac{\mathrm{D}S^{\mu\nu}}{\mathrm{D}\tau} &=& 0\,,\label{eqn:eom4smunu}
\eena
where $\mathrm{D}/\mathrm{D}\tau=(\mathrm{d}x^{\mu}/\mathrm{d}\tau)\nabla_{\mu}$ is the covariant differentiation along the world-line  
with respect to the full connection. In other words, the net spin undergoes parallel transport by the full connection along its trajectory.\footnote{If an external non-gravitational force acts on a spinning test particle, it will undergo
Fermi-Walker transport along its world-line.  This situation is beyond the interest of a satellite experiment, so it will be
neglected in the present paper.}

In analog to the precession of spin, we will work out the implications of the assumption that the rotational angular momentum 
also precesses by parallel transport along the free-fall trajectory using the full connection.

\subsection{World line of the center of mass}

In GR, test particles move along well-defined trajectories -- \emph{geodesics}.  In the presence of torsion, things might be
different.  The idea of \emph{geodesics} originates from two independent concepts: \emph{autoparallels} and \emph{extremals}
\footnote{This terminology follows Hehl {\etal} \cite{hehl}.}.  Autoparallels, or affine geodesics, are curves along which the
velocity vector $\mathrm{d}x^{\mu}/\mathrm{d}\lambda$ is transported parallel to itself by the full connection
$\Gamma^{\rho}_{\phantom{\mu}\mu\nu}$.  With an affine parameter $\lambda$, the geodesic equation is
\begin{equation}\label{eqn:autopa}
\frac{\mathrm{d}^2 x^{\rho}}{\mathrm{d}\lambda ^2}+\Gamma^{\rho}_{\phantom{\mu}(\mu\nu)} \frac{\mathrm{d}x^{\mu}}{\mathrm{d}\lambda} \frac{\mathrm{d}x^{\nu}}{\mathrm{d}\lambda} =0\,.
\end{equation}
Extremals, or metric geodesics, are curves of extremal spacetime interval with respect to the metric $g_{\mu\nu}$.  
Since $\mathrm{d}s=[-g_{\mu\nu}(x)\mathrm{d}x^{\mu}\mathrm{d}x^{\nu}]^{1/2}$ does not depend on the full connection, the geodesic differential equations derived from $\delta \int \mathrm{d}s=0$ state that the 4-vector is parallel transported by the Levi-Civita connection.  That is, with the parameter $\lambda$ properly chosen, 
\begin{equation}\label{eqn:extremal}
\frac{\mathrm{d}^2 x^{\rho}}{\mathrm{d}\lambda ^2}+\left\{ \begin{array}{c} \rho \\ \mu\nu \end{array} \right\} \frac{\mathrm{d}x^{\mu}}{\mathrm{d}\lambda} \frac{\mathrm{d}x^{\nu}}{\mathrm{d}\lambda} =0\,.
\end{equation}
In Riemann spacetime where torsion identically vanishes, Eqs.(\ref{eqn:autopa}) and (\ref{eqn:extremal}) coincide.  In a
Riemann-Cartan spacetime, however, these two curves coincide if and only if the torsion is totally antisymmetric in all three
indices \cite{hehl}.  This is because the symmetric part of the full connection can be written from \Eq{eqn:fullconn1} as follows:
\beq{eqn:symm-conn}
\Gamma^{\rho}_{\phantom{\rho}(\mu\nu)}\equiv\frac{1}{2}(\Gamma^{\rho}_{\phantom{\rho}\mu\nu}+\Gamma^{\rho}_{\phantom{\rho}\nu\mu})= \left\{ \begin{array}{c} \rho \\ \mu\nu \end{array} \right\} + S_{\phantom{\rho}\mu\nu}^{\rho}+S_{\phantom{\rho}\nu\mu}^{\rho}\,.
\eeq

Photons are expected to follow extremal world lines because 
the gauge invariance of the electromagnetic part of the Lagrangian, well established by
numerous experimental upper bounds on the photon mass, prohibits torsion from coupling to the electromagnetic field to lowest order \cite{hehl}. 
As a consequence, the classical path of a light ray is at least to leading order determined by the metric alone as an extremal 
path, or equivalently as an autoparallel curve with respect to the Levi-Civita connection, independent of whether there is torsion.

On the other hand, the trajectory of a rotating test particle is still an open question in theory. 
Papapetrou \cite{papapetrou} claims that, even in GR, a gyroscope will deviate from the metric geodesic, albeit slightly. 
In torsion gravity theories, the equations of motion for the orbital 4-momentum differs more strongly between different 
approaches \cite{hehl,sy2,NSH,Hojman,Hojman2,cognola,kopczynski,pereira1}, and it is an open question to what extent
they are consistent with all classical GR tests (deflection of light rays, gravitational redshift, precession of the
perihelion of Mercury, Shapiro time delay, binary pulsars, etc.).
To bracket the uncertainty, we will examine the two extreme assumption in turn -- that world lines are
autoparallels and extremals, respectively.

Only the autoparallel scheme, not the extremal scheme, is theoretically consistent, for two reasons.  The first reason is based on the equivalence of the two approaches using the two alternative quantities $S^{\mu}$ and $S^{\mu\nu}$ to describe the angular momentum.  The equivalence is automatic in GR.  In a torsion theory, however, \Eq{eqn:eom4smu} and (\ref{eqn:eom4smunu}) can be simultaneously valid only if the trajectory is autoparallel.  This can be seen by taking the covariant differentiation of \Eq{eqn:ss2}. 
Note that $\mathrm{D}\bar{\epsilon}^{\,\mu\nu\rho\sigma}/\mathrm{D}\tau=0$.  One finds
\ben
\bar{\epsilon}^{\,\mu\nu\rho\sigma}\frac{\mathrm{D}u_{\nu}}{\mathrm{D}\tau}S_{\rho\sigma}=0\,.
\een
This equation is satisfied if $\mathrm{D}u_{\nu}/\mathrm{D}\tau=0$, \ie if the gyroscope world line is autoparallel. If an extremal world line is assumed, then
one has to make an \emph{a priori} choice between $S^{\mu}$ and $S^{\mu\nu}$, since the precession rates calculated using the two quantities will differ.

The second reason is that for $S^{\mu}$, the condition $S^{\mu}u_{\mu} = 0$ (\Eq{eqn:condition-smu}) must be satisfied anywhere along the world line.  Taking the covariant differentiation for both sides of \Eq{eqn:condition-smu}, one finds 
\beq{eqn:smu-derived}
S^{\mu}\mathrm{D}u_{\mu}/\mathrm{D}\tau = 0\,,
\eeq
assuming $\mathrm{D}S^{\mu}/\mathrm{D}\tau = 0$.  Obviously, autoparallels are consistent with \Eq{eqn:smu-derived}, while extremals are not.  The same argument applies for $S^{\mu\nu}$, \ie taking the covariant differentiation of both sides of \Eq{eqn:condition-smunu}.

Despite the fact that the extremal scheme is not theoretically consistent in this sense, the inconsistencies are numerically small for the linear regime $m/r \ll 1$.  
They are therefore of interest as an approximate phenomenological prescription that might at some 
time in the future be incorporated into a consistent theory.  We therefore include results also for this case below.

\subsection{Newtonian limit}

In \Sec{sec:param}, we parametrized the metric, torsion and connection of Earth, including an arbitrary parameter $m$ with units of mass.  To give $m$ a physical interpretation, the Newtonian limit of a test particle's orbit should be evaluated.  Obviously, the result depends on whether the autoparallel or extremal scheme is assumed.

In the remainder of this paper, we denote an arbitrary parameter with units of mass as $m_0$ and the physical mass as $m$.  Metric and torsion parameters in accordance with $m_0$ are denoted with a superscript $(0)$, \ie  $\mathcal{H}^{(0)},\mathcal{F}^{(0)},\mathcal{G}^{(0)},t_1^{(0)},t_2^{(0)},w_1^{(0)}\ldots w_5^{(0)}$.

If an autoparallel world line is assumed, using the parametrization of equations (\ref{eqn:full-conn323}), it can be shown that the equation of motion to lowest order becomes
\beq{eqn:auto}
\frac{\mathrm{d}\vec{v}}{\mathrm{d}t}=-\left[t_1^{(0)}-\frac{\mathcal{H}^{(0)}}{2}\right]\frac{m_0}{r^2}\hat{e}_r\,.
\eeq
Therefore Newton's Second Law interprets the mass of the central gravitating body to be 
\beq{eqn:massauto}
m = \left[t_1^{(0)}-\frac{\mathcal{H}^{(0)}}{2}\right]m_0\,.\quad \textrm{(autoparallel scheme)}
\eeq
However, if $t_1^{(0)}-\mathcal{H}^{(0)}/2=0$, the autoparallel scheme fails totally.

Similarly, for a theory with extremal world-lines, the extremal equation in Newtonian approximation is
\beq{eqn:extrem}
\frac{\mathrm{d}\vec{v}}{\mathrm{d}t}=-\frac{[-\mathcal{H}^{(0)}]}{2}\frac{m_0}{r^2}\hat{e}_r\,.
\eeq
Therefore the physical mass of the body generating the gravity field is 
\beq{eqn:massextrem}
m = -\frac{\mathcal{H}^{(0)}}{2}m_0\,,\quad\textrm{(extremal scheme)}
\eeq
as long as $\mathcal{H}^{(0)}\ne 0$.  For the Schwarzschild metric ($\mathcal{H}^{(0)}=-2$), $m=m_0$.

After re-scaling $m$ from $m_0$, all metric and torsion parameters make the inverse re-scaling, \eg\ $t_1 = t_1^{(0)}(m_0/m)$ since the combination $t_1 m$ is the physical parameters during parametrization of metric and torsion.  This inverse scaling applies to $\mathcal{H}^{(0)},\mathcal{F}^{(0)},\mathcal{G}^{(0)},t_2^{(0)},w_1^{(0)}\ldots w_5^{(0)}$ as well.  A natural consequence of the re-scaling is an identity by definition:
\bena
t_1 - \mathcal{H}/2  & = & 1\,,\quad\textrm{(autoparallel scheme)}\\
\textrm{or}\quad\quad\quad \mathcal{H}    & = & -2 \,,\quad\textrm{(extremal scheme)}
\eena

\section{Precession of a gyroscope II: instantaneous rate}
\label{sec:instan-precess}

We now have the tools to calculate the precession of a gyroscope. Before proceeding, let us summarize the assumptions made so far:
\begin{enumerate}
\item A gyroscope can feel torsion through its rotational angular momentum, and the equation of motion is either 
$\mathrm{D}S^{\mu}/\mathrm{D}\tau=0$ or $\mathrm{D}S^{\mu\nu}/\mathrm{D}\tau=0$.
\item The world line of a gyroscope is either an autoparallel curve or an extremal curve.
\item The torsion and connection around Earth are parametrized by \Eq{eqn:w2-axi} and (\ref{eqn:full-conn323}).  
\end{enumerate}
With these assumptions, the calculation of the precession rate becomes straightforward except for one subtlety described below.  

\subsection{Transformation to the center-of-mass frame}

The precession rate $\mathrm{d}\vec{S}/\mathrm{d}t$ derived from a naive application of the equation of motion $\mathrm{D}S^{\mu}/\mathrm{D}\tau=0$ is the rate
measured by an observer at rest relative to the central gravitating body.  This rate is gauge-dependent and unphysical, 
since it depends on which coordinates the observer uses; for example,
isotropic spherical coordinates and standard spherical coordinates yield different precession rates. The physical observable is the precession rate
$\mathrm{d}\vec{S}_0/\mathrm{d}t$ measured by the observer co-moving with the center of mass of the gyroscope, \ie in the instantaneous local inertial frame.  

The methodology of transforming $\vec{S}$ to $\vec{S}_0$ was first established by Schiff \cite{schiff} in which he used the 4-tensor $S^{\mu\nu}$.  The basic idea
using the 4-vector $S^{\mu}$ is as follows. Since we are interested in the transformation only 
to leading order in $(v/c)^2$ and $m/r$, 
we are allowed to consider the coordinate
transformation and the velocity transformation separately and add them together in the end. We adopt standard spherical coordinates with the line element of
\Eq{eqn:metricaxisy}. The off-diagonal metric element proportional to $ma/r^2$ can be ignored for the purposes of this transformation. Consider a measuring rod in the
rest frame of the central body. It will be elongated by a factor of $(1+\mathcal{F}m/2r)$ in the radial direction measured by the observer in the center-of-mass
frame, but unchanged in the tangential direction.  The 4-vector $S^{\mu}$ transforms as $\mathrm{d}x^{\mu}$; thus its radial component is enlarged by a factor of
$(1+\mathcal{F}m/2r)$ and the tangential components are unchanged.  This can be compactly written in the following form:
\ben
\vec{S}_0=\vec{S}+\mathcal{F}\frac{m}{2r^3}(\vec{S}\cdot\vec{r})\vec{r}\,.
\een
Now consider the velocity transformation to the center-of-mass frame by boosting the observer along the $x$-axis, say, with velocity $v$.  
We have the Lorentz boost from
$S^{\mu}=(S^0,S^x,S^y,S^z)$ to $S_0^{\mu}=\left(S_0\,^0,S_0\,^x,S_0\,^y,S_0\,^z\right)$ as follows:
\bena
S_0\,^0 &=& \gamma(S^0-v\,S^x)\,,\\
S_0\,^x &=& \gamma(S^x-v\,S^0)\,,\\
S_0\,^y &=& S^y\,,\\
S_0\,^z &=& S^z\,,
\eena
where $\gamma=1/\sqrt{1-v^2}\approx 1+v^2/2$.  The condition \mbox{$S^{\mu}u_{\mu}=0$} gives 
\[ S^0 = \vec{v}\cdot\vec{S}=v\,S^x\,,\]
which verifies that $S_0\,^0=0$ in the center-of-mass frame.  The spatial components can be written compactly as 
\ben
\vec{S}_0=\vec{S}-\frac{1}{2}(\vec{S}\cdot\vec{v})\vec{v}\,.
\een
Combining the coordinate transformation and the velocity transformation, we find the following transformation from standard spherical coordinates 
to the center-of-mass frame:
\beq{eqn:xform-smu}
\vec{S}_0=\vec{S}+\mathcal{F}\frac{m}{2r^3}(\vec{S}\cdot\vec{r})\vec{r}-\frac{1}{2}(\vec{S}\cdot\vec{v})\vec{v}\,.
\eeq
The time derivative of \Eq{eqn:xform-smu} will lead to the expression for  \emph{geodetic precession} to leading order , \ie to order $(m/r)v$. 
To complete the discussion of transformations, note that the off-diagonal metric element proportional to $ma/r^2$ could add a term of order $ma/r^2$ 
to \Eq{eqn:xform-smu}, which leads to a precession rate proportional to $(ma/r^2)v$.  Since the leading term of the \emph{frame dragging} 
effect is of the order $ma/r^2$, the leading frame-dragging effect is invariant under these transformations, so we are allowed to ignore the 
off-diagonal metric element in the transformation.

The transformation law obtained using the 4-tensor $S^{\mu\nu}$ is different from using $S^{\mu}$ 
 --- this is not surprising because both descriptions coincide only in the rest frame of the gyroscope's center of mass. Schiff \cite{schiff} gave the transformation law from standard spherical coordinates to the center-of-mass frame, using $S^{\mu\nu}$: 
\begin{eqnarray}
\vec{S}_0&=&\vec{S}'+\mathcal{F}\frac{m}{2r}[\vec{S}'-(\vec{r}/r^2)(\vec{r}\cdot\vec{S}')]\nonumber\\
& & -\frac{1}{2}[v^2 \vec{S}'-(\vec{v}\cdot\vec{S}')\vec{v}]\,.\label{eqn:xform}
\end{eqnarray}

In taking the time derivative of \Eq{eqn:xform-smu} or (\ref{eqn:xform}), one encounters terms proportional to $\mathrm{d}\vec{v}/\mathrm{d}t$.   \Eq{eqn:auto} or (\ref{eqn:extrem}) should be applied, depending on whether autoparallel  or extremal scheme, respectively, is assumed.

\subsection{Instantaneous rates}

\subsubsection{Autoparallel scheme and using $S^{\mu}$}
\label{subsubsec:smu-auto}

Now we are now ready to calculate the precession rate.  In spherical coordinates $x^{\mu}=(t,r,\theta,\phi)$, we expand the rotational angular momentum 
vector in an orthonormal basis:
\[\vec{S}=S_r\hat{e}_r+S_{\theta}\hat{e}_{\theta}+S_{\phi}\hat{e}_{\phi}\,.\]
In terms of the decomposition coefficients, the 4-vector is
\[ S^{\mu}=(S^0,S^1,S^2,S^3)=(S^0,S_r,S_{\theta}/r,S_{\phi}/r\sin\theta)\,.\]
Applying the equation of motion $\mathrm{D}S^{\mu}/\mathrm{D}\tau=0$, transforming $\vec{S}$ to $\vec{S}_0$ by \Eq{eqn:xform-smu} and taking the time derivative
using autoparallels (Eq.~\ref{eqn:auto}), we obtain the following instantaneous gyroscope precession rate:
\bena
\frac{\mathrm{d}\vec{S}_0}{\mathrm{d}t} &=& \vec{\Omega}\times\vec{S}_0\,, \label{eqn:instn-precession1}\\
\textrm{where}\quad \vec{\Omega}\phantom{G} &=& \vec{\Omega}_G + \vec{\Omega}_F\,,\label{eqn:instn-precession2}\\
\vec{\Omega}_G &=& \left( \frac{\mathcal{F}}{2}-\frac{\mathcal{H}}{4}+t_2+\frac{t_1}{2}\right)\frac{m}{r^3}(\vec{r}\times\vec{v})\,, \label{eqn:GP1}\\
\vec{\Omega}_F &=& \frac{\mathcal{G}I}{r^3}\left[ -\frac{3}{2}(1+\mu_1)(\vec{\omega}_E \cdot \hat{e}_r)\hat{e}_r \right.\nonumber\\
& & \left. +\frac{1}{2}(1+\mu_2)\vec{\omega}_E \right]\,. \label{eqn:FD1}
\eena
Here $I\omega_E=ma$ is the angular momentum of Earth, where $I$ is Earth's moment of inertia about its poles and $\omega_E$ is its angular velocity.  
The new effective torsion constants are defined so that they represent the torsion-induced correction to the GR prediction:
\bena
\mu_1 &\equiv & (w_1-w_2-w_3+2w_4+w_5)/(-3\mathcal{G})\,,\label{eqn:mu1new}\\
\mu_2 &\equiv & (w_1-w_3+w_5)/(-\mathcal{G})\,,
\eena
Since $t_1-\mathcal{H}/2=1$ in the autoparallel scheme, \Eq{eqn:GP1} simplifies to 
\beq{eqn:GP11}
\vec{\Omega}_G = \left( 1+ \mathcal{F}+2t_2\right)\frac{m}{2r^3}(\vec{r}\times\vec{v})\,.
\eeq   

In the literature, the precession due to $\Omega_G$ is called \emph{geodetic precession}, and that due to $\Omega_F$ is called \emph{frame dragging}.  
From \Eq{eqn:GP1}, it is seen that geodetic precession depends on the mass of Earth and not on whether Earth is spinning or not.  It is of order $mv$. The
frame-dragging effect is a unique effect of Earth's rotation and highlights the importance of the GPB experiment, since GPB will be the first to
accurately measure the effect of the off-diagonal metric element that lacks a counterpart in Newtonian gravity.  The frame dragging effect is of order
$ma$, so it is independent of whether the gyroscope is moving or static.  In the presence of torsion, we term $\Omega_G$ the ``generalized geodetic precession'',
and $\Omega_F$ the ``generalized frame-dragging''.

\subsubsection{Extremal scheme and using $S^{\mu}$}

We now repeat the calculation of \Sec{subsubsec:smu-auto}, but assuming an extremal trajectory (\Eqnopar{eqn:extrem}) 
when taking the time derivative of \Eq{eqn:xform-smu}, obtaining the following instantaneous gyroscope precession rate:
\bena
\frac{\mathrm{d}\vec{S}_0}{\mathrm{d}t} &=& \vec{\Omega}\times\vec{S}_0-t_1\frac{m}{r^3}(\vec{S}_0\cdot\vec{v})\vec{r}\,, \label{eqn:instn-precession21}\\
\textrm{where}\quad \vec{\Omega}\phantom{G} &=& \vec{\Omega}_G + \vec{\Omega}_F\,.\nonumber\\
\vec{\Omega}_G &=& \left( \frac{\mathcal{F}}{2}-\frac{\mathcal{H}}{4}+t_2\right)\frac{m}{r^3}(\vec{r}\times\vec{v})\,, \label{eqn:GP-extremal_smu}
\eena
and $\vec{\Omega}_F$ is the same as in \Eq{eqn:FD1}.
Since $\mathcal{H}=-2$ in the extremal scheme, \Eq{eqn:GP-extremal_smu} is simplified to formally coincide with \Eq{eqn:GP11}.

\subsubsection{Extremal scheme and using $S^{\mu\nu}$}

In spherical coordinates, $S^{\mu\nu}$ satisfies
\ben
S^{12}=\frac{1}{r}S^{\,'}_{\phi}\,,\:\: S^{23}=\frac{1}{r^2 \sin\theta}S^{\,'}_{r}\,,\:\: S^{31}=\frac{1}{r\sin\theta}S^{\,'}_{\theta}\,,\label{eqn:ssrel}
\een
where $S^{\,'}_r,\,S^{\,'}_{\theta},\,S^{\,'}_{\phi}$ are the components of $\vec{S}^{\,'}$ in spherical coordinates, i.e. $\vec{S}^{\,'} = S^{\,'}_r\hat{e}_r+S^{\,'}_{\theta}\hat{e}_{\theta}+S^{\,'}_{\phi}\hat{e}_{\phi}\,.$
We now repeat the calculation of \Sec{subsubsec:smu-auto} assuming an extremal trajectory (\Eqnopar{eqn:extrem}) and the $S^{\mu\nu}$-based precession of
\Eq{eqn:xform} when taking the time derivative of \Eq{eqn:xform-smu}, obtaining the following instantaneous gyroscope precession rate:
\bena
\frac{\mathrm{d}\vec{S}_0}{\mathrm{d}t} &=& \vec{\Omega}\times\vec{S}_0+t_1\frac{m}{r^3}\vec{r}\times(\vec{v}\times\vec{S}_0)\,, \label{eqn:instn-precession31}\\
\textrm{where}\quad \vec{\Omega}\phantom{G} &=& \vec{\Omega}_G + \vec{\Omega}_F\,.\nonumber
\eena
$\vec{\Omega}_G$ and $\vec{\Omega}_F$ are the same as in equations (\ref{eqn:GP-extremal_smu}) and (\ref{eqn:FD1}), respectively.

In both cases using extremals, the precession rates have anomalous terms proportional to $t_1$;
see \Eq{eqn:instn-precession21}) and~\ref{eqn:instn-precession31}).  We call these terms the ``anomalous geodetic precession''.  
These anomalies change the angular precession rate of a gyroscope, since their
contributions to $\mathrm{d}\vec{S}_0/\mathrm{d}t$ are not perpendicular to $\vec{S}_0$.  This is a phenomenon that GR does not predict. Meanwhile, $t_2$
contributes to modify only the magnitude and not the direction of $\vec{\Omega}_G$. We therefore term $t_1$ the anomalous geodetic torsion and $t_2$ the normal
geodetic torsion.  The torsion functions $w_1$,\ldots,$w_5$ contribute to the generalized frame-dragging effect via the two 
combinations $\mu_1$ and $\mu_2$, and we therefore term them ``frame-dragging torsions''.  

\subsubsection{Autoparallel scheme and using $S^{\mu\nu}$}

Repeating the calculation of \Sec{subsubsec:smu-auto}
using the $S^{\mu\nu}$-based precession rule of
\Eq{eqn:xform} gives the exact same instantaneous precession rate as in \Sec{subsubsec:smu-auto}.
This is expected since these two precession rules are equivalent in the autoparallel scheme.

\section{Precession of a gyroscope III: moment analysis}
\label{sec:mom-ana}

\begin{figure}
\centerline{\epsfxsize=\figsiz\epsffile{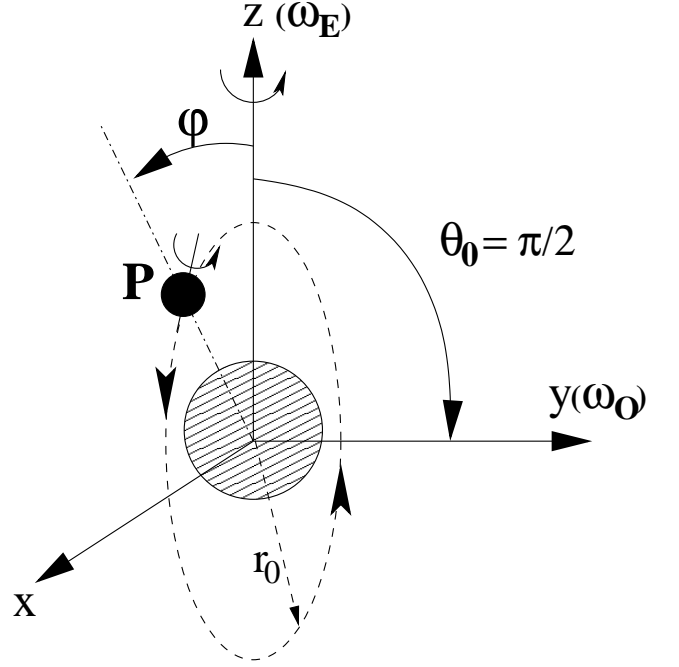}}
\caption[1]{\label{gpb2}\footnotesize%
A Gravity Probe B gyroscope moves around Earth along a circular polar orbit with $\theta_0=\pi/2$. 
$\omega_O$ is its orbital angular velocity and $\omega_E$ is Earth's rotational angular velocity around the $z$-axis.    }
\end{figure}

GPB measures the rotational angular momentum $\vec{S}_0$ of gyroscopes and therefore the precession rate 
$\mathrm{d}\vec{S}_0/\mathrm{d}t$ essentially continuously.  This
provides a wealth of information and deserves careful data analysis. Here we develop a simple but sensitive analysis method based on Fourier transforms.

\subsection{Fourier transforms}

The Gravity Probe B satellite has a circular polar orbit to good approximation\footnote{The actual GPB orbit has an orbital eccentricity of 0.0014 and 
an inclination of $90.007^\circ$ according to the Fact Sheet on the GPB website.  These deviations from the ideal orbit should cause negligible ($\lesssim 10^{-5}$) relative errors in our estimates above.}, \ie the inclination
angle of the orbital angular velocity $\vec{\omega}_O$ with respect
to the Earth's rotation axis (z-axis) is $\theta_0=\pi/2$.  Hence
the orbital plane is perpendicular to the equatorial plane. Let the $y$-axis
point along the vector $\vec{\omega}_O$ and let the $x$-axis be perpendicular
to the $y$-axis in the equatorial plane so that the three axes $\{x,y,z\}$ form a
right-handed coordinate basis as illustrated in \fig{gpb2}. A gyroscope at a point $P$ is marked by
the monotonically increasing angle $\varphi$ with respect to z axis.
The polar angle of the point $P$ can be regarded as a periodic
function of $\varphi$:
\beq{eqn:thetafn}
\theta(\varphi)=\left\{\begin{array}{lcl}\varphi &, & 0\le\varphi\le\pi\\ 
2\pi-\varphi&, & \pi\le\varphi\le 2\pi \end{array}\right.
\eeq 
So for a particular circular polar orbit, $\mathrm{d}\vec{S}_0/\mathrm{d}t(\vec{r},\vec{v})$ can be regarded as a periodic function of $\varphi$, 
where $r_0$ is the fixed radius, allowing us to write $\mathrm{d}\vec{S}_0/\mathrm{d}t(\vec{r},\vec{v})\equiv\mathrm{d}\vec{S}_0/\mathrm{d}t(\varphi)$.

Now define the Fourier \emph{moments} of the precession rate as 
\bena
\vec{a}_0 &=&
\frac{1}{2\pi}\int_0^{2\pi}\frac{\mathrm{d}\vec{S}_0}{\mathrm{d}t}(\varphi)\mathrm{d}\varphi
=\left<\frac{\mathrm{d}\vec{S}_0}{\mathrm{d}t}(\varphi)\right>,\\
\vec{a}_n &=& \frac{1}{2\pi}\int_0^{2\pi}\frac{\mathrm{d}\vec{S}_0}{\mathrm{d}t}(\varphi)\cos
 n\varphi \mathrm{d}\varphi\,,\\
\vec{b}_n &=& \frac{1}{2\pi}\int_0^{2\pi}\frac{\mathrm{d}\vec{S}_0}{\mathrm{d}t}(\varphi)\sin
 n\varphi \mathrm{d}\varphi\,,
\eena
where $n=1,2,\ldots$, so that we can write 
\beq{eqn:dsdt-fourier}
\frac{\mathrm{d}\vec{S}_0}{\mathrm{d}t}(\varphi)=\vec{a}_0+2\sum_{n=1}^{\infty}(\vec{a}_n\cos
n\varphi+\vec{b}_n\sin n\varphi)\,.
\eeq

\subsection{Average precession}
\label{subsec:ave_precession}

We now write equations (\ref{eqn:instn-precession1}), (\ref{eqn:instn-precession2}), (\ref{eqn:FD1}), (\ref{eqn:GP11}), (\ref{eqn:instn-precession21}) and (\ref{eqn:instn-precession31})
 explicitly in terms of $\varphi$ and perform the Fourier transforms.  The average precession in the three calculation schemes above can be compactly written as follows:
\ben
\vec{a}_0  \equiv  \left<\frac{\mathrm{d}\vec{S}_0}{\mathrm{d}t}(\varphi)\right> = \vec{\Omega}_{\textrm{eff}}\times \vec{S}_0\label{eqn:moment1}\,.
\een

The angular precession rate is
\ben
\vec{\Omega}_{\textrm{eff}} = b_t \frac{3m}{2r_0}\vec{\omega}_O+b_\mu\frac{I}{2r_0^3}\vec{\omega}_E\,,\label{eqn:moment2}
\een
where $\vec{\omega}_O=\omega_O\hat{y}$ is the orbital angular velocity and $\vec{\omega}_E=\omega_E \hat{z}$ is the rotational angular velocity of Earth. 
Here the ``biases'' relative to the GR prediction are defined by
\bena
b_t &\equiv & \frac{1}{3}(1+\mathcal{F}+2t_2+|\eta| t_1)\,,\label{eqn:btdef}\\
b_\mu &\equiv & \frac{\mathcal{(-G)}}{2}(1+3\mu_1-2\mu_2)\,,\label{eqn:bmudef}\\
      &=& \frac{\mathcal{(-G)}}{2}[1+(w_1+w_2-w_3-2w_4+w_5)/\mathcal{G}]\,,\nonumber
\eena
where the constant $\eta$ reflects the different assumptions that we have explored, and takes the following values:
\beq{eqn:etadef}
\eta=\left\{ \begin{array}{rl}  0 & \textrm{using autoparallels}\\
                                +1 & \textrm{using $S^{\mu\nu}$ and extremals}\\
                                -1 & \textrm{using $S^{\mu}$ and extremals}
\end{array}\right.
\eeq
From the above formulas, we see that the three schemes give identical results when $t_1=0$.

For comparison, GR predicts the average precession rate
\bena
\vec{a}_0 &\equiv & \left<\frac{\mathrm{d}\vec{S}_0}{\mathrm{d}t}(\varphi)\right> = \vec{\Omega}_{\textrm{eff}}\times \vec{S}_0\label{eqn:moment1GR}\,,\nonumber\\
\textrm{where}\quad\vec{\Omega}_{\textrm{eff}} &=& \frac{3m}{2r_0}\vec{\omega}_O+\frac{I}{2r_0^3}\vec{\omega}_E\,,\label{eqn:moment2GR}
\eena
\ie, $b_t=b_\mu=1$.

It is important to note that torsion contributes to the \emph{average} precession above only via {\it magnitudes} of the precession rates, 
leaving the precession axes intact. The geodetic torsion parameters $t_1$ and $t_2$ are degenerate, entering only in the linear combination corresponding 
to the bias $b_t$. 
The frame-dragging torsion parameters $w_1,\ldots,w_5$ are similarly  degenerate, entering only in the linear combination corresponding to the bias $b_\mu$. 
If for technical reasons, the average precession rate is the only quantity that GPB can measure, then only these biases can be constrained.

\subsection{Higher moments}
\label{subsec:moment}

Interestingly, all higher Fourier moments vanish except for $n=2$:
\begin{eqnarray}
\vec{a}_2 &=& \frac{-3\mathcal{G}I\omega_E}{8r_0^3}(1+\mu_1)\hat{z}\times\vec{S}_0 +\eta\, t_1\frac{m}{4r_0}\omega_O(S_0\,^x\hat{z}+S_0\,^z\hat{x}) \,,\nonumber\\
\vec{b}_2 &=& \frac{-3\mathcal{G}I\omega_E}{8r_0^3}(1+\mu_1)\hat{x}\times\vec{S}_0 +\eta\, t_1\frac{m}{4r_0}\omega_O(S_0\,^x\hat{x}-S_0\,^z\hat{z}) \,.\nonumber\\ 
\label{eqn:n2moments}
\end{eqnarray}
Here we use the notation $S_0\,^i\equiv \vec{S}_0\cdot\hat{i}$, where $i$ denotes the $x$, $y$ and $z$ axes.

For comparison, GR predicts the following second moments (moments with $m=1$ and $m>2$ vanish):
\bena
\vec{a}_2 &=& \frac{3I\omega_E}{4r_0^3}\hat{z}\times\vec{S}_0 \,,\\
\vec{b}_2 &=& \frac{3I\omega_E}{4r_0^3}\hat{x}\times\vec{S}_0 \,.
\eena
Technically, it may be difficult to measure these second moments because of the extremely small precession rate per orbit. 
However, if they \emph{could} be measured, they could break the degeneracy between $t_1$ and $t_2$: $|t_1|$ could be measured through the \emph{anomalous} $n=2$
precession moment (the second term in \Eq{eqn:n2moments}). The sign ambiguity of $t_1$ is due to the relative sign difference between the two schemes using
extremals and $S^{\mu\nu}$ versus $S^{\mu}$. The degeneracy between $w_1,\ldots,w_5$ could be alleviated as well, since the linear combination $\mu_1$ (defined in
\Eq{eqn:mu1new}) could be measured through the correction to the \emph{normal} $n=2$ precession moment (the first term in \Eq{eqn:n2moments}).  
By ``anomalous'' or ``normal'', we mean the term whose precession axis has not been or already been, respectively, predicted by GR.
In addition, the anomalous second-moment terms cannot be expressed as the cross product of $\vec{S}_0$ and an angular velocity vector.

\section{Constraining torsion parameters with Gravity Probe B}
\label{sec:general-torsion-constraints}

The parametrized Post-Newtonian (PPN) formalism has over the past decades demonstrated its success as a theoretical framework of testing GR, by
embedding GR in a broader parametrized class of metric theories of gravitation.  This idea can be naturally generalized by introducing more general departures
from GR, \eg\ torsion. For solar system tests, the seven torsion parameters derived in \Sec{sec:param} define the torsion extension of the PPN
parameters, forming a complete set that parametrizes all observable signatures of torsion to lowest order.

However, most of existing solar system tests cannot constrain the torsion degrees of freedom.  Photons are
usually assumed to decouple from the torsion to preserve gauge invariance (we return below to the experimental basis of this), 
in which case tests using electromagnetic signals (\eg\ Shapiro time delay and the deflection of light) can only
constrain the metric, \ie the PPN parameter $\gamma$, as we explicitly calculate in Appendix \ref{appendix:shapiro} and
Appendix \ref{subsec:deflec}.  Naively, one might expect that Mercury's perihelion shift could constrain torsion parameters if Mercury's orbit is an autoparallel curve,
but calculations in Appendix \ref{subsec:advance-perih-auto} and Appendix \ref{subsec:advance-perih-extreme} show that to lowest
order, the perihelion shift is nonetheless only sensitive to the metric. Moreover, PPN calculations
\cite{Will:2005va} show that a complete account of the  perihelion shift must involve second-order parameters in $m/r$ (\eg\ the PPN
parameter $\beta$), which are beyond our first-order parametrization, as well as the first-order ones. We therefore neglect the constraining power of
Mercury's perihelion shift here.  In contrast, the results in \Sec{subsec:ave_precession} show that Gravity Probe B will be very
sensitive to torsion parameters even if only the average precession rates can be measured.  

We may also constrain torsion with experimental upper bounds on the photon mass, since the ``natural'' extension of Maxwell Lagrangian
($\partial_\mu \to \nabla_\mu$ using the full connection) breaks gauge invariance and introduces anomalous electromagnetic forces and a
quadratic term in $A_\mu$ that may be identified with the photon mass. In Appendix \ref{appendix:photon-mass}, we estimate the constraints
on the torsion parameters $t_1$ and $t_2$ from the measured photon mass limits, and show that these ground-based experiments can constrain
$t_1$ or $t_2$ only to a level of the order unity, \ie, not enough to be relevant to this paper.  

\begin{figure*}
\centering
\begin{displaymath}
\begin{array}{cc}
\includegraphics[width=0.5\textwidth]{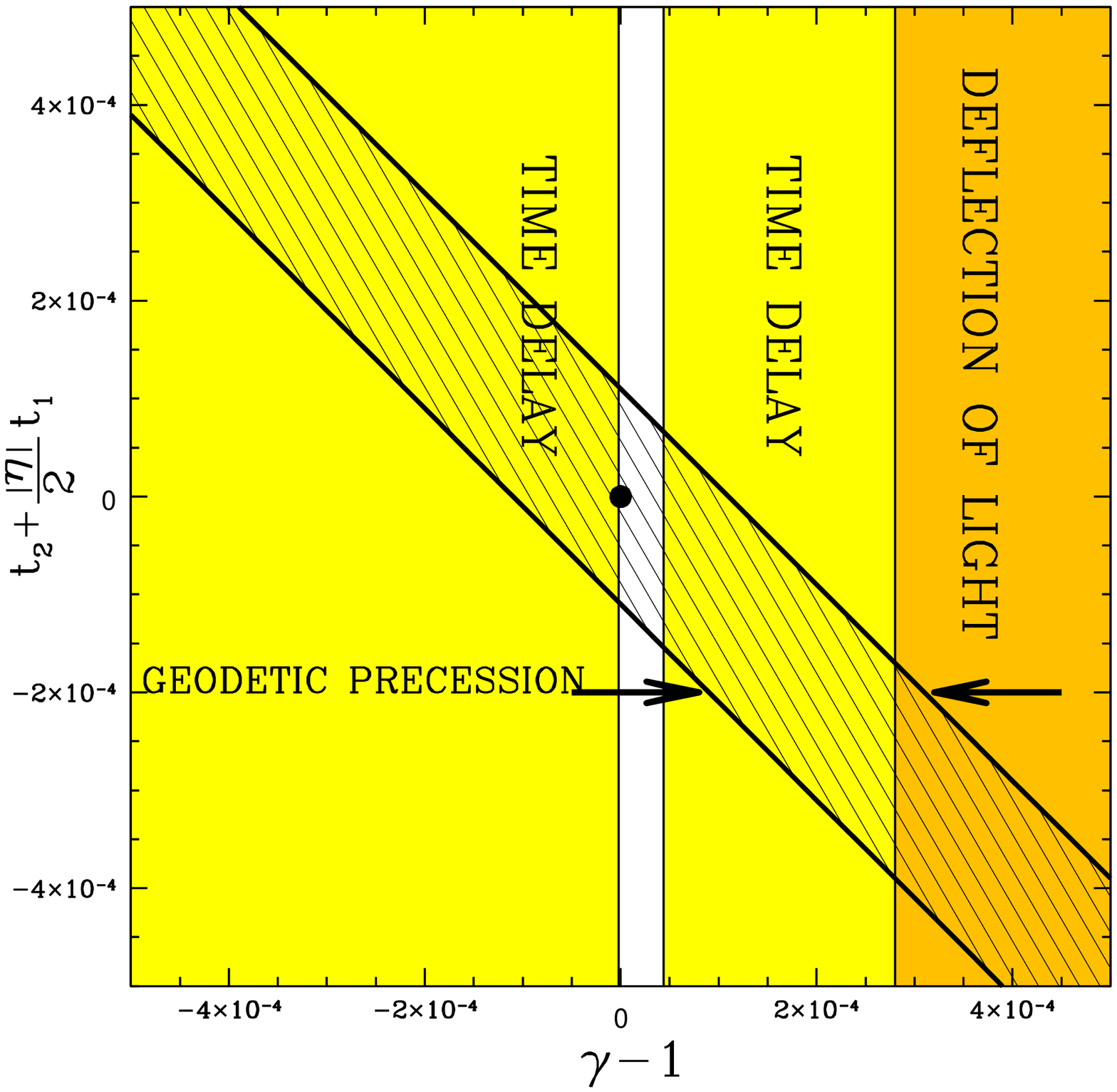} & 
\includegraphics[width=0.5\textwidth]{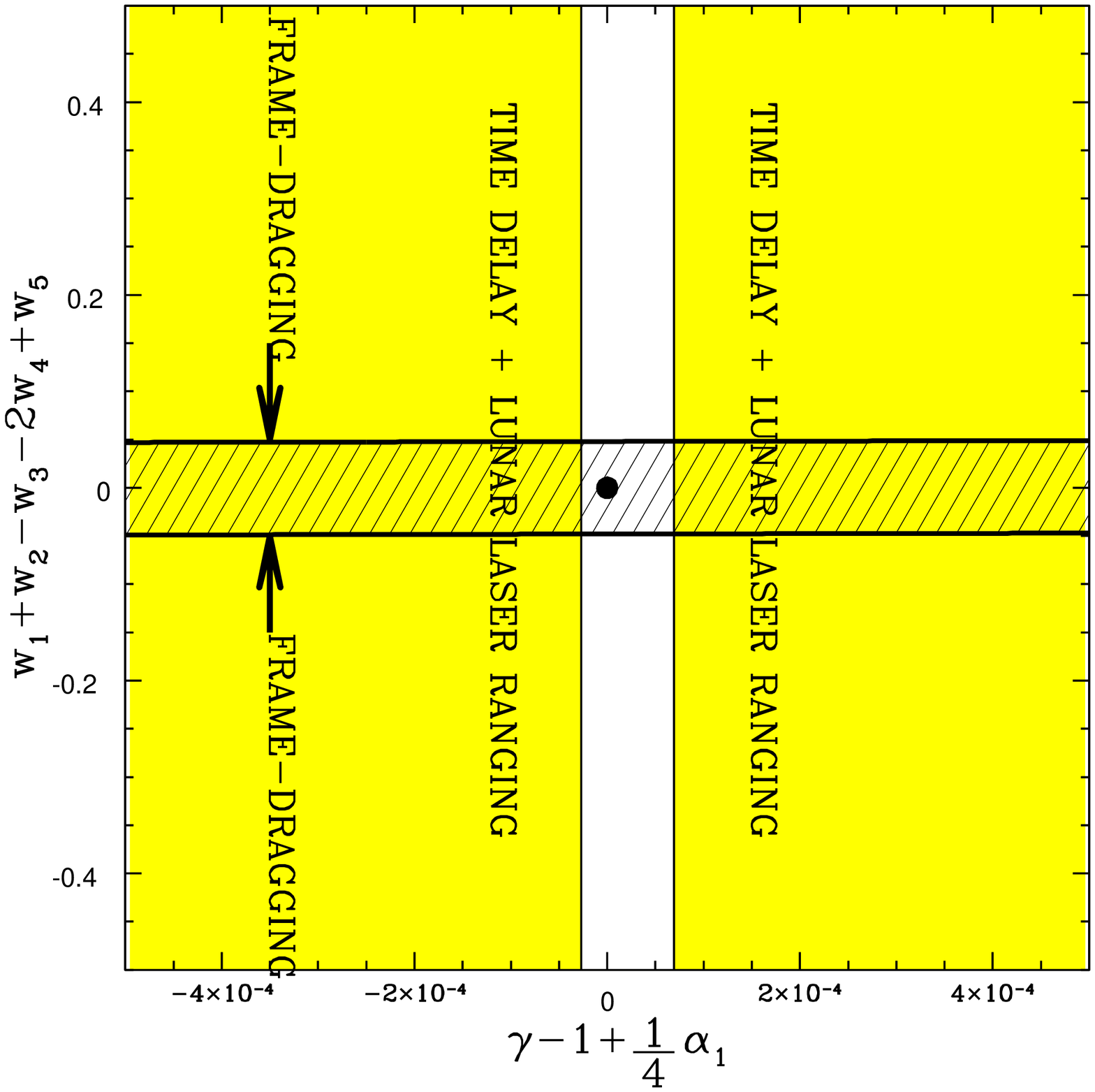} 
\end{array}
\end{displaymath}
\caption{constraints on the PPN parameters ($\gamma$, $\alpha_1$) and torsion parameters 
($t_1$, $t_2$, $w_1$ \ldots $w_5$) from solar system tests.  General Relativity corresponds to the black dot 
($\gamma -1=\alpha_1=$ all torsion parameters $=0$).  
Left panel: the shaded regions in the parameter space have already been ruled out by the deflection of light (orange/grey) 
and Shapiro time delay (yellow/light grey).  Gyroscope experiments are sensitive to torsion parameters.  
If the geodetic precession measured by Gravity Probe B is consistent with GR, this will rule out everything 
outside the hatched region, implying that $-1.5\times 10^{-4} <  t_2+\frac{|\eta|}{2}t_1 < 1.1\times 10^{-4}$
(assuming a target angle accuracy of 0.5 milli-arcseconds).  The unpublished preliminary results of Gravity Probe B have 
confirmed the geodetic precession to better than 1\%, giving a constraint
 $| t_2+\frac{|\eta|}{2}t_1 | \lesssim 0.01$.  
   Right panel: the shaded regions in the parameter space have already been ruled out by Shapiro time delay combined with lunar laser ranging experiment (yellow/light grey).  Lunar laser ranging constrains $|\alpha_1|< 10^{-4}$ \cite{Will:2005va}.
  If the frame-dragging effect measured by Gravity Probe B is consistent with GR, this will rule out everything outside the hatched region, implying that $|w_1+w_2-w_3-2w_4+w_5| < 4.8\times 10^{-2}$.}
\label{fig:extendPPN-constraint}
\end{figure*}

In Appendix \ref{appendix:solar-tests}, we confront solar system tests with the predictions from GR generalized with our torsion parameters.
In general, it is natural to assume that all
metric parameters take the same form as in PPN formalism \footnote{This may not be completely true in some particular theories, \eg\
$\mathcal{H} \ne -2$ in Einstein-Hayashi-Shirafuji theories in the autoparallel scheme, shown in Table \ref{table:EHS1}.}, \ie
\cite{Will:2005va}
\bena
\mathcal{H} &=& -2\,,\label{eqn:H=-2}\\
\mathcal{F} &=& 2\gamma\,,\\
\mathcal{G} &=& -(1+\gamma+\frac{1}{4}\alpha_1)\,. 
\eena
Therefore, Shapiro time delay and the deflection of light share the same multiplicative bias factor $(\mathcal{F}-\mathcal{H})/4 = (1+\gamma)/2$ relative to the GR prediction.  
The analogous bias for gravitational redshift is unity since $(\Delta\nu/\nu)/(\Delta\nu/\nu)^{(GR)}=-\mathcal{H}/2=1$. 
In contrast, both the geodetic precession and the frame-dragging effect have a non-trivial multiplicative bias in Eqs.(\ref{eqn:btdef}) and (\ref{eqn:bmudef}):
\bena
b_t &=& \frac{1}{3}(1+2\gamma)+\frac{1}{3}(2t_2+|\eta|t_1) \,,\\
b_\mu &=& \frac{1}{2}(1+\gamma+\frac{1}{4}\alpha_1)-\frac{1}{4}\left(w_1+w_2-w_3\right.\nonumber\\
 & & \left. -2w_4+w_5\right)\,.
\eena
We list the observational constraints that solar system tests can place on the PPN and torsion parameters in Table \ref{tab:general-torsion-constraint}
and plot the constraints in the degenerate parameter spaces in Figure \ref{fig:extendPPN-constraint}.
We see that GPB will 
ultimately
 constrain the linear combination $t_2+\frac{|\eta|}{2}t_1$ (with $\eta$ depending on the parallel transport scheme) 
at the $10^{-4}$ level and the combination $w_1+w_2-w_3-2w_4+w_5$ at the $1\%$ level. 
The unpublished preliminary results of GPB have confirmed the geodetic precession to less than 1\% level.  This imposes a constraint on $|t_2+\frac{|\eta|}{2}t_1|\lesssim 0.01$.  The combination $w_1+w_2-w_3-2w_4+w_5$ cannot be constrained by frame-dragging until GPB will manage to improve the accuracy to the target level of less than 1 milli-arcsecond.  

\begin{table*}
\noindent 
\footnotesize{
\begin{center}
\begin{tabular}{|l||l|c|p{3.5cm}|}
\hline
Effects  & Torsion Biases  &  Observ. Constraints & Remarks \\ \hline
Shapiro time delay & $\Delta t/\Delta t ^{(GR)}=(1+\gamma)/2$ &  $\gamma-1 = (2.1\pm 2.3)\times 10^{-5}$ & Cassini tracking \cite{Bertotti:2003rm}  \\\hline
Deflection of light & $\delta/\delta^{(GR)}=(1+\gamma)/2 $   & $\gamma-1 = (-1.7\pm 4.5)\times 10^{-4}$ &  VLBI \cite{Shapiro2004} \\\hline
Gravitational redshift & $(\Delta\nu/\nu)/(\Delta\nu/\nu)^{(GR)}=1$ & no constraints &  \\\hline
Geodetic Precession & $\Omega_G/\Omega_G^{(GR)}=b_t$ & $\left|(\gamma-1)+(t_2+\frac{|\eta|}{2}t_1)\right|< 1.1\times 10^{-4}$ & Gravity Probe B \\ \hline
Frame-dragging & $\Omega_F/\Omega_F^{(GR)}=b_\mu$ & $\left|(\gamma-1+\frac{1}{4}\alpha_1)-\frac{1}{2}(w_1+w_2-w_3-2w_4+w_5)\right|<0.024$ & Gravity Probe B \\ \hline
\end{tabular}
\caption{Constraints of PPN and torsion parameters with solar system tests.  The observational constraints on PPN parameters are taken from Table 4 of \cite{Will:2005va}. 
Unpublished preliminary results of Gravity Probe B have confirmed geodetic precession to better than 1\%, 
giving a constraint $|(\gamma-1)+(t_2+\frac{|\eta|}{2}t_1)| \lesssim 0.01$.  
The full GPB results are yet to be released, so whether the frame dragging
will agree with the GR prediction is not currently known.  The last two rows 
show the limits that would correspond to a GPB result consistent with GR, assuming an angle accuracy of 0.5 milli-arcseconds.}
\label{tab:general-torsion-constraint}
\end{center}
}
\end{table*}

\section{Linearized Kerr solution with torsion in Weitzenb\"ock spacetime}
\label{sec:counterexample}

So far, we have used only symmetry principles to derive the most general torsion possible around Earth to lowest order. 
We now turn to the separate question of whether there is any gravitational Lagrangian that actually produces torsion around Earth.
We will show that the answer is yes by exploring the specific example of the 
Hayashi-Shirafuji Lagrangian \cite{HS1} in Weitzenb\"ock spacetime, showing that it populates a certain subset of
the torsion degrees of freedom that we parametrized above and that this torsion mimics the Kerr metric to lowest order even though the Riemann curvature of 
spacetime vanishes.
We begin with a
brief review of Weitzenb\"ock spacetime and the Hayashi-Shirafuji Lagrangian, then give the linearized solution in terms of the seven parameters
$t_1,t_2,w_1,\ldots,w_5$ from above. The solution we will derive is a particular special case of 
what the symmetry principles allow, and is for the particularly simple case where the Riemann curvature vanishes (Weitzenb\"ock spacetime). 
 Later in \Sec{subsec:linearinterpol}, 
we will give a more general Lagrangian producing both torsion and curvature, effectively interpolating between the Weitzenb\"ock case below and standard GR.

We adopt the convention only here in \Sec{sec:counterexample} 
 and \Sec{subsec:linearinterpol} 
that Latin letters are indices for the internal basis, whereas Greek letters are 
spacetime indices, both running from 0 to 3.  

\subsection{Weitzenb\"ock spacetime}

We give a compact review of Weitzenb\"ock spacetime and Hayashi-Shirafuji Lagrangian here and in \Sec{subsec:HS} respectively. 
We refer the interested reader to their original papers \cite{weitzenbock,HS1} for a complete survey of these subjects. 

Weitzenb\"ock spacetime is a Riemann-Cartan spacetime in which the Riemann curvature tensor, defined in \Eq{eqn:curvature}, vanishes identically: 
\beq{eqn:vanishingR}
R^{\rho}_{\phantom{\rho}\lambda\nu\mu}(\Gamma)=0\,.
\eeq
\Fig{fig:spaces} illustrates how Weitzenb\"ock spacetime is related to other spacetimes.

Consider a local coordinate neighborhood of a point $p$ in a Weitzenb\"ock manifold with local coordinates $x^{\mu}$.  Introduce the coordinate basis $\left\{\bar{E}_{\,\mu}\right\}=\{\left(\partial/\partial x^{\mu}\right)_p\}$ and the dual basis $\left\{\bar{E}^{\,\mu}\right\}=\{\left(d x^{\mu}\right)_p\}$.  A vector $\bar{V}$ at $p$ can be written as $\bar{V}=V^{\mu}\bar{E}_{\,\mu}$.  The manifold is equipped with an inner product; the metric is the inner product of the coordinate basis vectors,
\[
g(\bar{E}_{\,\mu},\bar{E}_{\,\nu})=g(\bar{E}_{\,\nu},\bar{E}_{\,\mu})=g_{\mu\nu}\,.
\]
There exists a quadruplet of orthonormal vector fields $\bar{e}_{\,k}(p)$, where $\bar{e}_{\,k}(p)=e_{\,k}^{\,\phantom{1}\mu}(p)\bar{E}_{\,\mu}$, such that
\ben
g(\bar{e}_{\,k},\bar{e}_{\,l})=g_{\mu\nu}e_{\,k}^{\,\phantom{1}\mu}e_{\,l}^{\,\phantom{1}\nu}=\eta_{kl}\,,
\een
where $\eta_{kl}=\mathrm{diag}(-1,1,1,1)$.
There also exists a dual quadruplet of orthonormal vector fields $\bar{e}^{\,k}(p)$, where $\bar{e}^{\,k}(p)=e^{\,k}_{\,\phantom{1}\mu}(p)\bar{E}^{\,\mu}$, such that
\ben
e_{\,k}^{\,\phantom{1}\mu}e^{\,k}_{\,\phantom{1}\nu}=\delta_{\phantom{1}\nu}^{\mu}\,,
\qquad  e_{\,k}^{\,\phantom{1}\mu} e^{\,l}_{\,\phantom{1}\mu}=\delta_{k}^{\phantom{1}l}\,.\label{eqn:bbg1}
\een
This implies that
\ben
\eta_{kl}e^{\,k}_{\,\phantom{1}\mu}e^{\,l}_{\,\phantom{1}\nu}=g_{\mu\nu}\,.\label{eqn:bbg2}
\een
which is often phrased as the $4\times 4$ matrix ${\bf e}$ (a.k.a.~the {\it tetrad} or {\it vierbein}) 
being ``the square root of the metric''.

An alternative definition of Weitzenb\"ock spacetime that is equivalent to that of \Eq{eqn:vanishingR} is the requirement that the Riemann-Cartan spacetime 
admit a quadruplet of linearly independent \emph{parallel vector fields} $e_k^{\phantom{1}\mu}$, defined by\footnote{Note that 
Hayashi and Shirafuji \cite{HS1} adopted a convention where the \mbox{order} of the lower index placement in the connection is opposite 
to that in \Eq{eqn:parallel-vector}.}
\beq{eqn:parallel-vector}
\nabla_{\mu}e_k^{\phantom{1}\nu}=\partial_{\mu}e_k^{\phantom{1}\nu}+\Gamma^{\nu}_{\phantom{1}\mu\lambda}e_k^{\phantom{1}\lambda}=0\,.
\eeq
Solving this equation, one finds that
\beq{eqn:conn-b}
\Gamma^{\lambda}_{\phantom{1}\mu\nu}=e_k^{\phantom{1}\lambda}\partial_{\mu}e^k_{\phantom{1}\nu}\,,
\eeq
and that the torsion tensor
\beq{eqn:tor-b}
S_{\mu\nu}^{\phantom{12}\lambda}=\frac{1}{2}e_k^{\phantom{1}\lambda}(\partial_{\mu}e^k_{\phantom{1}\nu}-\partial_{\nu}e^k_{\phantom{1}\mu})\,.
\eeq
This property of allowing globally parallel basis vector fields was termed ``teleparallelism'' by Einstein, since it allows unambiguous parallel transport,
and formed the foundation of the torsion theory he termed ``new general relativity'' \cite{Einstein:1928-30,Einstein:1930b,moller:1961,Pellegrini:1962,moller:1978,Unzicker:2005in,Obukhov:2004hv,Vargas:1992ab,Vargas:1992ac,Mueller-Hoissen:1983vc,Mielke:1992te,Kreisel:1979kh,Treder:1978vf,Kreisel:1980kb,Pimentel:2004bp,Maluf:2001ef}.

A few additional comments are in order:
\begin{enumerate}

\item It is easy to verify that the first definition of Weitzenb\"ock spacetime (as curvature-free, \ie via \Eq{eqn:vanishingR}) 
follows from the second definition --- one simply uses the the explicit expression for the connection (\Eqnopar{eqn:conn-b}). It is
also straightforward to verify that $\nabla_{\mu}g_{\nu\rho}=0$ using \Eq{eqn:bbg2} and (\ref{eqn:parallel-vector}).

\item \Eq{eqn:conn-b} is form invariant under general (spacetime) coordinate transformations due to the nonlinear transformation law (\Eq{eqn:gamma-law}) of
the connection, provided that $e_k^{\phantom{1}\mu}$ and $e^k_{\phantom{1}\mu}$ transform as a contravariant vector and a covariant vector, respectively.

\item The Weitzenb\"ock spacetime preserves its geometry under \emph{global} proper orthochronous Lorentz transformations, \ie a new equivalent quadruplet of
parallel vector fields $\underline{e}'$ is obtained by a global proper orthochronous Lorentz transformation,
${e'}_k^{\phantom{1}\mu}=\Lambda^l_{\phantom{1}k} e_l^{\phantom{1}\mu}$.  

\end{enumerate}

\subsection{Hayashi-Shirafuji Lagrangian}\label{subsec:HS}

The Hayashi-Shirafuji Lagrangian \cite{HS1} is a gravitational Lagrangian density constructed in the geometry of Weitzenb\"ock
spacetime\footnote{The Hayashi-Shirafuji theory differs from the teleparallel gravity theory decribed in \cite{Arcos:2004zh,Arcos:2005ec}, which is argued to be fully equivalent to GR.}.  It is a Poincar\'e gauge theory in that the parallel vector fields $\underline{e}\,_k$ (rather than the metric or
torsion) are the basic entities with respect to which the action is varied to obtain the gravitational field equations.  

First, note that the torsion tensor in \Eq{eqn:tor-b} is \emph{reducible} under the group of global Lorentz transformation.  
It can be decomposed into three irreducible parts under this Lorentz group \cite{Hayashi-Bregman}\footnote{Note 
that we denote the irreducible parts (\ie  $t_{\lambda\mu\nu},v_{\mu},a_{\mu}$) by 
the same letters as in \cite{HS1}, but that these quantities here are only one half as large as in \cite{HS1}, 
due to different conventions in the definition of torsion.  
Similarly, the quantities $c_1,c_2,c_3$ in \Eq{eqn:HSlag} are four times as large as in \cite{HS1}.},
\ie into parts which do not mix under a global Lorentz transformation:
\bena
t_{\lambda\mu\nu} &=& \frac{1}{2}(S_{\nu\mu\lambda}+S_{\nu\lambda\mu})+\frac{1}{6}(g_{\nu\lambda}v_{\mu}+g_{\nu\mu}v_{\lambda})\nonumber\\
& & -\frac{1}{3}g_{\lambda\mu}v_{\nu}\,,\label{eqn:decomp-torsion-t}\\
v_{\mu}&=& S_{\mu\lambda}^{\phantom{12}\lambda}\,,\\
a_{\mu}&=& \frac{1}{6}\bar{\epsilon}_{\mu\nu\rho\sigma}S^{\sigma\rho\nu}\,,\label{eqn:decomp-torsion-a}
\eena
Here $\bar{\epsilon}_{\mu\nu\rho\sigma}=\sqrt{-g}\epsilon_{\mu\nu\rho\sigma}$ and
$\bar{\epsilon}\,^{\mu\nu\rho\sigma}=\epsilon\,^{\mu\nu\rho\sigma}/\sqrt{-g}$ are 4-tensors, and the Levi-Civita symbol is
normalized such that $\epsilon_{0123}=-1$ and $\epsilon^{0123}=+1$. The tensor $t_{\lambda\mu\nu}$ satisfies
$t_{\lambda\mu\nu}=t_{\mu\lambda\nu}$, $g^{\mu\nu}t_{\lambda\mu\nu}=g^{\lambda\mu}t_{\lambda\mu\nu}=0$, and
$t_{\lambda\mu\nu}+t_{\mu\nu\lambda}+t_{\nu\lambda\mu}=0$. Conversely, the torsion can be written in terms of its 
irreducible parts as 
\beq{eqn:decomp-torsion-combination} S_{\nu\mu\lambda} =
\frac{2}{3}(t_{\lambda\mu\nu}-t_{\lambda\nu\mu})+\frac{1}{3}(g_{\lambda\mu}v_{\nu}-g_{\lambda\nu}v_{\mu})+\bar{\epsilon}_{\lambda\mu\nu\rho}a^{\rho}\,.
\eeq

In order that the field equation be a second-order differential equation in $\underline{e}\,_k$ (so that torsion can propagate),
the Lagrangian is required to be quadratic in the torsion tensor. In addition, the Lagrangian should be invariant under the group of
general coordinate transformations, under the global proper orthochronous Lorentz group, and under parity reversal in the
internal basis ($\underline{e}\,_{0}\to \underline{e}\,_{0},\,\underline{e}\,_{a}\to -\underline{e}\,_{a}$). Hayashi and
Shirafuji suggested the gravitational action of the following form \cite{HS1}:
\beqa{eqn:HSlag}
I_G &=& \int \mathrm{d}^4 x \sqrt{-g}\:[\:\frac{1}{2\kappa}R\left(\{\,\}\right)+c_1\, t^{\lambda\mu\nu}t_{\lambda\mu\nu}\nonumber\\
    & & +c_2\, v^{\mu}v_{\mu}+c_3\, a^{\mu} a_{\mu}]\,,
\eeqa
where $c_1,c_2,c_3$ are three free parameters, 
$R\left(\{\,\}\right)$ is the scalar curvature calculated using the Levi-Civita connection and $\kappa=8\pi G/c^4$.  
The \emph{vacuum} field equations are obtained by varying this action with respect to 
the tetrad
$e^{k}_{\phantom{1}\nu}$ and then multiplying by 
$\eta^{kj} e_{j}^{\phantom{1}\mu}$.  
Note that in Hayashi-Shirafuji theory, the torsion (or equivalently, the connection) is not an independent variable as in some standard torsion theories
\cite{hehl}.  Instead, the torsion is exclusively determined by the tetrad via \Eq{eqn:tor-b}.  The resultant field equation is 
\beq{eqn:HSfieldeq}
\frac{1}{2\kappa}G^{\mu\nu}(\{\,\})+\nabla_{\lambda}F^{\mu\nu\lambda}+v_{\lambda}F^{\mu\nu\lambda}+H^{\mu\nu}-\frac{1}{2}g^{\mu\nu}L_2=0\,.
\eeq
Here the first term denotes the Einstein tensor calculated using the Levi-Civita connection,  
but the field equation receives important non-Riemannian contributions from torsion through the other terms.  The other tensors in \Eq{eqn:HSfieldeq} are defined
as follows:
\bena
F\,^{\mu\nu\lambda} &=& c_1(t^{\mu\nu\lambda}-t^{\mu\lambda\nu})+c_2(g^{\mu\nu}v^{\lambda}-g^{\mu\lambda}v^{\nu})\nonumber\\
& & -\frac{1}{3}c_3\bar{\epsilon}^{\mu\nu\lambda\rho}a_{\rho}\,,\\
H^{\mu\nu} &=& 2S^{\mu\sigma\rho}F_{\rho\sigma}^{\phantom{12}\nu}-S^{\sigma\rho\nu}F^{\mu}_{\phantom{1}\rho\sigma}\,,\label{eqn:eom-HS-Hterm}\\
L_2 &=& c_1\, t^{\lambda\mu\nu}t_{\lambda\mu\nu} +c_2\, v^{\mu}v_{\mu}+c_3\, a^{\mu} a_{\mu}\,.
\eena
Since torsion is the first derivative of the tetrad as per \Eq{eqn:tor-b}, the field equation is a nonlinear second-order differential equation of the tetrad.  
Consequently, the tetrad (hence the torsion) can propagate in the vacuum.

\subsection{Static, spherically and parity symmetric vacuum solution}\label{subsubsec:sph-sym-HS}

Hayashi and Shirafuji derived the exact static, spherically and parity symmetric $R_{\mu\nu\rho\sigma}=0$ 
vacuum solutions for this Lagrangian in \cite{HS1}.  The parallel vector fields take the following form in isotropic 
rectangular coordinates (here Latin letters are spatial indices) \cite{HS1}:
\bena
e_{0}^{\phantom{1}0} &=& \left(1-\frac{m_0}{pr}\right)^{-p/2} \left(1+\frac{m_0}{qr}\right)^{q/2}\,,\nonumber\\
e_{0}^{\phantom{1}i} &=& e_{a}^{\phantom{111}0}=0\,,\nonumber\\
e_{a}^{\phantom{1}i} &=&  \left(1-\frac{m_0}{pr}\right)^{-1+p/2} \left(1+\frac{m_0}{qr}\right)^{-1-q/2}\delta^{i}_a\,,
\eena
where $m_0$ is a parameter with units of mass and will be related to the physical mass of the central gravitating body in \Sec{sec:constrain-torsion}.  
The new parameters $p$ and $q$ are functions of a dimensionless parameter $\epsilon$:
\bena
\epsilon &\equiv & \frac{\kappa (c_1+c_2)}{1+\kappa (c_1+4c_2)}\,,\label{eqn:epsilondef}\\
p & \equiv & \frac{2}{1-5\epsilon}\{ [ (1-\epsilon)(1-4\epsilon)]^{1/2}-2\epsilon\}\,,\\
q &\equiv & \frac{2}{1-5\epsilon}\{ [ (1-\epsilon)(1-4\epsilon)]^{1/2}+2\epsilon\}\,.
\eena
Here $\kappa = 8\pi G$.

The line element in the static, spherically and parity symmetric field takes the exact form \cite{HS1}
\beqa{eqn:sphe-line-element}
ds^2 &=& -\left(1-\frac{m_0}{pr}\right)^{p}\left(1+\frac{m_0}{qr}\right)^{-q}dt^2\nonumber\\
     & & +\left(1-\frac{m_0}{pr}\right)^{2-p} \left(1+\frac{m_0}{qr}\right)^{2+q}dx^i dx^i\,.
\eena
In order to generalize this solution to the axisymmetric case, we transform the parallel 
vector fields into standard spherical coordinates and keep terms to first order in $m_0/r$ (the subscript ``sp'' stands for ``spherical''):
\begin{widetext}
\ben
{e_{(\textrm{sp})}}\,_k^{\phantom{1}\mu} = 
\begin{array}{cl}        &   \qquad \to \mu \\
   \begin{array}{c} \downarrow \\ k \end{array} 
   & \begin{pmatrix} 1+\frac{m_0}{r} & 0 & 0 & 0 \\ 
    0 & \left[1-\frac{m_0}{r}\left(1+\frac{1}{q}-\frac{1}{p}\right) \right]\sin\theta\,\cos\phi & \frac{\cos\theta\,\cos\phi}{r} & -\frac{\csc\theta\,\sin\phi}{r} \\
    0 & \left[1-\frac{m_0}{r}\left(1+\frac{1}{q}-\frac{1}{p}\right) \right]\sin\theta\,\sin\phi & \frac{\cos\theta\,\sin\phi}{r} & \frac{\csc\theta\,\cos\phi}{r} \\
    0 & \left[1-\frac{m_0}{r}\left(1+\frac{1}{q}-\frac{1}{p}\right) \right]\cos\theta & -\frac{\sin\theta}{r} & 0 
    \end{pmatrix}
  \end{array} 
\een
\end{widetext}

A particularly interesting solution is that for the parameter choice $c_1=-c_2$ so that $\epsilon=0$ and $p=q=2$.  \Eq{eqn:sphe-line-element} shows that the resultant metric coincides with the Schwarzschild metric around an object of mass $m_0$.  The parameter $c_3$ is irrelevant here because of the static, spherically and parity symmetric field.  
When $c_1+c_2$ is small but nonzero, we have $\epsilon \ll 1$ and 
\bena
p &=& 2+\epsilon + \mathcal{O}(\epsilon^2)\,,\\
q &=& 2+9\epsilon + \mathcal{O}(\epsilon^2)\,.
\eena

By using equations (\ref{eqn:bbg2}), (\ref{eqn:conn-b}) and (\ref{eqn:tor-b}), we find that the linearized metric and torsion match our parametrization in \Sec{subsec:spher-symm}.  When $\epsilon \ll 1$, the line element is
\ben
ds^2 = -\left[1-2\frac{m_0}{r}\right]dt^2+\left[1+2(1-2\epsilon)\frac{m_0}{r}\right]dr^2+r^2 d\Omega^2\,,
\een 
and the torsion is
\beqa{eqn:torsion-HS-sp}
S_{tr}^{\phantom{01}t} &=& -\frac{m_0}{2r^2}\,,\\
S_{r\theta}^{\phantom{12}\theta}&=& S_{r\phi}^{\phantom{13}\phi} = -(1-2\epsilon)\frac{m_0}{2r^2}\,,
\eena
both to linear order in $m_0/r$.

\subsection{Solution around Earth}
\label{subsubsec:the axisym-sol2-HS}

We now investigate the field generated by a uniformly rotating spherical body to first order in $\varepsilon_a$.  
It seems reasonable to assume that to first order the metric coincides with the Kerr-like metric, \ie
\beq{eqn:kerr-like-HS}
g_{t\phi}=\mathcal{G}_0(m_0a/r)\sin ^2 \theta\,,
\eeq
around an object of specific angular momentum $a$ in the linear regime $m_0/r\ll 1$ and $a/r \ll 1$.  
Since the Kerr-like metric automatically satisfies 
$G(\{\,\})=0$ in vacuum, the vacuum field equation reduces to
\beq{eqn:reducedfield}
\nabla_{\lambda}F^{\mu\nu\lambda}+v_{\lambda}F^{\mu\nu\lambda}+H^{\mu\nu}-\frac{1}{2}g^{\mu\nu}L_2=0\,.
\eeq
We now employ our parametrization with ``mass'' in \Eq{eqn:w2-axi} replaced by $m_0$, where $m_0$ is the parameter in accordance with \Sec{subsubsec:sph-sym-HS}.  In \Sec{sec:constrain-torsion}, 
we will apply the Kerr solution $\mathcal{G}=-2$ after re-scaling $m_0$ to correspond to the physical mass.  
Imposing the no-curvature condition $R_{\mu\nu\rho\sigma}=0$, we find that this condition and \Eq{eqn:reducedfield} are satisfied to lowest order in $m_0/r$ and $a/r$ if
\beqa{eqn:torsion-HS-kerr-beta}
w_{1}^{(0)} &=& \mathcal{G}_0-\alpha_0,\,,\nonumber\\
w_{2}^{(0)} &=& -2(\mathcal{G}_0-\alpha_0)\,,\nonumber\\
w_{3}^{(0)} &=& w_{4}^{(0)}=\alpha_0\,,\nonumber\\
w_{5}^{(0)} &=& 2\alpha_0.
\eena
Here a superscript $(0)$ indicates the parametrization with $m_0$ in place of $m$. 
$\alpha_0$ is an undetermined constant and should depend on the Lagrangian parameters $c_1$, $c_2$ and $c_3$.  
This parameter has no effect on the precession of a gyroscope or on any of the other observational constraints that we consider, so its value is irrelevant to the present paper.

The parallel vector fields that give the Kerr metric, the connection and the torsion (including the spherically symmetric part) via equations (\ref{eqn:bbg1})--(\ref{eqn:bbg2}) and (\ref{eqn:conn-b})--(\ref{eqn:tor-b}) 
take the following form to linear order:
\begin{eqnarray}
  & & \quad e_k^{\phantom{1}\mu} = {e_{(\textrm{sp})}}\,_k^{\phantom{1}\mu}+ \nonumber\\
  & &  + \begin{array}{cl}
 &  \qquad \to\mu \\ \begin{array}{c} \downarrow \\ k \end{array} & 
\begin{pmatrix} 0 & \:\:\:0\:\:\: & \:\:\:0\:\:\: & -\alpha_0 \frac{m_0a}{r^3} \\ -(\mathcal{G}_0-\alpha_0)\frac{m_0a\sin\theta\sin\phi}{r^2} & 0 & 0 & 0 \\ (\mathcal{G}_0-\alpha_0)\frac{m_0a\sin\theta\cos\phi}{r^2} & 0 & 0 & 0 \\ 0 &  0 & 0 & 0 \end{pmatrix}\end{array}\nonumber\\
\end{eqnarray}

\begin{table*} 
\noindent 
\begin{center}
\begin{tabular}{|l|l|c|c||l|}
\hline
\multicolumn{2}{|l|}{  } & Hayashi-Shirafuji with $m_0$ & EHS with $m_0$ & Definitions \\
\hline
metric     & $\mathcal{H}^{(0)}$ & -2 & -2 & $g_{tt}=-1-\mathcal{H}^{(0)}m_0/r+\mathcal{O}(m_0/r)^2$ \\
\cline{2-4}
parameters & $\mathcal{F}^{(0)}$ & $2(1-2\epsilon)$ & $2(1-2\tau)$ & $g_{rr}=1+\mathcal{F}^{(0)} m_0/r+\mathcal{O}(m_0/r)^2$\\
\hline
geodetic & $t_{1}^{(0)}$ & $-1$ & $-\sigma$ & anomalous, $S_{tr}^{\phantom{01}t} = t_{1}^{(0)}\,m_0/2r^2$ \\
\cline{2-4}
torsions & $t_{2}^{(0)}$ & $-(1-2\epsilon)$ & $-\sigma(1-2\tau)$ & normal, $S_{r\theta}^{\phantom{12}\theta} = S_{r\phi}^{\phantom{13}\phi} = t_{2}^{(0)}\,m_0/2r^2$\\
\hline
         & $w_{1}^{(0)}$ & $\mathcal{G}_0-\alpha_0$ & $\sigma(\mathcal{G}_0-\alpha_0)$ & $S_{r\phi}^{\phantom{0i}t} = w_{1}^{(0)}\,(m_0a/2r^2)\sin^2\theta$ \\
\cline{2-4}
frame-   & $w_{2}^{(0)}$ & $-2(\mathcal{G}_0-\alpha_0)$ & $-2\sigma(\mathcal{G}_0-\alpha_0)$ & $S_{\theta\phi}^{\phantom{0i}t} = w_{2}^{(0)}\,(m_0a/2r)\sin\theta\cos\theta$ \\
\cline{2-4}
dragging & $w_{3}^{(0)}$ & $\alpha_0$ & $\sigma\alpha_0$ & $S_{t\phi}^{\phantom{0i}r} = w_{3}^{(0)}\,(m_0a/2r^2)\sin^2\theta$ \\
\cline{2-4}
torsions & $w_{4}^{(0)}$ & $\alpha_0$ & $\sigma\alpha_0$ & $S_{t\phi}^{\phantom{0i}\theta} = w_{4}^{(0)}\,(m_0a/2r^3)\sin\theta\cos\theta$\\
\cline{2-4}
         & $w_{5}^{(0)}$ & $2\alpha_0$ & $2\sigma\alpha_0$ & $S_{tr}^{\phantom{0i}\phi} = w_{5}^{(0)} \,m_0a/2r^4$\\
\hline
\end{tabular}
\caption{Summary of metric and torsion parameters for General Relativity, Hayashi-Shirafuji gravity 
and Einstein-Hayashi-Shirafuji (EHS) theories.  The subscript 0 indicates all parameter values are normalized by an arbitrary constant $m_0$ (with the units of mass) that is not necessarily the physical mass of the body generating the gravity.  The parameter $\alpha_0$ in frame-dragging torsions is an undetermined constant and should depend on the Hayashi-Shirafuji Lagrangian parameters $c_1$, $c_2$ and $c_3$.  The parameter $\tau$, defined in \Eq{eqn:epsilondef} and assumed small, is an indicator of how close the emergent metric is to the Schwarzschild metric.  The values in the column of Einstein-Hayashi-Shirafuji interpolation are those in the Hayashi-Shirafuji \emph{times} the interpolation parameter $\sigma$.}
\label{table:HS1}
\end{center}
\end{table*}

\section{A toy model: linear interpolation in Riemann-Cartan Space between GR and Hayashi-Shirafuji Lagrangian}
\label{subsec:linearinterpol}

We found that the Hayashi-Shirafuji Lagrangian admits both the Schwarzschild metric and (at least to linear order) the Kerr metric, but
in the Weitzenb\"ock spacetime where there is no Riemann curvature and all spacetime structure is due to torsion.
This is therefore an opposite extreme of GR, which admits these same metrics in Riemann spacetime with all curvature and no torsion.
Both of these solutions can be embedded in Riemann-Cartan spacetime, 
and we will now present a more general two-parameter family of Lagrangians that interpolates between these two extremes, always allowing the
Kerr metric and generally explaining the spacetime distortion with a combination of curvature and torsion.  
After the first version of this paper was submitted, Flanagan and Rosenthal showed that the Einstein-Hayashi-Shirafuji Lagrangian 
has serious defects \cite{Flanagan:2007dc}, while leaving open the possibility that there may be other viable Lagrangians
in the same class (where spinning objects generate and feel propagating torsion).
This Lagrangian should therefore not be viewed as a viable physical model, but as a pedagogical toy model 
admitting both curvature and torsion, 
giving concrete illustrations of the various effects and constraints that we discuss.

This family of theories, which we will term 
Einstein-Hayashi-Shirafuji (EHS) theories, have 
an action in in Riemann-Cartan space of the form
\beqa{eqn:HSlag-interp}
I_G &=& \int \mathrm{d}^4 x \sqrt{-g}\:[\:\frac{1}{2\kappa}R\left(\{\,\}\right)+\sigma^2\,c_1\, t^{\lambda\mu\nu}t_{\lambda\mu\nu}\nonumber\\
    & & +\sigma^2\,c_2\, v^{\mu}v_{\mu}+\sigma^2\,c_3\, a^{\mu} a_{\mu}]\,
\eeqa
where $\sigma$ is a parameter in the range $0\le\sigma\le 1$. 
Here the tensors $t_{\lambda\mu\nu}$, $v_{\mu}$ and $a_{\mu}$ are the decomposition 
(in accordance with  Eqs.\ref{eqn:decomp-torsion-t}---\ref{eqn:decomp-torsion-a})
of $\sigma^{-1}S_{\nu\mu\lambda}$,
which is independent of $\sigma$ and depends only on $e^i_{\phantom{1}\mu}$ as per \Eq{eqn:tor-b-interp}.  The function $\sigma^2$ associated with the coefficients $c_1$, $c_2$ and $c_3$ in \Eq{eqn:HSlag-interp} may be replaced by any other regular function of $\sigma$ that approaches to zero as $\sigma\to 0$.
The metric in the EHS theories is defined in \Eq{eqn:bbg2}.  
Similar to the Hayashi-Shirafuji theory, the field equation for EHS theories is obtained by varying the action with respect to the tetrad.  The resultant field equation is  identical to that
for the Hayashi-Shirafuji Lagrangian (Eq.~\ref{eqn:HSfieldeq}) 
except for the replacement $c_{1,2,3}\to \sigma^2 c_{1,2,3}$. 
Also, the $S^{\mu\sigma\rho}$ in \Eq{eqn:eom-HS-Hterm} is replaced by $\sigma^{-1}S^{\mu\sigma\rho}$.  
Thus the EHS Lagrangian admits the same solution for 
$e_k^{\phantom{1}\mu}$.  
Since the metric is independent of the parameter $\sigma$, the EHS Lagrangian admits both the spherically symmetric metric in \Eq{eqn:sphe-line-element} and the Kerr-like metric in \Eq{eqn:kerr-like-HS}, at least to the linear order.  
For the spherically symmetric metric, the parameter $\epsilon$ in Hayashi-Shirafuji theory is 
generalized to a new parameter $\tau$ in EHS theories, defined by the replacement $c_{1,2}\to \sigma^2 c_{1,2}$:
\beq{eqn:epsdefEHS}
\tau \equiv  \frac{\kappa \sigma^2(c_1+c_2)}{1+\kappa \sigma^2(c_1+4c_2)}\,.
\eeq

The torsion around Earth is linearly proportional to $\sigma$, given by the parameter $\sigma$ times the solution in \Eq{eqn:torsion-HS-sp} and (\ref{eqn:torsion-HS-kerr-beta}):
\beq{eqn:tor-b-interp}
S_{\mu\nu}^{\phantom{12}\lambda}\equiv\frac{\sigma}{2}e_k^{\phantom{1}\lambda}(\partial_{\mu}e^k_{\phantom{1}\nu}-\partial_{\nu}e^k_{\phantom{1}\mu})\,.  
\eeq
By virtue of \Eq{eqn:fullconn1} 
(the metric compatibility condition), 
it is straightforward to show that the connection is of the form
\ben
\Gamma^{\rho}_{\phantom{\rho}\mu\nu} = (1-\sigma) \left\{ \begin{array}{c} \rho \\ \mu\nu \end{array} \right\} 
+ \sigma\,e_k^{\phantom{1}\rho}\partial_{\mu}e^k_{\phantom{1}\nu}\,.
\een
EHS theory thus interpolates smoothly between metric gravity \eg\ GR $(\sigma=0)$ and the all-torsion Hayashi-Shirafuji theory ($\sigma=1$).  
If $\sigma\ne 1$, it is straightforward to verify that the curvature calculated by the full connection does not vanish.  Therefore, the EHS theories live in
neither Weitzenb\"ock space nor the Riemann space, but in the Riemann-Cartan space that admits both torsion and curvature.   

It is interesting to note that since the Lagrangian parameters $c_1$ and $c_2$ are independent of the torsion parameter $\sigma$, the effective
parameter $\tau$ is not necessarily equal to zero when $\sigma=0$ (\ie, $\sigma^2 c_1$ or $\sigma^2 c_2$ can be still finite).
In this case ($\sigma=0$ and yet $\tau \ne 0$), obviously this EHS theory is an extension to GR without adding torsion.  In
addition to the extra terms in the Lagrangian of \Eq{eqn:HSlag-interp}, the extension is subtle in the symmetry of the
Lagrangian.   In the tetrad formalism of GR, \emph{local} Lorentz transformations are symmetries in the internal space of
tetrads.  Here in this $\sigma=0, \tau\ne 0$ EHS theory, the allowed internal symmetry is \emph{global} Lorentz
transformations as in the Weitzenb\"ock spacetime, because $t_{\lambda\mu\nu}$, $v_\mu$ and $a_\mu$ contain the partial
derivatives of tetrads (see Eq.~\ref{eqn:tor-b-interp}).  So the $\sigma=0$ and $\tau \ne 0$ EHS theory is a tetrad theory in
Riemann spacetime with less gauge freedom.  

Since GR is so far consistent with all known observations, it is interesting to explore (as we will below) what observational upper limits can be placed on both $\sigma$ and $\tau$
.
\begin{table*} 
\noindent 
\begin{center}
\begin{tabular}{|l|l|c|c|c||l|}
\hline
\multicolumn{2}{|l|}{  } & GR &  EHS with autoparallels & EHS with extremals & Definitions \\
\hline
mass & $m$ & $m=m_0$ & $m=(1-\sigma)m_0$ & $m=m_0$ & set by Newtonian limit \\
\hline
metric     & $\mathcal{H}$ & -2 & $-2/(1-\sigma)$ & $-2$ & $g_{tt}=-1-\mathcal{H}m/r+\mathcal{O}(m/r)^2$ \\
\cline{2-5}
parameters & $\mathcal{F}$ & 2 & $2(1-2\tau)/(1-\sigma)$ & $2(1-2\tau)$ & $g_{rr}=1+\mathcal{F}m/r+\mathcal{O}(m/r)^2$\\
\cline{2-5}
           & $\mathcal{G}$ & -2 & -2 & -2 & $g_{t\phi}=\mathcal{G}(ma/r)\sin^2\theta$\\
\hline
geodetic & $t_1$ & 0 & $-\sigma/(1-\sigma)$ & $-\sigma$ & anomalous, $S_{tr}^{\phantom{01}t} = t_1\,m/2r^2$ \\
\cline{2-5}
torsions & $t_2$ & 0 & $-\sigma(1-2\tau)/(1-\sigma)$ & $-\sigma(1-2\tau)$ & normal, $S_{r\theta}^{\phantom{12}\theta} = S_{r\phi}^{\phantom{13}\phi} = t_2\,m/2r^2$\\
\hline
         & $w_1$ & 0 & $\sigma(\mathcal{G}-\alpha)$ & $\sigma(\mathcal{G}-\alpha)$ & $S_{r\phi}^{\phantom{0i}t} = w_1\,(ma/2r^2)\sin^2\theta$ \\
\cline{2-5}
frame-   & $w_2$ & 0 & $-2\sigma(\mathcal{G}-\alpha)$ & $-2\sigma(\mathcal{G}-\alpha)$ & $S_{\theta\phi}^{\phantom{0i}t} = w_2\,(ma/2r)\sin\theta\cos\theta$ \\
\cline{2-5}
dragging & $w_3$ & 0 & $\sigma\alpha$ & $\sigma\alpha$ & $S_{t\phi}^{\phantom{0i}r} = w_3\,(ma/2r^2)\sin^2\theta$ \\
\cline{2-5}
torsions & $w_4$ & 0 & $\sigma\alpha$ & $\sigma\alpha$ & $S_{t\phi}^{\phantom{0i}\theta} = w_4\,(ma/2r^3)\sin\theta\cos\theta$\\
\cline{2-5}
         & $w_5$ & 0 & $2\sigma\alpha$ & $2\sigma\alpha$ & $S_{tr}^{\phantom{0i}\phi} = w_5 \,ma/2r^4$\\
\hline
\hline
effective  & $\mu_1$ & 0 & $-\sigma$ & $-\sigma$ & $\mu_1 = (w_1-w_2-w_3+2w_4+w_5)/(-3\mathcal{G})$ \\ \cline{2-5}
torsions   & $\mu_2$ & 0 & $-\sigma$ & $-\sigma$ & $\mu_2 = (w_1-w_3+w_5)/(-\mathcal{G})$ \\ 
\hline
\hline
bias & $b_t$ & 1 & $1-4\tau/3$ & $1-\sigma-4\tau/3$ & $b_t=(1+\mathcal{F}+2t_2+|\eta|t_1)/3$\\ \cline{2-5}
     & $b_\mu$ & 1 & $(-\mathcal{G}/2)(1-\sigma)$ & $(-\mathcal{G}/2)(1-\sigma)$ & $b_\mu = (-\mathcal{G}/2)(1+3\mu_1-2\mu_2)$ \\ \hline
\end{tabular}
\caption{Summary of metric and torsion parameters for Einstein-Hayashi-Shirafuji (EHS) theories of interpolation parameter $\sigma$ in autoparallel scheme and in extremal scheme.  All parameter values are normalized by the physical mass $m$ of the body generating the gravity.  The parameter $\mathcal{G}$ and $\alpha$ are related to $\mathcal{G}_0$ and $\alpha_0$ in Table \ref{table:HS1} by $\mathcal{G}=\mathcal{G}_0/(1-\sigma)$ and $\alpha=\alpha_0/(1-\sigma)$ in autoparallel scheme, $\mathcal{G}=\mathcal{G}_0$ and $\alpha=\alpha_0$ in extremal scheme.  The value for $\mathcal{G}$ is set to $-2$ by the Kerr metric in linear regime $m/r \ll 1$ and $a/r \ll 1$. 
}
\label{table:EHS1}
\end{center}
\end{table*}

\begin{table*}
\noindent 
\footnotesize{
\begin{center}
\begin{tabular}{|l|c|c|c|}
\hline
            & General Relativity & EHS with autoparallels & EHS with extremals \\ \hline
Averaged Geodetic Precession & $(3m/2r_0)\vec{\omega}_O\times\vec{S}_0$ & $(1-4\tau/3)(3m/2r_0)\vec{\omega}_O\times\vec{S}_0$ & $(1-\sigma-4\tau/3)(3m/2r_0)\vec{\omega}_O\times\vec{S}_0$ \\ \hline
Averaged Frame-dragging & $(I/2r_0^3)\vec{\omega}_E\times\vec{S}_0$ & $(-\mathcal{G}/2)(1-\sigma)(I/2r_0^3)\vec{\omega}_E\times\vec{S}_0$ & $(-\mathcal{G}/2)(1-\sigma)(I/2r_0^3)\vec{\omega}_E\times\vec{S}_0$ \\ \hline
Second moment $\vec{a}_2$ & $(3I\omega_E/4r_0^3)\hat{z}\times\vec{S}_0$ & $(-3\mathcal{G}I\omega_E/8r_0^3)(1-\sigma)\hat{z}\times\vec{S}_0$ & $(-3\mathcal{G}I\omega_E/8r_0^3)(1-\sigma)\hat{z}\times\vec{S}_0 -\eta \sigma m\omega_O (S_0^x\hat{z}+S_0^z\hat{x})/4r_0$ \\ \hline
Second moment $\vec{b}_2$ & $(3I\omega_E/4r_0^3)\hat{x}\times\vec{S}_0$ & $(-3\mathcal{G}I\omega_E/8r_0^3)(1-\sigma)\hat{x}\times\vec{S}_0$ & $(-3\mathcal{G}I\omega_E/8r_0^3)(1-\sigma)\hat{x}\times\vec{S}_0 -\eta \sigma m\omega_O (S_0^x\hat{x}-S_0^z\hat{z})/4r_0$ \\ \hline
\end{tabular}
\caption{Summary of the predicted Fourier moments of the precession rate for General Relativity and the Einstein-Hayashi-Shirafuji (EHS) theories in autoparallel scheme and in extremal scheme.  $\eta=+1$ for extremal scheme using $S^{\mu\nu}$, and $-1$ for extremal scheme using $S^{\mu}$.  Other multiple moments vanish.  Here $m$ and $I\omega_E$ are the Earth's mass and rotational angular momentum, respectively.}
\label{table:EHS2}
\end{center}
}
\end{table*}

\begin{table*}
\noindent 
\footnotesize{
\begin{center}
\begin{tabular}{|l||l|c|c|p{3cm}|}
\hline
Effects  & Torsion Biases  &   EHS in autoparallel scheme &  EHS in extremal scheme & PPN biases\\ \hline
Shapiro time delay & $\Delta t/\Delta t ^{(GR)}=(\mathcal{F}-\mathcal{H})/4$ &  $1+\sigma-\tau$ & $1-\tau$ & $(1+\gamma)/2$ \\\hline
Deflection of light & $\delta/\delta^{(GR)}=(\mathcal{F}-\mathcal{H})/4$ & $1+\sigma-\tau$ & $1-\tau$ &  $(1+\gamma)/2$ \\\hline
Gravitational redshift & $(\Delta\nu/\nu)/(\Delta\nu/\nu)^{(GR)}=-\mathcal{H}/2$ & $1+\sigma$ & 1 & $1+\alpha$ \\\hline
Geodetic Precession & $\Omega_G/\Omega_G^{(GR)}=b_t$ & $1-\frac{4}{3}\tau$ & $1-\sigma-\frac{4}{3}\tau$ & $(1+2\gamma)/3$ \\ \hline
Frame-dragging & $\Omega_F/\Omega_F^{(GR)}=b_\mu$ & $1-\sigma$ & $1-\sigma$ & $(1+\gamma+\alpha_1/4)/2$ \\ \hline
\end{tabular}
\caption{Summary of solar system experiments (1): the biases relative to GR predictions for the Einstein-Hayashi-Shirafuji (EHS) theories.  Both parameters $\tau$ and $\sigma$ are assumed small.  The biases in the PPN formalism are also listed for comparison, taken from \cite{Will:2005va}.}
\label{tab:solar-constraint1}
\end{center}
}
\end{table*}

\begin{table*}
\noindent 
\footnotesize{
\begin{center}
\begin{tabular}{|l||c|c|c|p{4cm}|}
\hline
Effects  &  PPN  &   EHS in autoparallel scheme &  EHS in extremal scheme & Remarks \\\hline
Shapiro time delay & $\gamma-1 = (2.1\pm 2.3)\times 10^{-5}$ & $\sigma-\tau = (1.1\pm 1.2)\times 10^{-5}$ &  $\tau = (-1.1\pm 1.2)\times 10^{-5}$ & Cassini tracking \cite{Bertotti:2003rm} \\\hline
Deflection of light & $\gamma-1 = (-1.7\pm 4.5)\times 10^{-4}$ & $\sigma-\tau = (-0.8\pm 2.3)\times 10^{-4}$ &  $\tau = (0.8\pm 2.3)\times 10^{-4}$ & VLBI \cite{Shapiro2004} \\\hline
Gravitational redshift & $|\alpha|<2\times 10^{-4}$ & $|\sigma|<2\times 10^{-4}$ & no constraints & Vessot-Levine rocket \cite{vessot1980} \\\hline
Geodetic Precession & $\left|\gamma-1\right|< 1.1\times 10^{-4}$ & $|\tau|<5.7\times 10^{-5}$ &  $|\sigma+4\tau/3|<7.6\times 10^{-5}$ & Gravity Probe B \\\hline
Frame-dragging & $\left|\gamma-1+\frac{1}{4}\alpha_1\right|<0.024$ & $|\sigma|<0.012$ & $|\sigma|<0.012$ & Gravity Probe B \\\hline
\end{tabular}
\caption{Summary of solar system experiments (2): constraints on the PPN and EHS parameters.  The constraints on PPN parameters are taken from Table 4 and Page 12 of \cite{Will:2005va}.   
The full results of Gravity Probe B are yet to be released, so whether the frame dragging 
will agree with the GR prediction is not currently known.  The last two rows show the limits that would correspond to a GPB result consistent with GR,
assuming an angle accuracy of 0.5 milli-arcseconds.}
\label{tab:solar-constraint2}
\end{center}
}
\end{table*}

\section{Example: testing Einstein Hayashi-Shirafuji theories with GPB and other solar system experiments}
\label{sec:constrain-torsion}

Above we calculated the observable effects that arbitrary Earth-induced torsion, if present, would have on GPB.
As a foil against which to test GR, let us now investigate the observable effects that would result for the explicit 
Einstein-Hayashi-Shirafuji class of torsion theories that we studied in \Sec{subsubsec:the axisym-sol2-HS} and \ref{subsec:linearinterpol}.

There are four parameters $c_1$, $c_2$, $c_3$ and $\sigma$ that define an EHS theory via the action in \Eq{eqn:HSlag-interp}.
We will test EHS theories with GPB and other solar system experiments.  For all these weak field experiments, only two EHS parameters ---  $\tau$ (defined in \Eq{eqn:epsdefEHS}) and $\sigma$, both assumed small --- that are functions of the said four are relevant and to be constrained below.  

The predicted EHS metric and torsion parameters, studied in \Sec{subsec:linearinterpol}, are listed in Table \ref{table:HS1}.  Below, we will test both the autoparallel and extremal calculation schemes.  In each scheme, the physical mass $m$ will be determined by the Newtonian limit.  All metric and torsion parameters are converted in accordance with $m$ and listed in Table \ref{table:EHS1}.  Then the parameter space ($\tau$, $\sigma$) will be constrained by solar system experiments.

\begin{figure}
\centering
\includegraphics[width=0.5\textwidth]{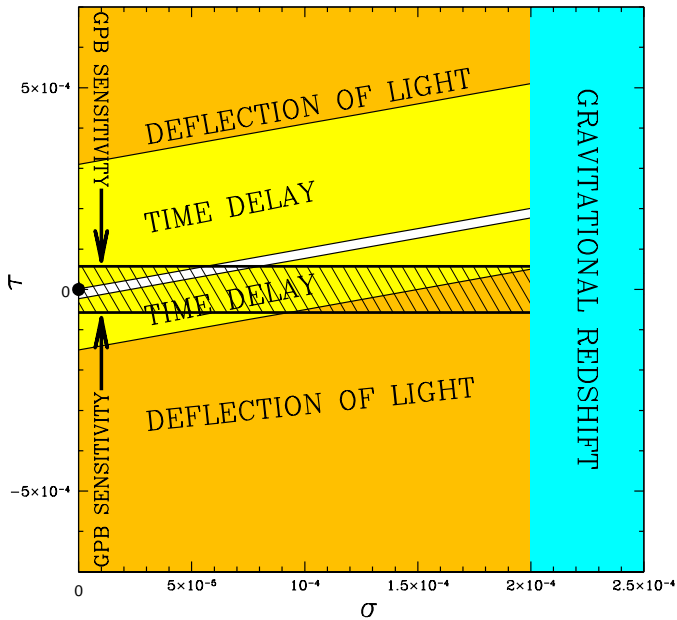}
\caption{Constraints on the EHS parameters $(\sigma,\tau)$ from solar system tests in the autoparallel scheme.  
General Relativity corresponds
to the black dot ($\sigma=\tau=0$). The shaded regions in the parameter space have already been ruled out by Mercury's perihelion
shift (red/dark grey), the deflection of light (orange/grey), Shapiro time delay (yellow/light grey) 
and gravitational redshift (cyan/light grey).  
If the geodetic
precession and frame-dragging measured by Gravity Probe B are consistent with GR to the target accuracy of 0.5 milli-arcseconds,
this will rule out everything outside the hatched region, implying that
$0\le \sigma<8.0\times 10^{-5}$ and $-2.3\times 10^{-5} <\tau< 5.7\times 10^{-5}$.
Preliminary result of Gravity Probe B have only confirmed the geodetic precession to about 1\%, thus 
bringing no further constraints beyond those from
gravitational redshift.
}
\label{fig:all-constraints-auto}
\end{figure}

\begin{figure}
\centering
\includegraphics[width=0.5\textwidth]{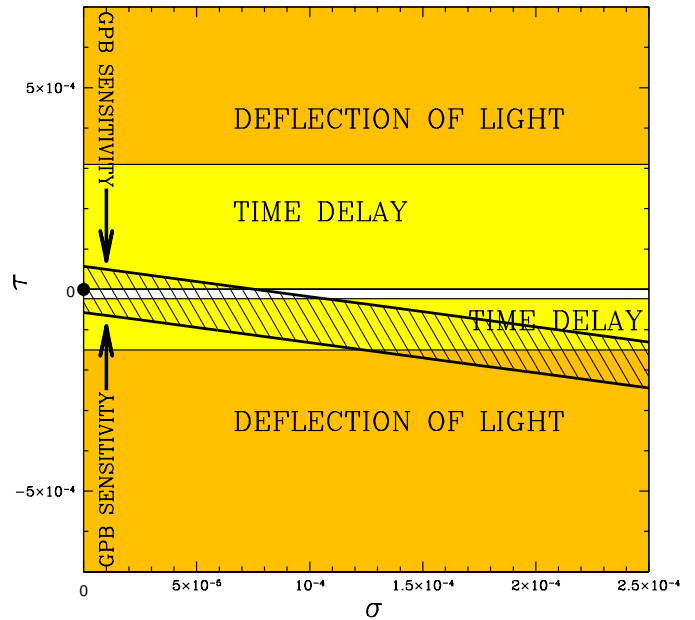}
\caption{Constraints on EHS parameters $(\sigma,\tau)$ from solar system tests in the extremal scheme. 
General Relativity corresponds to the black dot
($\sigma=\tau=0$). The shaded regions have already been ruled out by Mercury's perihelion shift (red/dark grey), 
the deflection of light (orange/grey) and Shapiro time delay (yellow/light grey). 
If the geodetic
precession and frame-dragging measured by Gravity Probe B are consistent with GR to the target accuracy of 0.5 milli-arcseconds, this will rule out everything outside the hatched region, implying that
$0\le \sigma<1.1\times 10^{-4}$ and $-2.3\times 10^{-5} <\tau< 0.1\times 10^{-5}$.  
The preliminary result of Gravity Probe B have confirmed the geodetic precession only to about 1\%, 
implying that $\sigma < 0.01$. 
}
\label{fig:all-constraints-extremal}
\end{figure}

\subsection{Autoparallel scheme}

Hayashi-Shirafuji maximal torsion theory is inconsistent with the autoparallel scheme, since $t_1-\mathcal{H}/2=0$ (see $t_1$ and
$\mathcal{H}$ in Table \ref{table:HS1}).  By \Eq{eqn:auto}, this means that $\mathrm{d}\vec{v}/\mathrm{d}t=0 + \mathcal{O}(m/r)^2$. 
The violation of Newton's law rules out the application of the autoparallel scheme to the Hayashi-Shirafuji theory.  

However, the Einstein-Hayashi-Shirafuji theories can be consistent with this scheme.  Using Table \ref{table:HS1}, the Newtonian limit can be written as 
\beq{eqn:EHSnewtonian} 
\frac{d\vec{v}}{dt}=-(1-\sigma)\frac{m_0}{r^2}\hat{e}_r\,,
\eeq
so the physical mass of the central gravitating body is 
\ben
m=(1-\sigma)m_0\,.
\een

Table \ref{table:EHS1} lists values of metric and torsion parameters in accordance with the physical mass $m$.  Using these parameters, the precession rates of gyroscopes in GPB orbit can be calculated via equations (\ref{eqn:moment2}),(\ref{eqn:btdef}),(\ref{eqn:bmudef}) and (\ref{eqn:n2moments}).  The results are listed in Table \ref{table:EHS2}.  For GPB, the average precession rates are the only experimentally accessible observables in practice.  GPB will measure the precession of gyroscopes with respect to two different axes: the orbital angular velocity $\vec{\omega}_O$ (geodetic precession) and the Earth's rotational angular velocity $\vec{\omega}_E$ (frame-dragging).  As indicated in Table \ref{table:EHS2}, the geodetic precession and frame-dragging rates are 
\bena
\Omega_G &=& (1-\frac{4}{3}\tau)\Omega_G^{(GR)}\,,\\
\Omega_F &=& \left(-\frac{\mathcal{G}}{2}\right)(1-\sigma)\Omega_F^{(GR)}\,,
\eena
where $\Omega_G^{(GR)}$ and $\Omega_F^{(GR)}$ are the geodetic precession and frame-dragging rate predicted by General Relativity, respectively.  

The existing solar system experiments, including Shapiro time delay, deflection of light, gravitational redshift, advance of Mercury's perihelion, can put constraints on the parameters $\tau$ and $\sigma$.  The derivation of these constraints essentially follow any standard textbook of General Relativity \cite{MTW-Weinberg-Islam:book} except for more general allowance of parameter values, so we leave the technical detail in Appendix \ref{appendix:solar-tests} with the results summarized in Table \ref{tab:solar-constraint1}.

It is customary that  biases of GR predictions are expressed in terms of PPN parameters on which observational constraints can be placed with solar system experiments.  In EHS theories, these biases are expressed in terms of the parameters $\tau$ and $\sigma$.  Thus we can place constraints on the EHS parameters $\tau$ and $\sigma$ by setting up the correspondence between PPN and EHS parameters via the bias expression.  Table \ref{tab:solar-constraint1} lists the biases in the PPN formalism for this purpose, and Table \ref{tab:solar-constraint2} lists the observational constraints on the EHS parameters $\tau$ and $\sigma$ with the existing solar system tests.

If GPB would see no evidence of the torsion induced precession effects, the ($\tau$,$\sigma$) parameter space can be further constrained.  Together with other solar system experiments, the observational constraints are listed in Table \ref{tab:solar-constraint2} and shown in \Fig{fig:all-constraints-auto}.

\begin{figure}
\centerline{\epsfxsize=\figsiz\epsffile{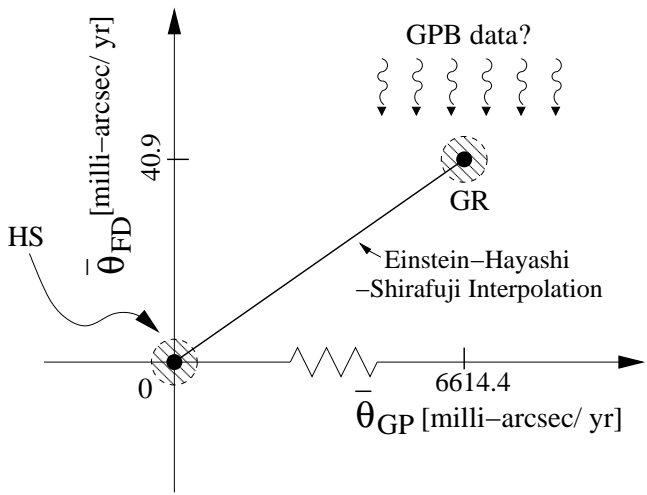}}
\caption[1]{\label{gpb4}\footnotesize%
Predictions for the \emph{average} precession rate by General Relativity, Hayashi-Shirafuji (HS) gravity and Einstein-Hayashi-Shirafuji
theories (for the case of $\tau=0$ and the Kerr solution $\mathcal{G}=-2$) that interpolate between these two extremes, in the extremal
scheme. $\bar{\theta}_{\textrm{GP}}$ is the geodetic precession rate around the orbital angular velocity vector $\vec{\omega}_O$ and
$\bar{\theta}_{\textrm{FD}}$ is the angular frame-dragging rate around Earth's rotation axis $\vec{\omega}_E$.  The shaded areas of
about 0.5 milli-arcseconds per year in radius are the approximate forecast GPB measurement uncertainties.  The two calculation
schemes using $S^\mu$ and $S^{\mu\nu}$ with extremals for the Hayashi-Shirafuji Lagrangian (labeled ``HS'' in the figure) agree on
the predicted average rates.
The unpublished preliminary results of Gravity Probe B have confirmed the geodetic precession to better 
than 1\%, so this already rules out the Hayashi-Shirafuji Lagrangian and most EHS theories in the extremal scheme in the sense 
that $\sigma < 0.01$. 
}
\end{figure}

\subsection{Extremal scheme}

Einstein-Hayashi-Shirafuji theories predict $\mathcal{H}=-2$ regardless of $\tau$ and $\sigma$.  By the Newtonian limit, therefore, the physical mass of the central gravitating body is just the mass parameter $m_0$, \ie $m=m_0$.  So the parameter values do not need rescaling and are re-listed in Table \ref{table:EHS1}.  By these parameters the precession rates can be calculated and listed in Table \ref{table:EHS2}.  As indicated in Table \ref{table:EHS2}, the geodetic precession and frame-dragging rates are
\bena
\Omega_G &=& (1-\sigma-\frac{4}{3}\tau)\Omega_G^{(GR)}\,,\\
\Omega_F &=& \left(-\frac{\mathcal{G}}{2}\right)(1-\sigma)\Omega_F^{(GR)}\,.
\eena

It is worth noting again that the extremal scheme is not a fully consistent framework from the theoretical point of view.  However, it serves perfectly to show the role of EHS theories as the bridge between no-torsion GR and Hayashi-Shirafuji maximal torsion theory.  \Fig{gpb4} illustrates this connectivity in terms of the
 predictions of GR, Hayashi-Shirafuji theory and the intermediate $0<\sigma<1$ EHS theories, taking $\tau=0$ and Kerr solution $\mathcal{G}=-2$, on the \emph{average} precession rate (the $\vec{a}_0$ in Table \ref{table:EHS2}).  The EHS theories are seen to connect the  extreme GR and HS cases with a straight line.  If the data released by GPB ends up falling within the shaded area corresponding to the GR prediction, the Hayashi-Shirafuji Lagrangian will thus have been ruled out with very high significance, and the GPB torsion constraints can be quantified as sharp upper limits on the $\sigma$-parameter.

More generally, Gravity Probe B will improve the constraints on the ($\tau$,$\sigma$) parameter space by its precise measurements of precession rates, in addition to the constraints put by existing solar system experiments.  These constraints are listed in Table \ref{tab:solar-constraint2} and shown in \Fig{fig:all-constraints-extremal}.  As before, the technical details are given in Appendix \ref{appendix:solar-tests}.

\subsection{Preliminary constraints from GPB's unpublished results}

In April 2007, Gravity Probe B team announced that, while they continued mining the data for the ultimately optimal accuracy, the geodetic precession was found to agree with GR at the 1\% level.  The frame-dragging yet awaits to be confirmed.  Albeit preliminary, these unpublished results, together with solar system tests, already place the first constraint on some torsion parameters to the 1\% level.  More quantitatively, $|t_2+\frac{|\eta|}{2}t_1|\lesssim 0.01$ in the model-independent framework, while  $w_1+w_2-w_3-2w_4+w_5$ is not constrained.  In the context of EHS theories, the constraint is scheme dependent.  In the autoparallel scheme, GPB's preliminary results place no better constraints than those from gravitational redshift ($\sim 10^{-4}$).  In the extremal scheme, however, the preliminary results give the constraint $\sigma < 0.01$.  The bottom line is that GPB has constrained torsion parameters to the 1\% level now and will probably reach the $10^{-4}$ level in the future.

\section{Conclusions and Outlook}
\label{sec:conclusion}

The PPN formalism has demonstrated that a great way to test GR is to embed it in a broader parametrized class of theories, 
and to constrain the corresponding parameters observationally. In this spirit, we have explored observational constraints on 
generalizations of GR including torsion. 

Using symmetry arguments, we showed that to lowest order, the torsion field around a uniformly rotating spherical mass such as Earth is
determined by merely seven dimensionless parameters.
We worked out the predictions for these seven torsion parameters for a two-parameter
Einstein-Hayashi-Shirafuji generalization of GR which includes as special cases both standard no-torsion GR ($\sigma=0$) 
and the no-curvature, all torsion ($\sigma=1$) Weitzenb\"ock spacetime.
We showed that classical solar system tests rule out a large class of these models, and that Gravity Probe B (GPB) can further improve
the constraints. 
GPB is useful here because this class of theories suggested that, depending on the Lagrangian, rotating objects can generate torsion observable with gyroscopes.
In other words, despite some claims in the literature to the contrary, the question of whether there is observable 
torsion in the solar system is one which ultimately can and should be tested experimentally.

Our results motivate further theoretical and experimental work.
On the theoretical side, it would be interesting to address in more detail the question of which Lagrangians make torsion couple to rotating objects.
A well-defined path forward would be to generalize the matched asymptotic expansion method of \cite{D'Eath1975a,D'Eath1975b} to match
two generalized EHS Kerr-like Solutions in the weak-field limit to obtain the laws of motion for two well-separated rotating objects, and determine which of the
three non-equivalent prescriptions above, if any, is correct. It would also be interesting to look for generalizations of the EHS Lagrangian that populate a large fraction 
of the seven torsion degrees of freedom that symmetry allows.
Finally, additional observational constraints can be investigated involving, \eg, binary pulsars, gravitational 
waves and cosmology.

On the experimental side, Gravity Probe B has now successfully completed its data taking phase. We have shown that the GPB data constitute a potential gold mine of information
about torsion, but that its utility for constraining torsion theories will depend crucially on how the data are analyzed and released. 
At a minimum, the average geodetic and frame dragging precessions can be compared with the predictions shown in \fig{gpb4}. However, if it is technically feasible for 
the GPB team to extract and publish also different linear combinations of the instantaneous precessions corresponding to the second moments of these precessions, 
this would enable looking for further novel effects that GR predicts should be absent.
In summary, although the nominal goal of GPB is to look for an effect that virtually everybody expects will be present (frame dragging), it also has the potential 
to either discover torsion or to build further confidence in GR by placing stringent limits on torsion theories.

The authors wish to thank Francis Everitt, Thomas Faulkner, Friedrich Hehl, Scott Hughes,  Erotokritos Katsavounidis, Barry Muhlfelder, Tom Murphy,  
Robyn Sanderson, Alexander Silbergleit, Molly Swanson, Takamitsu Tanaka and Martin White 
for helpful discussions and comments.  
This work is supported
by the U.S.~Department of Energy (D.O.E.) under cooperative research agreement DE-FC02-94ER40818, 
NASA grants NAG5-11099 and NNG06GC55G, NSF grants AST-0134999 and 0607597, and fellowships from the David and Lucile
Packard Foundation and the Research Corporation.

\appendix 

\section{Parametrization of torsion in the static, spherically and parity symmetric case}
\label{appendix:spher-symm}

In this appendix, we derive a parametrization of the most general static, spherically and parity
symmetric torsion  in isotropic 
rectangular and spherical
coordinates.  The symmetry conditions are described in Section
\ref{subsubsec:general-setup1} with the quantity $\mathcal{O}$ now being the torsion tensor $S_{\mu\nu}^{\phantom{12}\rho}$.  
Note that torsion (the antisymmetric part of the connection) is a tensor under general coordinate transformations 
even though the full connection is not.

First note that time translation invariance is equivalent to the independence of torsion on time. Then consider
time reversal, under which a component of torsion flips its sign once for every temporal index.
Invariance under time reversal therefore requires that non-zero torsion components have either zero or two
temporal indices. Together with the fact that torsion is antisymmetric in its first two indices, this restricts the non-zero
components of torsion to be $S_{0i}^{\phantom{0i}0}$ and $S_{jk}^{\phantom{0i}i}$ ($i=1,2,3$).

Now consider the symmetry under (proper or improper) rotation (see \Eq{xform:spherical}). 
The orthogonality of the matrix $\R$
enables one to write
\beq{eqn:R-xform-appen}
\frac{\partial x'^i}{\partial x^j}=R^{ij}\,,\qquad \frac{\partial x^i}{\partial x'^j}=R^{ji}\,,\qquad \frac{\partial t'}{\partial t}=\frac{\partial t}{\partial t'}=1\,.
\eeq
Thus formal functional invariance means that
\beq{eqn1:form-inv2}
\begin{array}{lclcl}
S_{\phantom{1}0i}^{'\phantom{0i}0}(x') &=& R^{ij}S_{0j}^{\phantom{0i}0}(x) &=& S_{0i}^{\phantom{0i}0}(x'),\\
S_{\phantom{1}jk}^{'\phantom{jk}i}(x') &=& R^{jm}R^{kn}R^{il}S_{mn}^{\phantom{mn}l}(x) &=& S_{jk}^{\phantom{0i}i}(x').
\end{array}
\een
\Eq{eqn1:form-inv2} requires that the torsion should be built up of  $x^i$ and quantities invariant under O(3), such as scalar
functions of radius and Kronecker $\delta$-functions, since $\delta
'_{i'j'}=R^{i'i}R^{j'j}\delta_{ij}=R^{i'i}R^{j'i}=R^{i'i}(R^{-1})^{ij'}=\delta_{i'j'}\,.$  
Note that we are interested in the
parity symmetric case, whereas the Levi-Civita symbol $\epsilon_{ijk}$ is a three-dimensional \emph{pseudo}-tensor 
under orthogonal transformations, where ``pseudo'' means that $\epsilon_{ijk}$ is a tensor under SO(3) but not under O(3), 
since $\epsilon '_{i'j'k'}=R^{i'i}R^{j'j}R^{k'k}\epsilon_{ijk}=\det R \times \epsilon_{i'j'k'}\,.$  
Therefore, $\epsilon_{ijk}$ is prohibited
from entering into the construction of the torsion tensor by \Eq{eqn1:form-inv2}.

Thus using arbitrary combinations of scalar functions of radius, $x^i$ and Kronecker $\delta$-functions, 
the most general torsion tensor that can be constructed takes the form 
\bena
S_{0i}^{\phantom{0i}0} &=& t_1\frac{m}{2r^3}x^i\, ,\\
S_{jk}^{\phantom{0i}i} &=& t_2\frac{m}{2r^3}(x^j \delta_{ki}-x^k \delta_{ji})\,,\label{eqn:t2-appen}
\eena
where the combinations $t_1 m$ and $t_2 m$ are arbitrary functions of radius.
Note that in \Eq{eqn:t2-appen}, terms proportional to $x^i x^j x^k$ or $x^i\delta_{jk}$ are forbidden by the antisymmetry of the torsion.  
We will simply treat the functions $t_1(r)$ and $t_2(r)$ as constants, since GPB orbits at a fixed radius.

Transforming this result to spherical coordinates, we obtain
\begin{eqnarray*}
S_{tr}^{\phantom{01}t} &=&  S_{ti}^{\phantom{01}t} \frac{\partial x^i}{\partial r}= t_1\frac{m}{2r^2},\\
S_{r\theta}^{\phantom{12}\theta} &=& S_{jk}^{\phantom{12}i}\frac{\partial x^j}{\partial r}\frac{\partial x^k}{\partial \theta}\frac{\partial \theta}{\partial x^i}= t_2\frac{m}{2r^2}\,,\\
S_{r\phi}^{\phantom{13}\phi} &=& S_{jk}^{\phantom{12}i}\frac{\partial x^j}{\partial r}\frac{\partial x^k}{\partial \phi}\frac{\partial \phi}{\partial x^i} = t_2\frac{m}{2r^2}\, .
\end{eqnarray*}
All other components not related by the antisymmetry vanish. In the above equations, the second equalities follow from the chain rule and the facts that $\partial
x^i/\partial r=\hat{x}^i=\hat{e}_r^i$, $\partial x^i/\partial \theta=r\hat{e}_\theta^i$, and $\partial x^i/\partial
\phi=r\sin\theta\hat{e}_\phi^i$, where $\hat{e}_r^i$, $\hat{e}_\theta^i$ and $\hat{e}_\phi^i$ are the $i$th-components of the
unit vectors in spherical coordinates.  To first order in the mass $m$ of the central object, we need not distinguish between isotropic and standard spherical coordinates.
\bigskip

\section{Parametrization in stationary and spherically axisymmetric case}
\label{appendix:axisymm}

Above we considered the 0$th$ order contribution to the metric and torsion corresponding to the static, spherically and parity symmetric case
of a non-rotating spherical source.
In this appendix, we derive a parametrization of the most general 1$st$ order correction (denoted by a superscript $(1)$) to this metric and torsion that could be caused by rotation
of the source, \ie corresponding to the stationary and spherically axisymmetric case.
The symmetry conditions are described in Section \ref{subsubsec:prob-setup-axisym}, with the quantity
$\mathcal{O}$ replaced by the metric $g_{\mu\nu}^{(1)}$ for Appendix \ref{appen-subsec:metric} and by the torsion $S_{\mu\nu}^{(1)\rho}$
for Appendix \ref{appen-subsec:torsion}.

\subsection{The Metric}\label{appen-subsec:metric}

The invariance under time translation makes the metric time independent.  Under time reversal $\J\to -\J$, and a component of the metric flips its sign once for every temporal index.  Thus, the formal functional invariance equation for time reversal reads
\beq{eqn:metric-form-inv-1}
\pm g^{(1)}_{\mu\nu}(x|\J) = g^{(1)}_{\mu\nu}(x|-\J)\,.
\eeq

The plus sign in \Eq{eqn:metric-form-inv-1} is for components with even numbers of temporal indices, and minus sign for those with odd numbers.  Since only terms linear in $J/r^2=\varepsilon_m \varepsilon_a$ are concerned, the minus sign in the argument $-\J$ can be taken out as an overall factor, implying that the non-vanishing components of metric can have only one temporal index.  Thus the only nonzero first-order correction to $g_{\mu\nu}$ in rectangular coordinates is $g_{ti}^{(1)}$ (i=1,2,3). 

Now consider the transformation property under (proper or improper) rotation. 
By the orthogonality of the matrix $\R$, the vector $\x$
transforms as $\x\to\x'\equiv \R\x$ (\Eq{eqn:R-xform-appen}).  Since $\J$ is invariant under parity, formally the transformation of $\J$ writes as
\ben
\J \to \J'=(det\R) \times \R\J \,.
\een
The formal functional invariance for rotation reads
\ben
g^{(1)'}_{ti}(x'|\J) = R^{ij}g^{(1)}_{tj}(x|\J) = g^{(1)}_{ti}(x'|\J')\,.
\een
That $\J$ is a pseudo-vector under improper rotation requires that the Levi-Civita symbol $\epsilon_{ijk}$, also a pseudo-tensor, appear once and only once (because $\J$ appears only once) in the metric so as to compensate the $det\R$ factor incurred by transformation of $\J$.  Other possible elements for construction of the metric include scalar functions of radius, $x^i$, $J^i$, $\delta_{ij}$.  Having known the elements, the only possible construction is therefore
\ben
g_{ti}^{(1)}=\frac{\mathcal{G}}{r^2}\epsilon_{ijk}J^j \hat{x}^k\,,
\een
where $\hat{x}^i=x^i/r$ is the unit vector of position vector and $\mathcal{G}$ is dimensionless.  Assuming that there is no new scale other than the angular momentum $\J$ built into the 1$st$ order of torsion theory, \ie no new dimensional parameter with units of length, $\mathcal{G}(r)$ must be a constant by dimensional analysis, since the factor $J^i$ has explicitly appeared. 

In spherical polar coordinates where the $z$-axis is parallel to $\J$, this first-order correction to the metric takes the form
\ben
g_{t\phi}^{(1)} = \mathcal{G}\frac{ma}{r}\sin^2\theta\,,
\een
where $ma=J$ is the magnitude of $\J$.  All other components vanish.

\bigskip

\subsection{The Torsion}\label{appen-subsec:torsion}

We follow the same methodology as for our parametrization of the metric above. 
Given the time-independence, the property that $\J$ reverses under time-reversal requires that the non-vanishing components of torsion 
have only one temporal index, so they are $S_{\phantom{1}ij}^{(1)\,t},\; S^{(1)}_{tij}$ (i,j=1,2,3) in rectangular coordinates. 
(The antisymmetry of torsion over its first two indices excludes the possibility of three temporal indices.)  
Under (proper or improper) rotation, the formal functional invariance equation reads
\[ \begin{array}{lclcl}
S_{\phantom{(1)}ij}^{(1)\,'\phantom{0}t}(x'|\J) &=& R^{ik}R^{jl}S_{\phantom{1}kl}^{(1)\,t}(x|\J) &=&  S_{\phantom{1}ij}^{(1)\,t}(x'|\J')\,,\\
S^{(1)\,'}_{tij}(x'|\J) &=& R^{ik}R^{jl}S^{(1)}_{tkl}(x|\J) &=&  S^{(1)}_{tij}(x'|\J')\,.\\
\end{array} \]
Again, in building the torsion, one should use the Levi-Civita symbol $\epsilon_{ijk}$ once and only once to cancel the $det\R$ factor from the transformation of $\J$.
  The most general construction using scalar function of radius, $x^i$, $\delta_{ij}$, $J^i$ (also appearing once and only once) and $\epsilon_{ijk}$ is
\begin{eqnarray*}
S_{\phantom{0}ij}^{(1)\,t} &=& \frac{f_1}{2r^3}\epsilon_{ijk}J^k+\frac{f_2}{2r^3}J^k \hat{x}^l (\epsilon_{ikl}\hat{x}^j-\epsilon_{jkl}\hat{x}^i)\,,\\
S_{tij}^{(1)} &=& \frac{f_3}{2r^3}\epsilon_{ijk}J^k+\frac{f_4}{2r^3}J^k \hat{x}^l \epsilon_{ikl}\hat{x}^j+\frac{f_5}{2r^3}J^k \hat{x}^l \epsilon_{jkl}\hat{x}^i\,.
\end{eqnarray*}
By the same dimensional argument as in Appendix (\ref{appen-subsec:metric}), $f_1,\ldots,f_5$ must be dimensionless constants.

Transforming the above equations to spherical coordinates where the $z$-axis is parallel to $\J$, we obtain to first order
\begin{eqnarray*}
S_{\phantom{0}r\phi}^{(1)\phantom{0}t} &=& S_{ij}^{\phantom{0i}t}\frac{\partial x^i}{\partial r}\frac{\partial x^j}{\partial \phi} =  w_1\frac{ma}{2r^2}\sin^2\theta\,,\\
S_{\phantom{0}\theta\phi}^{(1)\phantom{0}t} &=& S_{ij}^{\phantom{0i}t}\frac{\partial x^i}{\partial \theta}\frac{\partial x^j}{\partial \phi} =  w_2\frac{ma}{2r}\sin\theta\cos\theta\,,\\
S_{\phantom{0}t\phi}^{(1)\phantom{0}r} &=& g^{rr} S_{tij}\frac{\partial x^i}{\partial \phi}\frac{\partial x^j}{\partial r} = w_3\frac{ma}{2r^2}\sin^2\theta\,, \\
S_{\phantom{0}t\phi}^{(1)\phantom{0}\theta} &=& g^{\theta\theta}S_{tij}\frac{\partial x^i}{\partial \phi}\frac{\partial x^j}{\partial \theta} = w_4\frac{ma}{2r^3}\sin\theta\cos\theta\,,\\
S_{\phantom{0}tr}^{(1)\phantom{0}\phi} &=& g^{\phi\phi} S_{tij}\frac{\partial x^i}{\partial r}\frac{\partial x^j}{\partial \phi} =w_5 \frac{ma}{2r^4}\,,\\
S_{\phantom{0}t\theta}^{(1)\phantom{0}\phi} &=& g^{\phi\phi}S_{tij}\frac{\partial x^i}{\partial \theta}\frac{\partial x^j}{\partial \phi} = -w_4\frac{ma}{2r^3}\cot\theta\,.\label{eqn:w6-axi}
\end{eqnarray*}
All other components vanish.  The constants are related by $w_1=f_1-f_2$, $w_2=f_1$, $w_3=f_4-f_3$, $w_4=-f_3$, $w_5=f_5+f_3$.

\section{Constraining torsion with solar system experiments}
\label{appendix:solar-tests}

\subsection{Shapiro time delay}
\label{appendix:shapiro}

For the electromagnetic field, if torsion is coupled to the vector potential $A_\mu$ by the ``natural'' extension, \ie, $\partial_\mu A_\nu \to \nabla_\mu A_\nu$ using the full connection, the Maxwell Lagrangian $-\frac{1}{4}F_{\mu\nu}F^{\mu\nu}$ will contain a quadratic term in $A_\mu$ that makes the photon massive and breaks gauge invariance in the conventional form.  Since the photon mass has been experimentally constrained to be $\lesssim 10^{-17}$ eV, we assume that $A_\mu$ does not couple to torsion.  Instead, we assume that the Maxwell field Lagrangian in the curved spacetime with torsion follows the extension $\partial_\mu A_\nu \to \nabla_\mu^{\{\}} A_\nu$ using the Levi-Civita connection.  Since the Levi-Civita connection depends on the metric and its derivatives only, light rays follow extremal curves (metric geodesics).  

In general, assume the line element in the field around a (physical) mass $m$ is 
\beq{eqn:metricDefappendix}
\mathrm{d}s^2 =  -\left[1+\mathcal{H}\frac{m}{r}\right]\mathrm{d}t^2 + \left[1+\mathcal{F}\frac{m}{r}\right]\mathrm{d}r^2+r^2 d\Omega^2\,.
\een
The effect of the rotation of the mass can be ignored when the rotation is slow.  

\begin{figure}
\centerline{\epsfxsize=\figsiz\epsffile{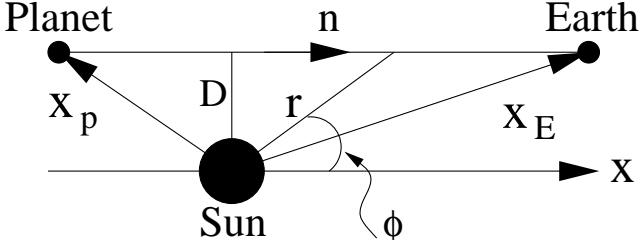}}
\caption[1]{\label{fig:deflection}\footnotesize%
Geometry of the Shapiro time delay measurement.}
\end{figure}

Light deflection angle is tiny for the solar system tests we consider, so a ray can be well approximated by a straight line.  Let us use coordinates where the Sun (of mass $m$), the Earth and a planet reflecting the light ray are all in the $x$-$y$ plane ($\theta=\pi/2$) and the $x$-axis points along the ray from the planet to Earth (see \Fig{fig:deflection}).  Let $D$ be the minimal distance of the ray from the Sun.  Then $r\sin\phi=D$, or $rd\phi=-\tan\phi dr$.  Since $\mathrm{d}s^2=0$ for a light ray,
\bena
dt^2 &=& (1+\mathcal{H}\frac{m}{r})^{-1} (1+\mathcal{F}\frac{m}{r}+\tan^2\phi)dr^2\nonumber\\
     &\approx & \frac{r^2 dr^2}{r^2-D^2}[1+(\mathcal{F}-\mathcal{H})\frac{m}{r}-\mathcal{F}\frac{mD^2}{r^3}]\,, \nonumber\\
dt &\approx & \frac{r |dr|}{\sqrt{r^2-D^2}}[1+(\mathcal{F}-\mathcal{H})\frac{m}{2r}-\mathcal{F}\frac{mD^2}{2r^3}]\,.
\eena
The round-trip travel time for an electromagnetic signal bouncing between Earth and the Planet in the gravitational field of the Sun is
\bena
T &=& 2\left[ \int_{r=D_p}^{r=D} dt + \int_{r=D}^{r=D_E} dt \right] \,,\nonumber\\
  &\approx & 2{[\sqrt{D_p^2-D^2}+\sqrt{D_E^2-D^2}]}+(\mathcal{F}-\mathcal{H})m\nonumber\\
  & & \times\ln\left[\frac{(\sqrt{D_p^2-D^2}+D_p)(\sqrt{D_E^2-D^2}+D_E)}{D^2}\right]\nonumber\\
  & & -\mathcal{F}m\left(\frac{\sqrt{D_p^2-D^2}}{D_p}+\frac{\sqrt{D_E^2-D^2}}{D_E}\right)\,.\label{eqn:T-appC}
\eena
If $D\ll D_E$ and $D\ll D_p$, the third term in \Eq{eqn:T-appC} is negligible compared to the second one.  The excess travel time $\Delta t$ of a round-trip light ray is 
\bena
\Delta t &\equiv & T - 2{[\sqrt{D_p^2-D^2}+\sqrt{D_E^2-D^2}]}\,, \nonumber\\
         &\approx & \left(\frac{\mathcal{F}-\mathcal{H}}{4}\right)\Delta t^{(GR)}\,,
\eena
where $\Delta t^{(GR)}$ is the excess time predicted by GR
\ben
\Delta t^{(GR)} = 4m\ln\left[\frac{(D_E+\vec{x}_E\cdot\hat{n})(D_p-\vec{x}_p\cdot\hat{n})}{D^2}\right]\,.
\een
Here $\vec{x}_E$ ($\vec{x}_p$) is the vector from the Sun to the Earth (the planet), and $\hat{n}$ is the unit vector from the planet to Earth (see \Fig{fig:deflection}).

For EHS theories in the autoparallel scheme, $(\mathcal{F}-\mathcal{H})/4=(1-\epsilon)/(1-\sigma)\approx 1+\sigma-\epsilon$, if $\sigma \ll 1$.  For EHS theories in the extremal scheme, $(\mathcal{F}-\mathcal{H})/4=1-\epsilon$.

\subsection{Deflection of light}
\label{subsec:deflec}

As discussed in Appendix \ref{appendix:shapiro}, we assume that a light ray follows an extremal curve (metric geodesic), taking the form 
\ben
\frac{D^{\{\}}u^{\mu}}{D\tau}=\frac{d^2 x^\mu}{d\tau^2}+\left\{\begin{array}{c} \mu \\ \nu\rho \end{array}\right\} \frac{dx^\nu}{d\tau}\frac{dx^\rho}{d\tau}=0\,.
\een
Here $D^{\{\}}/D\tau$ denotes the covariant differentiation using the Levi-Civita connection.

The $\mu=t$ component of the metric geodesic is 
\[ \frac{d^2 t}{d\tau^2}-\mathcal{H}\frac{m}{r^2}\frac{dt}{d\tau}\frac{dr}{d\tau}=0,\]
or, to order $\mathcal{O}(m/r)$, where $m$ is the mass of the Sun deflecting the light,
\[ \frac{d}{d\tau}\left[ (1+\mathcal{H}\frac{m}{r})\frac{dt}{d\tau}\right] = 0\,.\]
Integrating this gives a conserved quantity,
\beq{eqn:k-light}
k\equiv (1+\mathcal{H}\frac{m}{r})\frac{dt}{d\tau} = {\rm const}\,.
\een
The $\mu=\theta$ component of the metric geodesic admits the planar solution $\theta=\pi/2$.  The $\mu=\phi$ component of the metric geodesic, when $\theta=\pi/2$, is
\[ \frac{d^2 \phi}{d\tau^2}+\frac{2}{r}\frac{dr}{d\tau}\frac{d\phi}{d\tau}=0,\]
whose first integral gives another conserved quantity,
\beq{eqn:h-light}
h\equiv r^2 \frac{d\phi}{d\tau}= {\rm const}\,.
\een

For light rays in the equatorial plane $\theta=\pi/2$,
\beq{eqn:light-pi/2}
\frac{ds^2}{d\tau^2}=-\left[1+\mathcal{H}\frac{m}{r}\right]\left(\frac{dt}{d\tau}\right)^2 + \left[1+\mathcal{F}\frac{m}{r}\right]\left(\frac{dr}{d\tau}\right)^2+r^2 \left(\frac{d\phi}{d\tau}\right)^2=0\,.
\een
Note that the $\mu=r$ component of the metric geodesic is not independent of \Eq{eqn:light-pi/2}.  Rewriting $dt/d\tau$ and $d\phi/d\tau$ in terms of $k$ and $h$ via \Eq{eqn:k-light} and \Eq{eqn:h-light}, respectively, and using the fact that $dr/d\tau = (dr/d\phi)(d\phi/d\tau)$, one finds 
\ben
\frac{d^2 u}{d\phi^2}+u=\frac{3}{2}\mathcal{F}mu^2-\frac{k^2}{h^2}\frac{\mathcal{F}+\mathcal{H}}{2}m\,,
\een
where $u\equiv 1/r$.  The solution to order $\mathcal{O}(m)$ is 
\ben
u=\frac{\sin\phi}{D}+\frac{\mathcal{F}m}{2D^2}(1+C\cos\phi+\cos^2\phi)-\frac{k^2}{h^2}\frac{\mathcal{F}+\mathcal{H}}{2}m\,,
\een
where $D$ is the minimal distance of the ray to the Sun.  The $x$-axis is set up to be along the incoming direction of the ray.  $C$ is an arbitrary constant that can be determined at $\phi=\pi$ (incoming infinity).  As long as deflection angle $\delta\ll 1$,
\ben
\delta \simeq \frac{2\mathcal{F}m}{D}-\frac{k^2}{h^2}m(\mathcal{F}+\mathcal{H})D\,.
\een
Using 
\ben
\frac{h}{k}=r^2\frac{d\phi}{dt}(1-\mathcal{H}\frac{m}{r})\approx r^2\frac{d\phi}{dt}=D
\een
is the angular momentum of the light ray relative to the Sun, we finally obtain
\ben
\delta \simeq \frac{\mathcal{F}-\mathcal{H}}{4}\delta^{(GR)}\,,
\een
where $\delta^{(GR)}=4m/D$ is the deflection angle predicted by GR to lowest order.

\subsection{Gravitational Redshift}

As discussed above, we assume that the orbits of light rays are metric geodesics even when there is non-zero torsion.
Non-relativistically, the metric geodesic equation
for a test particle is
\ben
\frac{d\vec{v}}{dt}=-\frac{(-\mathcal{H})}{2}\frac{m}{r^2}\hat{e}_r\,.
\een
Effectively this introduces the gravitational potential $U$, defined by $d\vec{v}/dt=\vec{F}\equiv -\nabla U$,
\ben
U=-\frac{(-\mathcal{H})}{2}\frac{m}{r}\,.
\een
Thus the gravitational redshift of photons is 
\ben
\frac{\Delta\nu}{\nu}=\frac{(-\mathcal{H})}{2}\left(\frac{\Delta\nu}{\nu}\right)^{(GR)}\,,
\een
where $(\Delta\nu/\nu)^{(GR)}$ is the redshift predicted by GR
\ben
\left(\frac{\Delta\nu}{\nu}\right)^{(GR)}=-\frac{m}{c^2}(\frac{1}{r_1}-\frac{1}{r_2})\,.
\een
For EHS theories in the autoparallel scheme, $-\mathcal{H}/2=1/(1-\sigma)\approx 1+\sigma$ for $\sigma \ll 1$.  For EHS theories in extremal scheme,
$-\mathcal{H}/2=1$ exactly.

\subsection{Advance of Mercury's Perihelion in autoparallel scheme}
\label{subsec:advance-perih-auto}

In the autoparallel scheme, a massive test particle (\eg\ a planet in the field of the Sun) follows an autoparallel curve (\ie an affine geodesic). 
We now derive the advance of the perihelion when torsion is present. The autoparallel equation reads
\beq{eqn:auto-perih-appendix}
\frac{Du^{\mu}}{D\tau}=\frac{d^2 x^\mu}{d\tau^2}+\Gamma^\mu_{\phantom{1}\nu\rho} \frac{dx^\nu}{d\tau}\frac{dx^\rho}{d\tau}=0\,,
\een
where $D/D\tau$ is the covariant differentiation by the full connection.

The $\mu=t$ component of \Eq{eqn:auto-perih-appendix} reads
\[ \frac{d^2 t}{d\tau^2}+(t_1-\mathcal{H})\frac{m}{r^2}\frac{dt}{d\tau}\frac{dr}{d\tau}=0,\]
or, to order $\mathcal{O}(m/r)$, where $m$ is the mass of the central gravitating body (\eg\ the Sun),
\[ \frac{d}{d\tau}\left[ \left(1+(\mathcal{H}-t_1)\frac{m}{r}\right)\frac{dt}{d\tau}\right] = 0\,.\]
The integral gives a conserved quantity $k$,
\beq{eqn:k-light}
k\equiv \left(1+(\mathcal{H}-t_1)\frac{m}{r}\right)\frac{dt}{d\tau} = {\rm const}\,.
\een
The $\mu=\theta$ component of \Eq{eqn:auto-perih-appendix} admits the planar solution $\theta=\pi/2$.  
The $\mu=\phi$ component of \Eq{eqn:auto-perih-appendix}, when $\theta=\pi/2$, is
\[ \frac{d^2 \phi}{d\tau^2}+(\frac{2}{r}-t_2\frac{m}{r^2})\frac{dr}{d\tau}\frac{d\phi}{d\tau}=0,\]
whose first integral gives another conserved quantity $h$,
\beq{eqn:h-light}
h\equiv r^2 \frac{d\phi}{d\tau}(1+t_2\frac{m}{r})= {\rm const}\,.
\een
The path parameter $\tau$ can be chosen so that 
\beq{eqn:tau-time-appdenx}
ds^2/d\tau^2=g_{\mu\nu}\frac{dx^{\mu}}{d\tau}\frac{dx^{\nu}}{d\tau}=-1\,.
\een
\Eq{eqn:tau-time-appdenx} is consistent with the autoparallel scheme since $\nabla_\rho g_{\mu\nu}=0$ and $Du^\mu/D\tau=0$.  Note that the $\mu=r$ component of \Eq{eqn:auto-perih-appendix} is not independent of \Eq{eqn:tau-time-appdenx}.  For a test particle in the equatorial plane $\theta=\pi/2$, \Eq{eqn:tau-time-appdenx} reads
\beq{eqn:perih-pi/2}
-\left[1+\mathcal{H}\frac{m}{r}\right](\frac{dt}{d\tau})^2 + \left[1+\mathcal{F}\frac{m}{r}\right](\frac{dr}{d\tau})^2+r^2 (\frac{d\phi}{d\tau})^2=-1\,.
\een

Reusing the trick employed in Appendix \ref{subsec:deflec}, we find
\beqa{eqn:eomdefl1}
\frac{d^2 u}{d\phi^2}+u &=& \frac{3}{2}\mathcal{F}mu^2+\frac{m}{2h^2}\left[k^2(-\mathcal{H}-\mathcal{F}+\right.\nonumber\\
 & & \left. +2t_1+2t_2)+\mathcal{F}-2t_2\right],
\eena
to order $\mathcal{O}(mu)$, where $u\equiv 1/r$.  
Note that to lowest order $k\approx 1+\mathcal{O}(m,(velocity)^2)$, so the second term on the right hand side of \Eq{eqn:eomdefl1} becomes
$(t_1-\mathcal{H}/2)m/h^2$. Since $m$ \emph{is} the physical mass of the central gravitating body, the autoparallel scheme requires
$t_1-\mathcal{H}/2=1$.  Now \Eq{eqn:eomdefl1} becomes 
\ben
\frac{d^2 u}{d\phi^2}+u=\frac{m}{h^2}+\frac{3}{2}\mathcal{F}mu^2\,.
\een
Solve the equation perturbatively in the order of $\varepsilon\equiv (m/h)^2$, \ie use the ansatz $u=u_0+\varepsilon u_1$.  One finds
\bena
u_0 &=& \frac{m}{h^2}(1+e\cos\phi)\,\label{eqn:u0-appen}\\
u_1 &=& \frac{3\mathcal{F}m}{2h^2}\left[1+e\phi\sin\phi+\frac{e^2}{2}(1-\frac{1}{3}\cos 2\phi)\right]\label{eqn:u1-appen}
\eena
\Eq{eqn:u0-appen} gives the classical elliptical orbit with eccentricity $e$ and 
the semi-latus rectum $p\equiv a(1-e^2)=h^2/m$.  
The $\phi\sin\phi$ term in \Eq{eqn:u1-appen}  contributes to the advance of the perihelion, 
while the constant and $\cos 2\phi$ terms do not. Therefore
\beq{eqn:u-appen}
u\approx \frac{m}{h^2}\left\{1+e\cos\left[\phi\left(1-\frac{3\mathcal{F}m^2}{2h^2}\right)\right]\right\}\,.
\een
In \Eq{eqn:u-appen}, we used the fact that the second term inside the cosine 
is $\ll 1$.  The advance of the perihelion is now given by
\bena
\Delta\theta &=& \frac{2\pi}{1- \frac{3\mathcal{F}m^2}{2h^2}}-2\pi\nonumber\\
             &=& \frac{\mathcal{F}}{2}\Delta\theta^{(GR)}\,,\label{eqn:adv-perih-final}
\eena
where $\Delta\theta^{(GR)}=6\pi m^2/h^2 = 6\pi m/p$ is the perihelion advance predicted by GR.

\subsection{Advance of Mercury's Perihelion in extremal scheme}
\label{subsec:advance-perih-extreme}

The extremal scheme assumes that a test particle (\eg, a planet) follows the metric geodesic even though the torsion is present.  Following the same
algebra as in Appendix \ref{subsec:advance-perih-auto}, and noting that $\mathcal{H}=-2$ for the extremal scheme, we finds that the advance of the
perihelion in the extremal scheme has the same bias factor $\mathcal{F}/2$, \ie, \Eq{eqn:adv-perih-final} holds.  

\section{Constraining torsion parameters with the upper bounds on the photon mass } 
\label{appendix:photon-mass}

In this Appendix, we derive the contraints on torsion parameters that result from assuming that 
the ``natural'' extension $\partial_\mu \to \nabla_\mu$ (using the full connection) in the electromagnetic Lagrangian.
This breaks gauge invariance, and the photon generically gains a mass via an 
additional term of the form $-\frac{1}{2}m_\gamma^2 g^{\mu\nu}A_\mu A_\nu$ in the Lagrangian as we will now show. 
The assumption gives 
\ben
F_{\mu\nu}\equiv \nabla_\mu A_\nu - \nabla_\nu A_\mu=f_{\mu\nu}-2S_{\mu\nu}^{\phantom{12}\lambda}A_\lambda\,,
\een
where $f_{\mu\nu}\equiv\partial_\mu A_\nu - \partial_\nu A_\mu$.  
The Maxwell Lagrangian therefore becomes
\bena
\mathcal{L}_{\rm EM} &=& -\frac{1}{4}g^{\mu\alpha}g^{\nu\beta}F_{\mu\nu}F_{\alpha\beta}\,,\nonumber\\
&=& 
-\frac{1}{4}g^{\mu\alpha}g^{\nu\beta}f_{\mu\nu}f_{\alpha\beta}
-K^{\mu\nu}A_\mu A_\nu + S^{\mu\nu\lambda}A_\lambda f_{\mu\nu}\,,\nonumber\\
\eena
where $K^{\mu\nu}\equiv S_{\alpha\beta}^{\phantom{12}\mu}S^{\alpha\beta\nu}$\,.  The Euler-Lagrange equation for the action $S=\int d^4 x \sqrt{-g} \mathcal{L}_{\rm EM}$ yields the following equation of motion for $A_\mu$:
\beq{eqn:cov-A-eom}
\nabla^{\Gamma}_\mu f^{\mu\nu} = 2S_{\mu\lambda}^{\phantom{12}\mu}f^{\lambda\nu} 
+ 2 K^{\lambda\nu}A_\lambda + 2\nabla^{\{\}}_\mu (S^{\mu\nu\lambda}A_\lambda)\,.
\een
Here $\nabla^{\Gamma}_\mu$ and $\nabla^{\{\}}_\mu$ are the covariant derivative w.r.t. the full connection and the Levi-Civita connection, respectively.
Both the 2nd and 3rd terms on the right hand side of \Eq{eqn:cov-A-eom} contain the coupling to $A_\mu$.  To clarify this, 
\Eq{eqn:cov-A-eom} can be rewritten non-covariantly as
\beqa{eqn:noncov-A-eom}
\nabla^{\Gamma}_\mu f^{\mu\nu} &=& 2S_{\mu\lambda}^{\phantom{12}\mu}f^{\lambda\nu} 
+ 2 A_\lambda \left[ K^{\lambda\nu} + \partial_\mu S^{\mu\nu\lambda} \right.\nonumber\\
& & \left. + \left\{ \begin{array}{c} \alpha \\ \alpha\mu \end{array} \right\} S^{\mu\nu\lambda}\right] + 2 S^{\mu\nu\lambda}\partial_\mu A_\lambda\,,
\eena
in which the 2nd term on the right hand side is the direct coupling of $A_\mu$.  

The matrix $K^{\mu\nu}$ is symmetric. If it is also positive definitive up to the metric signature $(-+++)$,
the first term in the square bracket may be identified as the photon mass term. 
In the field of a non-rotating mass, using the parametrization 
(Eqs.~\ref{eqn:t1} and \ref{eqn:t2}), it can be shown that
\bena
K^{00} &=& -\frac{t_1^2 m^2}{2r^4}\,,\\
K^{0i} &=& 0\,,\\
K^{ij} &=& \frac{t_2^2 m^2}{2r^4} \left(\delta_{ij}-\frac{x^i x^j}{r^2}\right)\,.
\eena
The matrix $K$ has the eigenvalues $-\frac{t_1^2 m^2}{2r^4}$, 0 (with eigenvector $\hat{r}$) and 
$\frac{t_2^2 m^2}{2r^4}$ (with 2 degenerate eigenvectors).  Since the metric signature is $(-+++)$, 
all photon masses are positive or zero, The nonzero ones are of order
\ben
m_\gamma \simeq t \frac{m}{r^2} \,,
\een
or (with units reinserted)
\ben
m_\gamma c^2 \simeq t \frac{\hbar G}{c} \frac{m}{r^2} \,.
\een
Here $t=\max(|t_1|,|t_2|)$ and $r$ is the distance of the experiment location to the center of the 
mass $m$ that generates the torsion. For a ground-based experiment here on Earth, this gives 
\ben
t \simeq 4.64\times 10^{22} m_\gamma c^2 /(1\>{\rm eV})\,.
\een
The upper bound on the photon mass from ground-based experiments is $m_\gamma c^2 < 10^{-17}\,{\rm eV}$ \cite{Lakes:1998mi},
so the constraint that this bound places on the dimensionless torsion parameters is quite weak.

Experimentalists can also search for an anomalous electromagnetic force and translate the null results 
into photon mass bounds. To leading order, the anomalous force is 
$2\partial_\mu S^{\mu\nu\lambda} A_\lambda$, since the K-term is proportional to $S^2$, 
while the 2nd term in the square bracket of \Eq{eqn:noncov-A-eom} is proportional to $S$.  
In a field of a non-rotating mass $m$, 
\bena
(\partial_\mu S^{\mu\nu\lambda})^{00} &=& (\partial_\mu S^{\mu\nu\lambda})^{0i} = (\partial_\mu S^{\mu\nu\lambda})^{i0} =0\,,\\
(\partial_\mu S^{\mu\nu\lambda})^{ij} &=& t_2\frac{m}{2r^3}\left(-\delta_{ij}+3\frac{x^i x^j}{r^2}\right)\,,
\eena
which has eigenvalues $\frac{t_2 m}{2r^3}\times (0,-1,-1,2)$.  This 
cannot be identified as a mass term since there must be a negative ``mass squared'' regardless of 
the sign of $t_2$.  However, the anomalous electromagnetic force expressed as a photon mass 
can be estimated as
\ben
m_\gamma c^2 \simeq \sqrt{|t_2| \hbar^2 G \frac{m}{r^3}}\,,
\een
or 
\ben
\sqrt{|t_2|} \simeq 1.23\times 10^{18} m_\gamma c^2 /{\rm eV}\,.
\een
This implies that current  ground-based experimental upper bounds on the photon mass are too weak
(giving merely $|t|\simlt 10^2$, as compared to $|t|=1$ from Hayashi-Shirafuji gravity)
to place constraints on torsion parameters that are competitive with those from GPB.


\begin{thebibliography}{99}

\bibitem{Will:2005yc}
  C.~M.~Will,
  Annalen Phys.\  {\bf 15}, 19 (2005)

\bibitem{will2}
C.~M.~ Will, \emph{Theory and Experiment in Gravitational Physics}, Cambridge University Press (1993).

\bibitem{Will:2005va}
  C.~M.~Will,
  gr-qc/0510072.

\bibitem{Hulse:1974eb}
  R.~A.~Hulse and J.~H.~Taylor,
  Astrophys.~J. {\bf 195}, L51 (1975)

\bibitem{Weisberg:2002qg}
  J.~M.~Weisberg and J.~H.~Taylor,
  astro-ph/0205280.

\bibitem{Weisberg:2004hi}
  J.~M.~Weisberg and J.~H.~Taylor,
  astro-ph/0407149.

\bibitem{Champion:2004hc}
  D.~J.~Champion, \etal,
  Mon.\ Not.\ Roy.\ Astron.\ Soc.\  {\bf 350}, L61 (2004)

\bibitem{Peters:1963ux}
  P.~C.~Peters and J.~Mathews,
  Phys.\ Rev.\  {\bf 131}, 435 (1963)

\bibitem{Stairs:1997kz}
  I.~H.~Stairs {\it et al.},
  astro-ph/9712296.

\bibitem{Stairs:1999dv}
  I.~H.~Stairs, S.~E.~Thorsett, J.~H.~Taylor and Z.~Arzoumanian,
  astro-ph/9911198.

\bibitem{Hotan:2004ua}
  A.~W.~Hotan, M.~Bailes and S.~M.~Ord,
  Mon.\ Not.\ Roy.\ Astron.\ Soc.\  {\bf 355}, 941 (2004)

\bibitem{Hotan:2004ub}
  A.~W.~Hotan, M.~Bailes and S.~M.~Ord,
  Astrophys.\ J.\  {\bf 624}, 906 (2005)

\bibitem{vanStraten:2001zk}
 W.~van Straten,\etal,
  Nature {\bf 412}, 158 (2001)

\bibitem{Armitage:2004ga}
  P.~J.~Armitage,
  Astrophys.\ Space Sci.\  {\bf 308}, 89 (2004)

\bibitem{Chakrabarti:2004uz}
  S.~K.~Chakrabarti,
  astro-ph/0402562.

\bibitem{Merloni:2002gx}
  A.~Merloni,
  astro-ph/0210251.

\bibitem{Menou:2001ga}
  K.~Menou,
  astro-ph/0111469.

\bibitem{Peldan:1993hi}
  P.~Peldan,
  Class.\ Quant.\ Grav.\  {\bf 11}, 1087 (1994)

\bibitem{Sotiriou:2006hs}
  T.~P.~Sotiriou,
  gr-qc/0604028.

\bibitem{Sotiriou:2006qn}
  T.~P.~Sotiriou and S.~Liberati,
  gr-qc/0604006.

\bibitem{Sotiriou:2005cd}
  T.~P.~Sotiriou,
  Class.\ Quant.\ Grav.\  {\bf 23}, 1253 (2006)

\bibitem{Akbar:2006er}
  M.~Akbar and R.~G.~Cai,
  Phys.\ Lett.\ B {\bf 635}, 7 (2006)

\bibitem{Koivisto:2006ie}
  T.~Koivisto,
  Phys.\ Rev.\ D {\bf 73}, 083517 (2006)

\bibitem{Amarzguioui:2005zq}
  M.~Amarzguioui, O.~Elgaroy, D.~F.~Mota and T.~Multamaki,
  astro-ph/0510519.

\bibitem{Hwang:2001pu}
  J.~c.~Hwang and H.~Noh,
  Phys.\ Lett.\ B {\bf 506}, 13 (2001)

\bibitem{Meng:2003en}
  X.~H.~Meng and P.~Wang,
  Phys.\ Lett.\ B {\bf 584}, 1 (2004)

\bibitem{Esposito-Farese:1999pa}
  G.~Esposito-Farese,
  gr-qc/9903058.

\bibitem{Esposito-Farese:2004cc}
  G.~Esposito-Farese,
  AIP Conf.\ Proc.\  {\bf 736}, 35 (2004)

\bibitem{Damour:1996xx}
  T.~Damour,
  gr-qc/9606079.

\bibitem{Puetzfeld:2004yg}
  D.~Puetzfeld,
  New Astron.\ Rev.\  {\bf 49}, 59 (2005)

\bibitem{Kasper:1994xv}
  U.~Kasper, S.~Kluske, M.~Rainer, S.~Reuter and H.~J.~Schmidt,
  gr-qc/9410030.

\bibitem{Biswas:1999fa}
  S.~Biswas, A.~Shaw and D.~Biswas,
  gr-qc/9906074.

\bibitem{Mukherjee:2005zt}
  S.~Mukherjee, B.~C.~Paul, S.~D.~Maharaj and A.~Beesham,
  gr-qc/0505103.

\bibitem{Beesham:1987gd}
  A.~Beesham, N.~A.~Hassan and S.~D.~Maharaj,
  Europhys.\ Lett.\  {\bf 3} (1987) 1053.

\bibitem{Mahato:2006gi}
  P.~Mahato,
  gr-qc/0603134.


\bibitem{Blagojevic:book}
 M.~Blagojevi\'c, \emph{Gravitation and Gauge Symmetries}, Taylor \& Francis, 2001

\bibitem{hammond} 
R.~T.~Hammond, Rep. Prog. Phys. {\bf 65}, 599 (2002)

\bibitem{Gronwald:1995em}
  F.~Gronwald and F.~W.~Hehl,
  gr-qc/9602013.

\bibitem{Hehl:1997bz}
  F.~W.~Hehl,
  gr-qc/9712096.

\bibitem{DeAndrade}
V.~C.~De~Andrade, L.~C.~T.~Guillen and J.~G.~Pereira, gr-qc/0011087.

\bibitem{aldrovandi}
R.~Aldrovandi, J.~G.~Pereira, K.~H.~Vu, Braz. J. Phys. {\bf 34}, 1374 (2004)

\bibitem{hehl} 
F.~W.~Hehl, P.~von~der~Heyde, G.~D.~Kerlick and J.~M.~Nester, Rev. Mod. Phys {\bf 48}, 393 (1976)

\bibitem{nester}
J.~M.~Nester, Phys. Rev. {\bf D16}, 2396 (1977)

\bibitem{watanabe}
T.~Watanabe and M.~J.~Hayashi, gr-qc/0409029


\bibitem{capozziello}
  S.~Capozziello, G.~Lambiase and C.~Stornaiolo,
  Annalen Phys.\  {\bf 10}, 713 (2001)

\bibitem{Gasperini:1986mv}
  M.~Gasperini,
  Phys.\ Rev.\ Lett.\  {\bf 56}, 2873 (1986).

\bibitem{Shapiro:2001rz}
  I.~L.~Shapiro,
  Phys.\ Rept.\  {\bf 357}, 113 (2002)

\bibitem{Baekler:2006de}
  P.~Baekler and F.~W.~Hehl,
  gr-qc/0601063.

\bibitem{Poltorak:2004tz}
  A.~Poltorak,
  gr-qc/0407060.

\bibitem{Mielke:2004gg}
  E.~W.~Mielke,
  Phys.\ Rev.\ D {\bf 69}, 128501 (2004).

\bibitem{Kleyn:2004yj}
  A.~Kleyn,
  gr-qc/0405028.

\bibitem{Vassiliev:2003dk}
  D.~Vassiliev,
  Annalen Phys.\  {\bf 14}, 231 (2005)

\bibitem{Minkevich:2003it}
  A.~V.~Minkevich and Y.~G.~Vasilevski,
  gr-qc/0301098.

\bibitem{Obukhov:2002tm}
  Y.~N.~Obukhov and J.~G.~Pereira,
  Phys.\ Rev.\ {\bf D67}, 044016 (2003)

\bibitem{Tresguerres:1995un}
  R.~Tresguerres,
  Phys.\ Lett.\ A {\bf 200}, 405 (1995).

\bibitem{King:2000ha}
  A.~D.~King and D.~Vassiliev,
  Class.\ Quant.\ Grav.\  {\bf 18}, 2317 (2001)

\bibitem{Gronwald:1997bx}
  F.~Gronwald,
  Int.\ J.\ Mod.\ Phys.\ D {\bf 6}, 263 (1997)

\bibitem{Hehl:1999sb}
  F.~W.~Hehl and A.~Macias,
  Int.\ J.\ Mod.\ Phys.\ D {\bf 8}, 399 (1999)

\bibitem{Poberii:1994rz}
  E.~A.~Poberii,
  Gen.\ Rel.\ Grav.\  {\bf 26}, 1011 (1994).

\bibitem{Gruver:2001tt}
  C.~Gruver, R.~Hammond and P.~F.~Kelly,
  Mod.\ Phys.\ Lett.\ A {\bf 16}, 113 (2001)

\bibitem{Berthias:1993aa}
  J.~P.~Berthias and B.~Shahid-Saless,
  Class.\ Quant.\ Grav.\  {\bf 10}, 1039 (1993)

\bibitem{Lam:2002ve}
  Y.~Y.~Lam,
  gr-qc/0211009.

\bibitem{Saa:1993fx}
  A.~Saa,
  gr-qc/9309027.

\bibitem{Hehl:1994ue}
  F.~W.~Hehl, J.~D.~McCrea, E.~W.~Mielke and Y.~Neeman,
  Phys.\ Rept.\  {\bf 258}, 1 (1995)

\bibitem{Lord:1978qz}
  E.~A.~Lord,
  Phys.\ Lett.\ A {\bf 65}, 1 (1978).

\bibitem{Hehl:1976my}
  F.~W.~Hehl, G.~D.~Kerlick and P.~Von Der Heyde,
  Phys.\ Lett.\ B {\bf 63}, 446 (1976).

\bibitem{Rajaraman:2003st}
  A.~Rajaraman,
  astro-ph/0311160.

\bibitem{Vollick:2003ic}
  D.~N.~Vollick,
  Class.\ Quant.\ Grav.\  {\bf 21}, 3813 (2004)

\bibitem{Chiba:2003ir}
  T.~Chiba,
  Phys.\ Lett.\ B {\bf 575}, 1 (2003)

\bibitem{Antoniadis:1992ep}
  I.~Antoniadis and S.~D.~Odintsov,
  Mod.\ Phys.\ Lett.\ A {\bf 8}, 979 (1993)

\bibitem{Bytsenko:1993qn}
  A.~A.~Bytsenko, E.~Elizalde and S.~D.~Odintsov,
  Prog.\ Theor.\ Phys.\  {\bf 90} 677 (1993) 
 
\bibitem{Fabbri:2006xq}
  L.~Fabbri, 
  gr-qc/0608090.

\bibitem{Carroll:1994dq}
  S.~M.~Carroll and G.~B.~Field,
  Phys.\ Rev.\ D {\bf 50}, 3867 (1994)

\bibitem{weitzenbock} 
R.~Weitzenb\"ock, \emph{Invariantentheorie} (Groningen, P. Noordhoff, 1923); Chap. XIII, Sec.7

\bibitem{HS1}
K.~Hayashi and T.~Shirafuji, Phys. Rev. {\bf D19}, 3524 (1979)

\bibitem{schiff} 
L.~I.~Schiff, Phys. Rev. Lett., {\bf 4}, 215 (1960); Proc. Nat. Acad. Sci. USA,  {\bf 46}, 871 (1960)

\bibitem{Will:2002ma}
  C.~M.~Will,
  Phys.\ Rev.\ D {\bf 67}, 062003 (2003)

\bibitem{Adler:1999yt}
  R.~J.~Adler and A.~S.~Silbergleit,
  Int.\ J.\ Theor.\ Phys.\  {\bf 39}, 1291 (2000)

\bibitem{biemond:2004}
 J.~Biemond, physics/0411129 (2004)

\bibitem{Ashby:1990}
N.~Ashby and B.~Shahid-Saless, Phys. Rev. {\bf D42}, 1118 (1990)

\bibitem{Barker:1970zr}
  B.~M.~Barker and R.~F.~O'Connell,
  Phys.\ Rev.\ D {\bf 2}, 1428 (1970).

\bibitem{Moffat:2004cv}
  J.~W.~Moffat,
  gr-qc/0405091.

\bibitem{MTW-Weinberg-Islam:book}
See, \eg, C.~W.~Misner, K.~S.~Thorne, J.~A.~Wheeler, \emph{Gravitation} (W.~H.~Freeman, 1973); \nn Weinberg S, \emph{Gravitation and Cosmology : Principles and Applications of the General Theory of Relativity} (Wiley  1972). 

\bibitem{kerr} 
R.P.~Kerr, Phys. Rev. Lett {\bf 11}, 237 (1963)

\bibitem{boyer-lind} 
R.~H.~Boyer, R.~W.~Lindquist, J. Math. Phys. {\bf 8}, 265 (1967)

\bibitem{deSitter:1916}
W.~de~Sitter, M.~N.~Roy.~Astron.~Soc. {\bf 76}, 699 (1916)

\bibitem{Lense-Thirring:1918}
J.~Lense and H.~Thirring, Phys. Zeits. {\bf 19}, 156 (1918)

\bibitem{Einstein:1928-30}
 A.~Einstein, (a) Sitzungsber. Preuss. Akad. Wiss. 217 (1928); (b) 224 (1928); (c) 2 (1929); (d) 156 (1929); (e) 18 (1930); (f) 401 (1930).

\bibitem{Einstein:1930b}
A.~Einstein, W.~Mayer, Sitzungsber. Preuss. Akad. Wiss., 110 (1930)

\bibitem{moller:1961}
 C.~M\o ller, K. Dan. Vidensk. Selsk. Mat. Fys. Skr.,{\bf 1}, No.10 (1961).

\bibitem{Pellegrini:1962}
 C.~Pellegrini, J.~Plebanski, K. Dan. Vidensk. Selsk. Mat. Fys. Skr., {\bf 2}, No.4 (1962).

\bibitem{moller:1978}
 C.~M\o ller, K. Dan. Vidensk. Selsk. Mat. Fys. Skr.,{\bf 89}, No.13 (1978).

\bibitem{Unzicker:2005in}
  A.~Unzicker and T.~Case,
  physics/0503046.

\bibitem{Obukhov:2004hv}
  Y.~N.~Obukhov and J.~G.~Pereira,
  Phys.\ Rev.\ D {\bf 69}, 128502 (2004)

\bibitem{Vargas:1992ab}
  J.~G.~Vargas,
  Found.\ Phys.\  {\bf 22}, 507 (1992).

\bibitem{Vargas:1992ac}
  J.~G.~Vargas, D.~G.~Torr and A.~Lecompte,
  Found.\ Phys.\  {\bf 22}, 527 (1992).

\bibitem{Mueller-Hoissen:1983vc}
  F.~Muller-Hoissen and J.~Nitsch,
  Phys.\ Rev.\ D {\bf 28}, 718 (1983).

\bibitem{Mielke:1992te}
  E.~W.~Mielke,
  Annals Phys.\  {\bf 219}, 78 (1992).

\bibitem{Kreisel:1979kh}
  E.~Kreisel,
  Annalen Phys.\  {\bf 36}, 25 (1979).

\bibitem{Treder:1978vf}
  H.~J.~Treder,
  Annalen Phys.\  {\bf 35}, 377 (1978).

\bibitem{Kreisel:1980kb}
  E.~Kreisel,
  Annalen Phys.\  {\bf 37}, 301 (1980).

\bibitem{Pimentel:2004bp}
  B.~M.~Pimentel, P.~J.~Pompeia and J.~F.~da Rocha-Neto,
  Nuovo Cim.\  {\bf 120B}, 981 (2005)

\bibitem{Maluf:2001ef}
  J.~W.~Maluf and A.~Goya,
  Class.\ Quant.\ Grav.\  {\bf 18}, 5143 (2001)

\bibitem{Arcos:2004zh}
  H.~I.~Arcos and J.~G.~Pereira,
  Class.\ Quant.\ Grav.\  {\bf 21}, 5193 (2004)

\bibitem{Arcos:2005ec}
  H.~I.~Arcos and J.~G.~Pereira,
  Int.\ J.\ Mod.\ Phys.\ D {\bf 13}, 2193 (2004)

\bibitem{sy1} 
W.~R.~Stoeger, P.~B.~Yasskin, Gen. Rel. Grav., {\bf 11}, 427 (1979)

\bibitem{sy2} 
P.~B.~Yasskin, W.~R.~Stoeger, Phy. Rev. {\bf D21}, 2081 (1980)

\bibitem{NSH}
K.~Nomura, T.~Shirafuji and K.~Hayashi, Prog. Theor. Phys., {\bf 86}, 1239 (1991)

\bibitem{Kleinert:1996yi}
  H.~Kleinert and A.~Pelster,
  Gen.\ Rel.\ Grav.\  {\bf 31}, 1439 (1999)

\bibitem{Kleinert:1998as}
  H.~Kleinert and A.~Pelster,
  Acta Phys.\ Polon.\ B {\bf 29}, 1015 (1998)

\bibitem{D'Eath1975a}
P.~D.~D'Eath, Phys. Rev. {\bf D11}, 1387 (1975)

\bibitem{D'Eath1975b}
P.~D.~D'Eath, Phys. Rev. {\bf D12}, 2183 (1975)

\bibitem{Hojman} 
S.~Hojman, Phys. Rev. {\bf D18}, 2741 (1978)


\bibitem{Hojman2} 
S.~Hojman, M.~Rosenbaum and M.~P.~Ryan, Phys. Rev. {\bf D19}, 430 (1979)

\bibitem{cognola} 
G.~Cognola, R.~Soldati, L.~Vanzo, S.~Zerbini, Phys. Rev. {\bf D25}, 3109 (1982)

\bibitem{kopczynski} 
W.~Kopczynski, Phys. Rev. {\bf D34}, 352 (1986)

\bibitem{pereira1} 
H.~I.~Arcos, V.~C.~de~Andrade and J.~G.~Pereira, Int. J. Mod. Phys. {\bf D13}, 807 (2004)

\bibitem{robertson} 
H.~P.~Robertson, Ann. of Math., {\bf 33}, 496 (1932)

\bibitem{Hayashi-Bregman} 
K.~Hayashi and A.~Bregman, Ann. Phys., {\bf 75}, 562 (1973)

\bibitem{papapetrou} 
A.~Papapetrou, Proc. Roy. Soc. {\bf A209}, 248 (1951)

\bibitem{adamowicz-trautman}
W.~Adamowicz, A.~Trautman, Bull. Acad. Polon. Sci., Ser. Sci. Math. Astr. Phys. {\bf 23}, 339 (1975)

\bibitem{Bertotti:2003rm}
  B.~Bertotti, L.~Iess and P.~Tortora,
  Nature {\bf 425} 374 (2003).

\bibitem{Shapiro2004}
S.~S.~Shapiro, J.~L.~Davis, D.~E.~Lebach and J.~S.~Gregory, Phys. Rev. Lett., {\bf 92}, 121101 (2004)

\bibitem{vessot1980}
R.~F.~C.~Vessot \etal, Phys. Rev. Lett., {\bf 45}, 2081 (1980)

\bibitem{Lakes:1998mi}
  R.~Lakes, Phys.~Rev.~Lett.\  {\bf 80}, 1826 (1998).
  
\bibitem{Flanagan:2007dc}
  E.~E.~Flanagan and E.~Rosenthal,
  arXiv:0704.1447 [gr-qc].
  
\bibitem{Stelle:1979aj}
  K.~S.~Stelle and P.~C.~West,
  Phys.~Rev.~D {\bf 21}, 1466 (1980).

\bibitem{Pagels:1983pq}
  H.~R.~Pagels,
  Phys.~Rev.~D {\bf 29}, 1690 (1984).

\end{thebibliography}
\end{document}